\documentclass[showpacs,aps,prc,twocolumn,superscriptaddress,letterpaper]{revtex4}
\usepackage{graphicx}
\usepackage{ulem}
\usepackage{multirow}
\usepackage{rotating}
\usepackage{booktabs}
\usepackage{tabularx}
\usepackage[usenames,dvipsnames]{color}

\newcommand{\he}{\ensuremath{^3\mathrm{He\ }}}
\newcommand{\nm}{\ensuremath{\mathrm{N}_2\ }}
\newcommand{\hens}{\ensuremath{^3\mathrm{He}}}
\newcommand{\nmns}{\ensuremath{\mathrm{N}_2}}
\newcommand{\pc}{\ensuremath{\mathrm{pc}}}
\newcommand{\tc}{\ensuremath{\mathrm{tc}}}
\newcommand{\rb}{\ensuremath{\mathrm{Rb}}}
\newcommand{\kp}{\ensuremath{\mathrm{K}}}
\newcommand{\al}{\ensuremath{\mathrm{A}}}
\newcommand{\op}{\ensuremath{\mathrm{op}}}
\newcommand{\se}{\ensuremath{\mathrm{se}}}
\newcommand{\rf}{\ensuremath{\mathtt{rf} }}
\newcommand{\rbs}{\ensuremath{\mathrm{Rb}}\ }
\newcommand{\kps}{\ensuremath{\mathrm{K}}\ }
\newcommand{\etal}{{\it et al.\ }}
\newcommand{\pal}{\ensuremath{P_\al^\ell}}

\begin{document}

\title{The Development of High-Performance Alkali-Hybrid Polarized $\mathbf{^3}$He Targets for Electron Scattering}


\author{Jaideep T. Singh\footnote{present address: National Superconduting Cyclotron Laboratory, Michigan State University, East Lansing, MI 48824}}
\email[]{singhj@nscl.msu.edu}
\affiliation{Department of Physics, University of Virginia, Charlottesville, VA 22903}
\affiliation{Physics Division, Argonne National Laboratory, Argonne, IL 60439}
\affiliation{Technische Universit\"{a}t M\"{u}nchen, Exzellenzcluster Universe, 85748 Garching, Germany}
\author{P.A.M.~Dolph}
\affiliation{Department of Physics, University of Virginia, Charlottesville, VA 22903}
\author{W.A.~Tobias}
\affiliation{Department of Physics, University of Virginia, Charlottesville, VA 22903}
\author{T.D.~Averett}
\affiliation{Department of Physics, College of William and Mary, Williamsburg, VA 23187}
\author{A.~Kelleher}
\affiliation{Department of Physics, College of William and Mary, Williamsburg, VA 23187}
\author{K.E.~Mooney\footnote{Present address: Department of Radiation Oncology, Washington University, Saint Louis, MO 63110}}
\affiliation{Department of Physics, University of Virginia, Charlottesville, VA 22903}
\author{V.V.~Nelyubin}
\affiliation{Department of Physics, University of Virginia, Charlottesville, VA 22903}
\author{Yunxiao~Wang}
\affiliation{Department of Physics, University of Virginia, Charlottesville, VA 22903}
\author{Yuan~Zheng}
\affiliation{Department of Physics, University of Virginia, Charlottesville, VA 22903}
\author{G.D.~Cates}
\affiliation{Department of Physics, University of Virginia, Charlottesville, VA 22903}


\date{\today}

\begin{abstract}
We present the development of high-performance polarized \he targets for use in electron scattering experiments that utilize the technique of alkali-hybrid spin-exchange optical pumping.
We include data obtained during the characterization of 24 separate target cells, each of which was constructed while preparing for one of four experiments at Jefferson Laboratory in Newport News, Virginia.
The results presented here document dramatic improvement in the performance of polarized \he targets, as well as the target properties and operating parameters that made those improvements possible.
Included in our measurements were determinations of the so-called $X$-factors that quantify a temperature-dependent and as-yet poorly understood spin-relaxation mechanism 
that limits the maximum achievable $^3$He polarization to well under 100\%.
The presence of this spin-relaxation mechanism was clearly evident in our data.
We also present results from a simulation of the alkali-hydrid spin-exchange optical pumping process that was developed to provide guidance in the design of these targets.  
Good agreement with actual performance was obtained by including details such as off-resonant optical pumping.
Now benchmarked against experimental data, the simulation is useful for the design of future targets.
Included in our results is a measurement of the \kp-\he spin-exchange rate coefficient $k^\kp_\se = \left ( 7.46 \pm 0.62 \right )\!\times\!10^{-20}\ \mathrm{cm^3/s}$ over the temperature range 503 K to 563 K.
\end{abstract}


\maketitle


\section{Introduction\label{sec:introduction}}
Nuclear spin-polarized \he targets have proven to be exceptionally useful in electron scattering for measurements of spin-dependent observables involving the neutron.  
This is due to the fact that the ground-state \he nuclear wave function is dominated by a configuration in which the proton spins are antialigned, and the spin of the \he nucleus is, to a good approximation, given by the spin of its sole neutron.
Examples of the physics investigated using polarized \he include the spin structure of the neutron~\cite{PhysRevLett.71.959}, 
the $Q^2$ dependence of the generalized Gerasimov-Drell-Hearn (GDH) integral~\cite{ama02}, 
the electric form factor of the neutron~\cite{rio10}, 
and single-spin asymmetries in semi-inclusive deep inelastic scattering (SIDIS)~\cite{sidis}.  
These experiments, spanning almost two decades, have been made possible by significant advances in the performance of polarized \he targets.
These advances have been due to both an improved understanding of the underlying physics of the targets as well as technological advances including, for example, dramatic progress in the capabilities of commercially available lasers.

Spin-exchange optical pumping (SEOP) is one of two techniques that are widely used to spin polarize \he for use as a nuclear target~\cite{chu87,larson,new91}.
The other technique is metastability-exchange optical pumping~\cite{col63,nac85}, but is not the subject of this work.
First demonstrated in 1960~\cite{bou60}, SEOP is a two step process in which an alkali-metal (or ``alkali'' for short) vapor is polarized using optical pumping which subsequently polarizes noble-gas nuclei via spin-exchange collisions \cite{rmp-seop}.
Historically, a pure rubidium (\rb) vapor was used to polarize \he for nuclear targets.
However, both calculations~\cite{spinrot97} and measurements~\cite{bar98} have shown that potassium (\kp) is far more efficient than \rb\ at transferring its polarization to \he nuclei.
This led to the use of hybrid mixtures of \rbs and \kps for improving the efficiency of the polarization process, a technique we will refer to as alkali-hybrid spin exchange optical pumping (AHSEOP)~\cite{hybridpatent,PhysRevLett.91.123003}. 
In AHSEOP, the \rbs vapor is still polarized by optical pumping, but the \rbs polarization is then rapidly shared with the \kp.
The exchange of polarization between \rbs and \kps atoms is sufficiently fast that the polarizations of the two vapors are nearly identical.
If the alkali-hybrid mixture is chosen so that there is significantly more \kps than \rb, the spin-exchange efficiency is greatly improved even though it is still \rbs that is being optically pumped.
For a given amount of laser power, the higher efficiency means that the rate at which \he is polarized can be significantly increased. The use of AHSEOP has indeed had a dramatic effect on target performance.

Another factor that greatly improved performance was the introduction of spectrally-narrowed diode lasers \cite{chann-narrow},
something that for several reasons boosts the maximum alkali polarization that is achievable and thereby also reduces the required laser power.
We will discuss in some detail the impact on target performance of both AHSEOP and spectrally-narrowed diode lasers.

In this work, we present data collected while developing and characterizing 24 glass target cells, each of which was constructed in preparation for one of four polarized \he experiments performed in experimental Hall A at Jefferson Laboratory (JLab) in Newport News, Virginia.
Those experiments included the Small Angle GDH experiment (E97-110, referred to herein as saGDH, which ran in 2003)~\cite{Sulkosky:2009zza}, a measurement of the electric form factor of the neutron, $G_E^n$, at high $Q^2$  (E02-013, referred to herein as GEN, which ran in 2006)~\cite{rio10}, an experiment to measure single-spin asymmetries in semi-inclusive deep inelastic scattering (E06-010, referred to herein as Transversity, which ran in late 2008 and early 2009)~\cite{sidis}, and an experiment to measure the twist three matrix element $d_2^n$ (E06-014, referred to herein simply as $d_2^n$, which ran in 2009).
In all cases the target cells included two chambers: a pumping chamber, in which the $^3$He was polarized, and a target chamber, through which the electron beam passed.  
The two chambers were connected by a single ``transfer tube" through which the polarized $^3$He diffused.
All but the target cells made for the saGDH experiment utilized alkali-hybrid mixtures for improved performance.

The results presented here document dramatic improvement in the performance of polarized \he targets, as well as the target properties and operating parameters that made those improvements possible. 
The data include the \he polarization achieved under various operating conditions, the time constants associated with the polarization process, and data characterizing the properties of the target cell itself, such as pressure, the ratio of \kps to \rbs in the alkali-hybrid mixture, and spin-relaxation rates that are intrinsic to the cell.
In roughly half the cells studied, we also measured the polarization and density of the alkali vapor using Faraday rotation techniques.
The results presented here summarize several thousand hours of data taking, and provide a valuable basis upon which to design and build the next generation of \he targets.  

In addition to {\it direct} measurements of target-cell properties and target cell performance, it is possible to obtain a particular {\it derived} measurement of a cell property that is critical to target performance.
In 2006, Babcock \etal reported evidence for a previously unrecognized spin-relaxation mechanism that limits the polarization when using SEOP to polarize \he~\cite{PhysRevLett.96.083003}. 
This spin-relaxation mechanism is temperature dependent, and empirically, appears to be roughly proportional to the alkali density.  The authors thus characterized the newly recognized spin-relaxation mechanism by a dimensionless parameter referred to simply as $X$, and showed that the maximum polarization that can be achieved in a target is $1/(1+X)$. The parameter $X$ can vary significantly from cell to cell, so it was important to us to measure $X$ in our target cells.
We note that in the course of our studies we have observed what appears to be hints of a temperature dependence of the $X$ parameter,
which would imply that the temperature dependence of the spin-relaxation mechanism characterized by $X$ is not exactly the same as the temperature dependence of the spin-exchange rate,
a possibility that was indeed pointed out by Babcock \etal in Ref.~\cite{PhysRevLett.96.083003}. 

\begin{figure}[htbp]
\begin{center}
\includegraphics[width = 8.6cm]{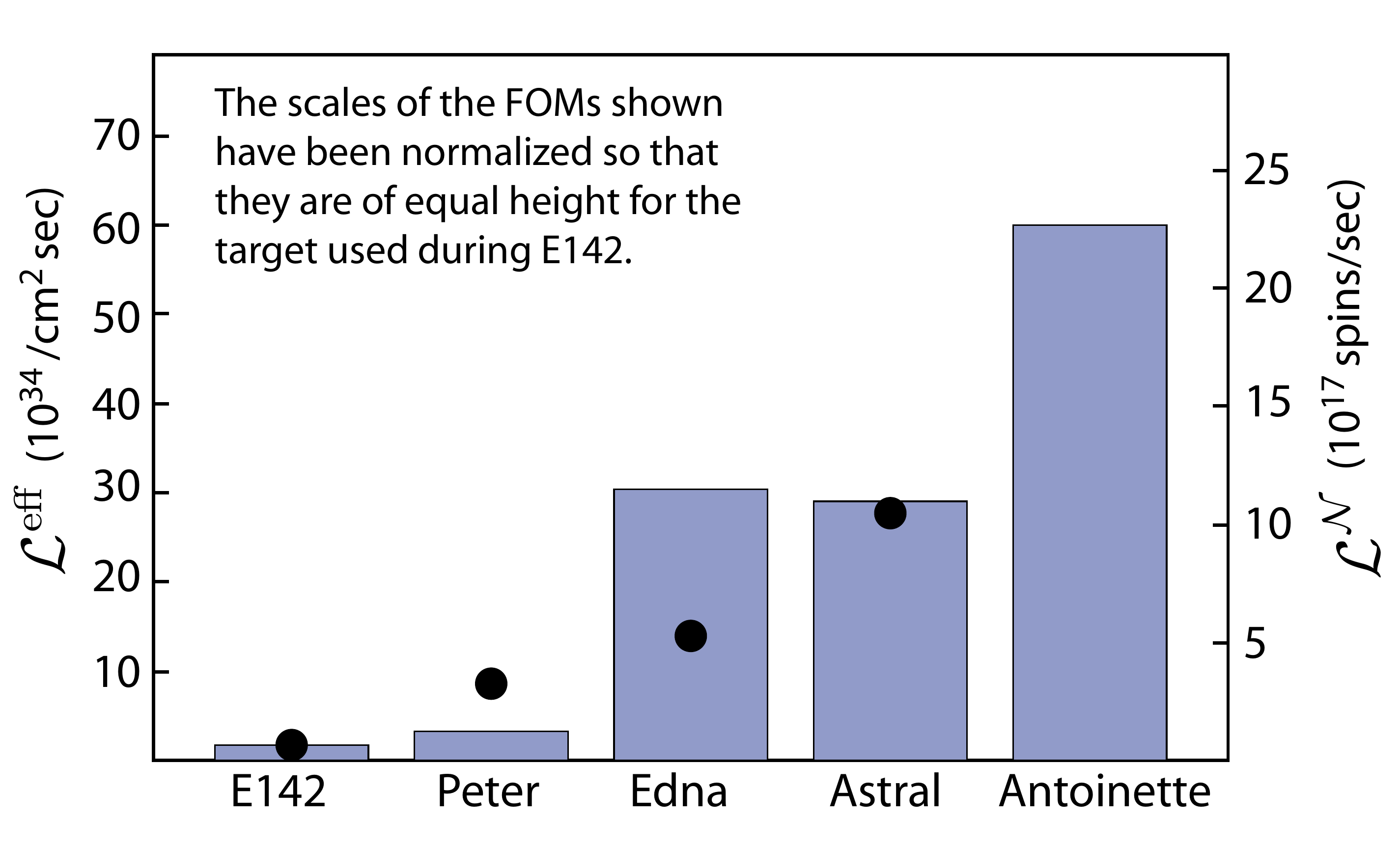}
\vskip -0.1truein
\caption{(color online) Shown are two figures of merit (FOM) for five polarized \he targets.
The solid circles (left-hand scale) indicate the luminosity weighted by \he polarization squared ($P_\mathrm{He}^2$) achieved in beam.
The shaded columns (right-hand scale) show a FOM proportional to the total number of spins polarized per second, again weighted by $P_\mathrm{He}^2$.
The target labeled E142 was used during the experiment reported in Ref.~\cite{PhysRevLett.71.959}.
The other targets are described in some detail in this paper.
The scales have been normalized so that the two FOMs have the same height for the cell marked E142.}
\label{fig:foms}
\end{center}
\end{figure}

An illustration of the improvements that have been achieved is given by Fig.~\ref{fig:foms}, in which two relevant figures of merit (FOM) are plotted for five different target cells. 
One FOM is the effective luminosity $\mathcal{L}^\mathrm{eff} = \mathcal{L}\,P_\mathrm{He}^2$,
where $\mathcal{L}$ is the usual luminosity for a fixed-target experiment (i.e. the product of beam current, target density, and interaction length) 
and $P_\mathrm{He}$ is the $^3$He polarization.
The luminosity $\mathcal{L}$ accounts for the actual number of scattering opportunites per unit time per unit area, whereas $P_\mathrm{He}^2$ accounts for the reduction in the statistical precision of some polarization-dependant asymmetry that is sought after.
The quantity that currently limits the number of scattering opportunities in the type of electron scattering experiments described earlier is the number of target spins that can polarized per unit time. 
In order to quantify the {\it potential} effective luminosity of a target, we define another FOM as $\mathcal{L}^\mathcal{N} \equiv\mathcal{N}\,\Gamma_\mathrm{s}\,P_\mathrm{He}^2$, where $\mathcal{N}$ is the total number of $^3$He atoms in the target, 
and $\Gamma_\mathrm{s}$, defined by Eqn.~(\ref{eqn:gammaS}), is a rate that characterizes the buildup of polarization.  
Fig.~\ref{fig:foms} suggests that the target Antoinette could tolerate even higher luminosities than have already been achieved, although this particular target was never used in beam.  
Most of the data used to construct Fig.~\ref{fig:foms} can be found in Sections~\ref{sec:expmeth} and \ref{sec:targetproperties}.
\footnote{In calculating $\mathcal{L}^\mathrm{eff}$ we have assumed, for the four cases shown,  beam currents of 3.3, 8, 8, and 12$\,\mu\rm A$, and polarizations of 0.30, 0.39, 0.47 and 0.55, respectively.}  

We note that the large values of $\mathcal{L}^\mathcal{N}$ observed in cells such as Antoinette, suggesting untapped potential, inspired the development of a new style of target cell, described recently by Dolph \etal~\cite{convection},  in which gas transfer between the pumping and target chambers is accomplished much more rapidly by using convection instead of diffusion.  Faster gas transfer is desirable because in experiments such as Transversity (which used the cell ``Astral" in Fig.~\ref{fig:foms}) the high luminosity caused significant ``polarization gradients''  ($\sim10\%$ relative) between the pumping and target chambers.  Without convection-based targets, the polarization gradients in several future approved experiments would be significantly more severe.

We begin in Section~\ref{sec:theory} by discussing the theory and rate equations underlying alkali-hybrid optical pumping, and a computer simulation that guided us in designing our targets.
The simulation, which incorporates several important effects that influence the ``photon demand'' in our targets, provides a fairly realistic description of the optical pumping process, and insight regarding how to best optimize performance.
It clearly shows, for example, the optimal range for the ratio of the \kps to \rbs number density and the relationship of laser power \& spectral width to target performance.
The simulation also provides a valuable link between the ``line-averaged'' alkali polarization, which we accessed experimentally, and the ``volume-averaged'' alkali polarization, which is the important quantity in determining the \he polarization.
In Section~\ref{sec:expmeth}, we describe our experimental methods for both the construction of our target cells, including how we prepare alkali-hybrid mixtures, and our measurements.
In Section~\ref{sec:targetproperties}, we present data on target performance and compare the observed trends to those evident in our simulations.
In Section~\ref{sec:kse}, we use our data to extract a value of the rate constant that quantifies \kp-\he spin exchange.
In Section~\ref{sec:X}, we present our studies of the \he polarization-limiting spin-relaxation mechanism that is characterized by the parameter $X$.
\section{Theory and Simulations\label{sec:theory}}
The spatial and spectral profile of the light used for optical pumping is modified nonlinearly as it propagates through the alkali vapor.
When insufficient laser intensity or spectrally broad lasers are used for optical pumping, the optical pumping rate and, consequently, the alkali polarization can vary dramatically throughout the pumping chamber.
Ultimately, because the spin-exchange rate is relatively slow compared to the \he diffusion rate, the \he polarization depends only on the volume-averaged alkali polarization.
A computer simulation was developed to better understand the influence of the various factors on the alkali polarization.
We describe the underlying theory, simulation, and results below.

\subsection{Alkali-Hybrid Optical Pumping\label{sec:ahop}}
In what follows, the two main simplifying assumptions are (1) the alkali nuclear spin is fully conserved during the optical pumping cycle and (2) the alkali excited state multipole moments are zero.
Violations of these assumptions are discussed in Sec.~\ref{sec:estXA}.
Under these two assumptions, the density matrix of the alkali vapor is described by the ground state \& excited state populations of the optically pumped alkali species and the ground state polarizations of the two alkali species.
The coupled differential equations that describe the time dependence of these quantities are:
\begin{equation}
\dot{p} = 
s R_{01} 
- p \left [ R_{01} + \frac{1}{\tau_{p0}} + \Gamma_q + \Gamma_m \right ]
+ d \Gamma'_m 
- s P_\mathrm{Rb} R_1 \label{eqn:pdiff}
\end{equation}
\begin{equation}
\dot{d} = 
s R_{02} 
+ p \Gamma_m
- d \left [ \frac{R_{02}}{2} + \frac{1}{\tau_{d0}} + \Gamma'_q + \Gamma'_m \right ]
+ \frac{s P_\mathrm{Rb} R_2}{2}
\end{equation}
\begin{widetext}
\begin{equation}
\dot{( s P_\mathrm{Rb} )} = 
s \left [ R_{1}(1-X'_a) -\frac{R_2}{2}\right ]
- p R_1 
+ \frac{d R_2}{4}
- s P_\mathrm{Rb} \left ( R_{01}(1+X_a) + R_{02} + \Gamma_\mathrm{Rb} + k_\mathrm{K}[\mathrm{K}] \right ) 
+ A_\mathrm{se} [\mathrm{K}] \left ( P_\mathrm{K} - s P_\mathrm{Rb} \right )
\end{equation}
\begin{equation}
\dot{P}_\mathrm{K} =  
\left [ K_{1} -\frac{K_2}{2}\right ]
- P_\mathrm{K} \left ( K_0 + \Gamma_\mathrm{K} + k'_\mathrm{Rb}[\mathrm{Rb}] \right ) 
+ A_\mathrm{se} [\mathrm{Rb}] \left ( sP_\mathrm{Rb} - P_\mathrm{K} \right ) \label{eqn:pkdiff}
\end{equation}
\end{widetext}
where $s=1-p-d$ is the population of the $5S_{1/2}$ ground state of Rb, 
$p$ ($d$) is the population of the $5P_{1/2}$ ($5P_{3/2}$) excited state of Rb, 
$P_\mathrm{Rb}$ ($P_\mathrm{K}$) is the Rb (K) ground state polarization,
$R_{01}$ \& $R_{02}$ ($K_{01} \& K_{02}$) are the Rb (K) D1 \& D2 unpolarized optical-pumping rates, 
$R_1$ \& $R_2$ ($K_1$ \& $K_2$) are the Rb (K) D1 \& D2 polarized optical-pumping rates, 
$K_0 = K_{01}+K_{02}$ ($R_0 = R_{01} + R_{02}$) is the total K (Rb) unpolarized optical-pumping rate, 
$\tau_{p0}$ ($\tau_{d0}$) is the natural radiative lifetime of the $5P_{1/2}$ ($5P_{3/2}$) excited state,
 $\Gamma_q$ ($\Gamma'_q$) is the \nm nonradiative quenching rate of the $5P_{1/2}$ ($5P_{3/2}$) excited state,
$\Gamma_m$ ($\Gamma'_m$) is the transfer rate from the $5P_{1/2}$ ($5P_{3/2}$) state to the $5P_{3/2}$ ($5P_{1/2}$) state,
$\Gamma_\rb$ ($\Gamma_\kp$) is the Rb (K) spin relaxation rate due to collisions with other Rb (K) atoms, \he atoms, \& \nm  molecules, 
$k_\kp$ ($k'_\rb$) is the Rb (K) spin relaxation rate constant due to collisions with K (Rb) atoms,
and $A_\mathrm{se}$ is the Rb-K spin-exchange rate constant.  
We have also allowed for the possibility of additional light-induced spin relaxation mechanisms by including the $X_a$ \& $X'_a$ terms. 
Sources of these terms, which we refer to as ``alkali $X$-factors,'' and their consequences are discussed in Sec.~\ref{sec:estXA}.
The equilibrium alkali polarizations, explicitly assumed to be unequal, are found by solving Eqns.~(\ref{eqn:pdiff})--(\ref{eqn:pkdiff}) to give:
\begin{widetext}
\begin{equation}
s P_\mathrm{Rb} = \frac{R_1(1-X'_a)-R_2/2 - A_{pd} +\eta_\mathrm{K}D\{K_1-K_2/2\}}{R_{01}(1+X_a) + R_{02} + B_{pd} + \Gamma_\mathrm{Rb} + k_\mathrm{K}[\mathrm{K}] + \eta_\mathrm{K}D\{K_0+\Gamma_\mathrm{K}+k'_\mathrm{Rb}[\mathrm{Rb}]\}} \label{eqn:sprb}
\end{equation}
\begin{equation}
 P_\mathrm{K} = \frac{K_1-K_2/2+(\eta_\mathrm{Rb}/D)\{R_1(1-X'_a)-R_2/2 - A_{pd}\}}{K_0 + \Gamma_\mathrm{K} + k'_\mathrm{Rb}[\mathrm{Rb}]+(\eta_\mathrm{Rb}/D)\{R_{01}(1+X_a) + R_{02} + B_{pd} + \Gamma_\mathrm{Rb} + k_\mathrm{K}[\mathrm{K}]\}} \label{eqn:spk}
\end{equation}
\end{widetext}
where $D = [\kp]/[\rb]$ is the ratio of the K to Rb vapor number densities,  
$\eta_\mathrm{Rb}$ ($\eta_\mathrm{K}$) is the Rb (K) spin-exchange efficiency with respect to collisions with K (Rb) atoms,
and $A_{pd}$ \& $B_{pd}$ are terms that arise due to the nonzero excited state populations of Rb.

We model the laser light as an incoherent mixture of right-circular ($\mathcal{R}$) and left-circular ($\mathcal{L}$) polarized photons with fluxes given by $\Phi_\mathcal{R}(\vec{r},\nu)$ and $\Phi_\mathcal{L}(\vec{r},\nu)$,
where $\vec{r}$ is the location inside the cell and $\nu$ is the frequency.
In this case, the polarized optical pumping rate $A_n$ for the D$n$ transition and the total unpolarized optical pumping rate $A_0 = A_{01} + A_{02}$ for alkali species A are:
\begin{eqnarray}
	& A_n(\vec{r})  =  \int_{0}^\infty \left [ \Phi_\mathcal{R}(\nu) - \Phi_\mathcal{L}(\nu) \right ] \cos(\theta) \sigma^\al_n(\nu)\,d\nu & \\
	& A_{0}(\vec{r})  =  \int_{0}^\infty \left [ \Phi_\mathcal{R}(\nu) + \Phi_\mathcal{L}(\nu)  \right ] \left [ \sigma^\al_{1}(\nu) + \sigma^\al_{2}(\nu) \right ]\,d\nu &
\end{eqnarray}
where the factor $\cos(\theta) = \hat{k}\cdot\hat{B}_0$ is due to the ``skew'' angle $\theta$ between the laser propagation direction $\hat{k}$ \& the magnetic field orienting the spins $\vec{B}_0$.
The absorption cross section for the D$n$ transition in the pressure-broadened limit for alkali species A is:
\begin{equation}
	\sigma^\mathrm{A}_n(\nu) = \pi r_e c f_n \left [ \frac{ \Gamma_n/(2\pi) }{ \Delta_n^2 + \Gamma_n^2/4 } \right ] g \left ( 2 \pi \Delta_n T_d \right ) \label{eqn:sigmaabs}
\end{equation}
where $r_e$ is classical electron radius, $c$ is the speed of light in vacuum, $f_n$ is the oscillator strength of the transition, $\Gamma_n$ is the pressure-broadened absorption linewidth, 
and $\Delta_n = \nu-\nu_n$ is the detuning from the pressure-shifted line center ($\nu_n)$.
As will be described more fully in Sec.~\ref{sec:estXA}, the function $g(x)$ and the parameter $T_d$ describe how the absorption lineshape is modified from that of a simple Lorentzian due to buffer gas collisions \cite{walkup,pbRb}.
If $T_d = 0\ \mathrm{ps}$, then $g(0)=1$ and the absorption lineshape is simply Lorentzian.

The ``pure'' alkali spin relaxation rates (i.e. due to collisions not involving the other alkali species) are:
\begin{eqnarray}
\Gamma_\mathrm{Rb} & = & k_\mathrm{Rb}[\mathrm{Rb}] + k_\mathrm{He}[^3\mathrm{He}] + k_\mathrm{N_2}[\mathrm{N_2}] + \Gamma^\rb_\se \\
\Gamma_\mathrm{K} & = & k'_\mathrm{K}[\mathrm{K}] + k'_\mathrm{He}[^3\mathrm{He}] + k'_\mathrm{N_2}[\mathrm{N_2}] + \Gamma^\kp_\se 
\end{eqnarray}
where $\Gamma^\al_\se = k^\al_\se [\hens]$ is the spin-exchange rate from \he to alkali species A and we've ignored the fact that spin exchange with \he 
is spin relaxing only to the extent that the \he polarization is less than the alkali polarizations.
The alkali-alkali spin exchange efficiencies are given by:
\begin{eqnarray}
\eta_\mathrm{Rb} & = & \frac{A_\mathrm{se}[\mathrm{K}]}{A_\mathrm{se}[\mathrm{K}] + R_0 + \Gamma_\mathrm{Rb} + k_\mathrm{K}[\mathrm{K}]} \label{eqn:etarb} \\
{\rm and} \nonumber \\
\eta_\mathrm{K}  & = & \frac{A_\mathrm{se}[\mathrm{Rb}]}{A_\mathrm{se}[\mathrm{Rb}] + K_0 + \Gamma_\mathrm{K} + k'_\mathrm{Rb}[\mathrm{Rb}]} \label{eqn:etak} \ \ .
\end{eqnarray}
The equilibrium populations of the Rb excited states are:
\begin{eqnarray} 
	p & = & \frac{\tau_p}{a} \left \{ R_{01}            + R_{02} M_d \tau_d - sP_\rb \left [ R_1            - \frac{R_2 M_d \tau_d}{2}\right ] \right \} \nonumber \\ 
{\rm and}	\\ 
	d & = & \frac{\tau_d}{a} \left \{ R_{01} M_p \tau_p + R_{02}            - sP_\rb \left [ R_1 M_p \tau_p - \frac{R_2}{2}\right ] \right \} \nonumber \\
\end{eqnarray}
where the effective lifetimes \& mixing rates are:
\begin{eqnarray}
 \tau_p & = & 1/ \left  ( 1/\tau_{p0} + \Gamma_q  +  \Gamma_m  + 2 R_{01} \right )\ \ , \\ 
 \tau_d & = & 1/\left  ( 1/\tau_{d0} + \Gamma'_q +  \Gamma'_m + 3 R_{02}/2 \right )\ \ , \\
 M_p & = & \Gamma'_m - R_{01}\ \ , \\
M_d & = & \Gamma_m - R_{02}
\end{eqnarray}
and $a=1-(M_p\tau_p)(M_d\tau_d)$.
Finally, the terms in the equilibrium alkali polarization due to these excited state populations are:
\begin{eqnarray}
	A_{pd} & = & \frac{\tau_p}{a} \left [ 2 R_1 - \frac{R_2}{2}\right ] \left [ R_{01} + R_{02} (M_d \tau_d) \right ] \nonumber \\ & & + \frac{\tau_d}{a} \left [ R_1 - \frac{3 R_2}{4} \right ] \left [ R_{01} (M_p \tau_p) + R_{02} \right ]  \\
	{\rm and} \nonumber \\
	B_{pd} & = & \frac{\tau_p}{a} \left [ 2 R_1 - \frac{R_2}{2}\right ] \left [ R_1    - \frac{R_2(M_d \tau_d)}{2} \right ] \nonumber \\ & & + \frac{\tau_d}{a} \left [ R_1 - \frac{3R_2}{4} \right ] \left [ R_1 (M_p \tau_p) - \frac{R_2}{2}  \right ]\ \ .
\end{eqnarray}
These terms, which account for stimulated emission, are only important when the optical pumping rates start limiting the excited state lifetimes (i.e. $\tau_p \approx 1/R_{01}$ and $\tau_d \approx 1/R_{02}$).
\begin{widetext}
The attenuation of the photon fluxes as the light propagates through the polarized alkali vapor is given by:
\begin{eqnarray}
\left [ \frac{1}{\Phi(\nu)} 
\frac{d \Phi(\nu) }{dz} \right ]_\mathcal{R,L} & = &  
- [\rb] \left \{ \left ( s - p \mp sP_\rb \cos(\theta) \right )  \sigma_1^\rb(\nu) + \left ( s - \frac{d}{2} \pm \frac{sP_\rb \cos(\theta)}{2} \right )  \sigma_2^\rb(\nu)  \right \} \nonumber \\
& & 
- [\kp] \left \{ \left ( 1 \mp P_\kp \cos(\theta) \right )  \sigma_1^\kp(\nu) + \left ( 1 \pm \frac{P_\kp \cos(\theta)}{2} \right )  \sigma_2^\kp(\nu)  \right \} \label{eqn:propeqn}
\end{eqnarray}
where the upper (lower) sign corresponds to right-circular (left-circular) polarized photons.
\end{widetext}

\subsection{Overview of Simulation\label{sec:oversim}}
Due to the high buffer gas pressures in the targets described in this work, the alkali diffusion rate is very slow compared to the alkali polarization rate.
This implies that the local alkali polarization is dependent only on the local photon flux \cite{rbpolimag} and we can safely ignore the effect of diffusion far from the chamber walls.
In this case, the simulation is simply a numerical integration of Eqn.~(\ref{eqn:propeqn}) over a discretized path through the pumping chamber. 
Assuming azimuthal symmetry, a spherical pumping chamber is divided into 100 radial bins from the center of the cell to the top of the cell, each with a different path length.
The path for each radial bin is divided into $z$-slices with a thickness of 100 $\mathrm{\mu m}$, which is chosen as a delicate balance between computational time and accuracy.
The initial flux of right- \& left- circular polarized photons is given by:
\begin{equation}
	\Phi_\mathcal{R,L}(r,z=0,\nu) = \sum_{k=1}^N \Phi_0^k(r,\nu) \left [ \frac{1 \pm P_\gamma^k}{2} \right ]
\end{equation}
where the sum is over a total of $N$ lasers, $r$ is the radial distance from the center of the cell, $z$ is the depth into the cell, 
and $P_\gamma^k$ \& $\Phi_0^k$ are the polarization \& total photon flux from the $k$-th laser.
Initially (at $z=0$) the photon flux from the $k$-th laser is assumed to have Gaussian spectral and transverse spatial profiles given by:
\begin{equation}
	\Phi^k_0 = \left ( \frac{2P^k_0/(h\nu)}{\pi w_k^2 (\sigma^k_\gamma \sqrt{2\pi}) } \right ) 
\exp\left [ -\frac{2r^2}{w_k^2} - \frac{(\nu-\nu^k_\gamma)^2}{2(\sigma_\gamma^k)^2} \right ] 
\end{equation}
where the $P^k_0$ is the laser power, $h$ is the Planck constant, $w_k$ is the beam radius, $\nu_\gamma^k$ is the central laser frequency, $\sigma_\gamma^k = \mathtt{FWHM}_k/\sqrt{8\log(2)}$, 
and $\mathtt{FWHM}_k$ is the full-width half maximum of the laser spectrum.
The attenuation of this incident photon flux due to a completely unpolarized ``diffusion'' layer of alkali vapor \cite{larson,wagshul} at the inner front surface of the pumping chamber 
is given by $\exp(-\sigma^\rb_1(\nu) [\rb] \sqrt{2D_\rb/R_1})$ \cite{rmp-seop} where $D_\rb$ is the \rb\ diffusion constant in \hens.
The spectral profile of the laser is binned by helicity and divided into 3000 frequency bins that represent a $\pm1500$ GHz ($\pm3.2$ nm) window centered at the Rb D1 pressure-shifted line center.
At the beginning of each $z$-slice, the optical pumping rates are calculated using the attenuated photon flux from the previous $z$-slice.
These rates are then used to calculate the alkali polarizations for the current $z$-slice, which are then used to further attenuate the photon flux.
Finally, the volume averaged alkali polarization is calculated by weighting the polarization in each radial bin \& $z$-slice by the fractional volume of that bin-slice.

\begingroup 
\squeezetable
\begin{table}
\centering
\begin{tabularx}{0.35\textwidth}{|X|c|c|}
\toprule \addlinespace
\hline
\bf parameter	&	\bf value	& \bf	units	\\ %
\hline
$[\hens]_\pc$	&	6.5	&	amg	\\
$[\nmns]/[\hens]$	&	0.01	&		\\
$1/\gamma_\se$	&	3	&	hrs	\\
\multirow{2}{*}{$D$}	&	0 (pure) 	 & \\
			&	6 (hybrid)	& \\
$2R_\pc$	&	3	&	in	 \\
$w/R_\pc$	&	1	 &  \\
$P_0$		&	75	&	W	\\
\multirow{2}{*}{$\mathtt{FWHM}$}	&	0.2 (narrowband) &	nm	\\
			&	2.0 (broadband)	&	nm	\\
$P_\gamma$		&	0.99	& \\
$\theta$		&	3	&	deg	\\
$T_d$ (Rb D1)					&	$-0.19$	&	ps	\\
$T_d$ (Rb D2)					&	$+0.10$	&	ps	\\
$T_d$ (K D1)					&	$+0.10$	&	ps	\\
$T_d$ (K D2)					&	$+0.10$	&	ps	\\
\hline
\end{tabularx}
\caption{Baseline Input Parameters to the Simulation. \label{tab:baseline}
The cell diameter is given by $2 R_\pc$.  The laser spectrum is taken to be centered on the pressure-shifted Rb D1 absorption line center.
}
\end{table}
\endgroup

\begingroup 
\squeezetable
\begin{table}
\centering
\begin{tabularx}{0.35\textwidth}{| X X X X |}
\hline
\bf parameter	 & \bf Rb	 & \bf K/Rb	 & \bf units	\\ 
\hline
$D$	& 0  & 6 	&  - 	\\
$D'$	&	0	&	4.58	&	-	\\
$[\rb]$	&	$13.7$	&	$2.46$	&	$10^{14}/\mathrm{cm}^3$	\\
$T_\mathrm{op}$	&	$210$	&	$260$	&	$\mathrm{^oC}$	\\
$\Gamma_\rb$	&	$1.08$	&	$0.692$	&	kHz	\\
$\Gamma_\kp$	&	-	&	$0.202$	&	kHz	\\
$2\bar{k}_\mathrm{sd}[\rb]$	&	-	&	$0.107$	&	kHz	\\
$1/\eta_\se$	&	$92.0$	&	$58.8$	&	-	\\
$1/\eta'_\se$	&	-	&	$22.5$	&	-	\\
$\left < 1/\eta_\se \right >$	&	$92.0$	&	$29.0$	&	-	\\
$1/\eta_\mathrm{ahse}$	&	$92.0$	&	$38.8$	&	-	\\
$\Gamma_\al$	&	$1.08$	&	$2.55$	&	kHz	\\
\hline
\end{tabularx}
\caption{Alkali Spin Relaxation Rates and Spin Exchange Efficiencies.  Values are calculated using the baseline input parameters for pure Rb and K-Rb alkali-hybrid SEOP.  
Although the effective alkali spin relaxation rate $\Gamma_\al$ is higher for AHSEOP, the alkali-\he spin exchange efficiency is much higher. \label{tab:alkali-rates}}
\end{table}
\endgroup

\begingroup 
\squeezetable
\begin{table}
\centering
\begin{tabularx}{0.35\textwidth}{|lXXX|}
\toprule \addlinespace
\hline
\bf parameter	 & \bf BB	 & \bf NB	 & \bf units \\ \addlinespace \midrule \addlinespace
\hline
$\mathtt{FWHM}$	&	$2.0$	&	$0.2$	&	nm	\\
$\Gamma_1^\rb$	&	$0.27$	& same	& nm	\\
$R_{01}$	&	$103$	&	$459$	&	kHz	\\
$100R_{02}/R_{01}$	&	$0.374$	&	$0.083$	&	-	\\
$100K_{01}/R_{01}$	&	$0.062$	&	$0.014$	&	-	\\
$100K_{02}/R_{01}$	&	$0.133$	&	$0.030$	&	-	\\
$A_\se [\rb]$	&	$249$	&  same	& kHz	\\
$\Gamma_p$,$\Gamma'_p$	&	$110$,$495$	&	same	&	GHz	\\
$\Gamma_m$,$\Gamma'_m$	&	$3.73$,$1.77$	&	same	&	GHz	\\
$\Gamma_q$,$\Gamma'_q$	&	$0.515$,$0.383$	&	same	&	GHz	\\
$\eta_\rb$	&	$0.935$	&	$0.765$	&	-	\\
$\eta_\kp$	&	$0.998$	&	$0.998$	&	-	\\
$p$	&	$5.42$	&	$11.9$	&	ppm	\\
$d$	&	$9.49$	&	$20.5$	&	ppm	\\
\hline
\addlinespace \bottomrule
\end{tabularx}
\caption{Optical Pumping and Excited State Parameters.  Values are calculated using $D=6$ baseline parameters for broadband (BB) and narrowband (NB) lasers at the front/center ($z=r=0$) of the cell. \label{tab:alkali-excited}}
\end{table}
\endgroup

The baseline parameters used in the simulation listed in Tab.~\ref{tab:baseline} were chosen to be representative of typical operating conditions during one of the experiments for which the target cells were originally built.
The laser power listed in the table, $\rm 75\,W$, is the amount actually incident on the pumping chamber. 
This value might be a bit optimistic since, in a practical situation, we have observed that an optical transport system can easily result in as much as a 50\% loss of power by the time the beams reach the pumping chamber.
In order to make fair comparisons, all of the calculations were done at constant spin-exchange rate $\gamma_\se$, where $\gamma_\se$ is the rate at which $^3$He nuclei are being polarized through collisions with alkali-metal atoms.
The alkali densities and the operating temperature are calculated from the alkali-\he spin-exchange rate and $D$ using the following equation:
\begin{equation}
	\gamma_\se = k_\se [\rb] \left ( 1 + D'\right ) 
\label{eq:gamma_se}
\end{equation}
where $D' = D\,k^\kp_\se/k^\rb_\se$ is the alkali-\he spin exchange rate ratio.
To provide a sense of scale, the various rates are listed in Tables~\ref{tab:alkali-rates}~\&~\ref{tab:alkali-excited} using the baseline parameters for both pure Rb and K-Rb vapors and also both narrowband and broadband light.
The rate constants and cross sections, along with their estimated temperature dependences, used to calculate these rates can be found in Appendix D of \cite{singh}.
Finally, the \he polarization is sensitive to the polarization-weighted alkali spin exchange rate given by:
\begin{equation}
	\left <P_\mathrm{A} \right > \gamma_\se \equiv \left < s P_\rb \right >_\pc k^\rb_\se [\rb] + \left < P_\kp \right >_\pc k^\kp_\se [\kp]
\end{equation}
where $\left < \cdots \right>_\pc$ refers to a volume average over the pumping chamber and the quantity $\left < P_\mathrm{A} \right>$ is the main result obtained from the simulation.

\subsection{Optimization of the K to Rb Density Ratio\label{sec:optD}}
\begin{figure}
\begin{center}
\includegraphics[width = 3.25truein]{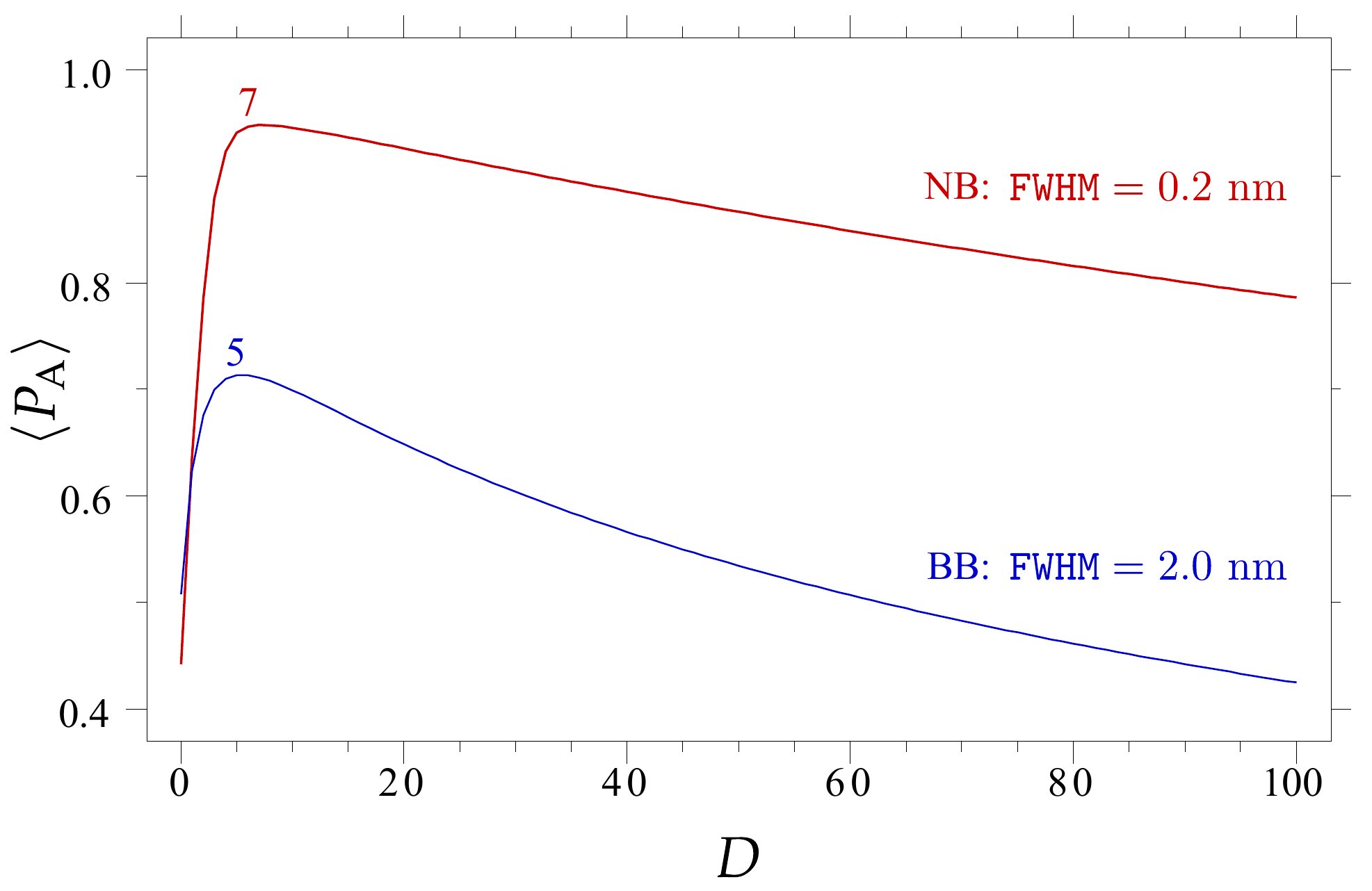}
\caption{(color online) Alkali Polarization vs. K to Rb Density Ratio $D$.  NB (BB) refers to a narrowband laser with 0.2 nm (2.0 nm) linewidth.  The optimal ratio for the NB (BB) laser considered is approximately 7 (5).}
\label{fig:PAD}
\end{center}
\end{figure}
Some of the important trends associated with alkali-hybrid optical pumping are clearly evident in Fig.~\ref{fig:PAD} which shows volume-averaged alkali polarization, $\langle P_A \rangle$, as a function of the ratio $D$. 
As one begins adding K to the alkali mixture, for a fixed amount of laser power ($P_0 = 75\ \mathrm{W}$) and a fixed spin-exchange rate ($\gamma_\se^{-1} = 3\ \mathrm{hrs}$), $\langle P_A \rangle$ sharply rises.
As the ratio $D$ gets sufficiently large, however, $\langle P_A \rangle$ starts to roll over and decrease.
The optimal ratio appears to be $6 \pm 1$ which is consistent with the range $5 \pm 2$ we report in Sec.~\ref{sec:IVA}.
This can be compared to the optimal range $4 \pm 2$ observed by Chen \etal~\cite{chen-2007}, which was for a different set of operating conditons such as lower \he densitiies.
It is not difficult to understand some of the dominant factors that influence these features in the $D$ dependence.

When both alkali-alkali spin exchange efficiencies are nearly unity $\eta_\rb,\eta_\kp \approx 1$, which as can seen from Tab.~\ref{tab:alkali-excited} is not too far from the case, the effective alkali spin relaxation rate due to collisions can be written as:
\begin{equation}
	\Gamma_\mathrm{A} = \Gamma_\rb + D \left \{ \Gamma_\kp + 2 \bar{k}_\mathrm{sd} [\rb]  \right \}
\label{eq:gammaA}
\end{equation}
where $\bar{k}_\mathrm{sd} = (k_\kp + k'_\rb)/2$ is the mean \rb-\kps spin relaxation rate constant.
Although $\Gamma_\kp$ is significantly smaller than $\Gamma_\rb$ due to the greater spin-exchange efficiency of \kp, its effect on the effective alkali spin-relaxation rate is enhanced due to $D$.
Ignoring all but the most dominant terms in Eqns.~(\ref{eqn:sprb}) \& (\ref{eqn:spk}), we can write the alkali polarization in the following more familiar form:
\begin{equation}
\label{eq:alkpolsimple}
P_\rb = P_\kp = P_\al = \frac{R_{1}}{R_{01} + \Gamma_\al}\ \ .
\end{equation}
It is clear from Eqn.~(\ref{eq:gammaA}), however, that $\Gamma_\al$ becomes arbitrarily large as $D$ is increased, increasingly limiting the alkali polarization $P_\al$, and causing much of the roll off evident in Fig.~\ref{fig:PAD}.
We note that there are additional factors that contribute to this roll off that will be discussed in the next section.

The fact that $\Gamma_\al$ becomes larger for any nonzero value of $D$ almost leaves us wondering where the benefits lie of having $D>0$.
Indeed, at the very front of the cell, the alkali polarization will be ever so slightly {\it lower} for any value of $D$ greater than zero.
The benefits, however, arise as the laser beam propagates through the cell.
The ``photon demand,'' the number of photons per second required to maintain a given alkali polarization throughout the pumping chamber, is proportional to $[\mathrm{Rb}]$, which for a fixed spin-exchange rate, is proportional to $1/(1+D')$.
With less \rbs to absorb the light, it takes fewer photons to keep the entire alkali vapor polarized.
This  makes it possible for the laser to penetrate further into the sample without significant decrease in intensity, thus causing the sharp rise in $\langle P_\al \rangle$ as \kps is added to the alkali mixture.
We discuss this argument in more detail in Sec.~\ref{sec:photonD}. 

\subsection{The Alkali X-Factors\label{sec:estXA}}
It is now well documented that much more laser power is required to achieve good performance than most simple optical pumping simulations suggest \cite{PhysRevLett.91.123003,chen-2007,fundamentals}.
For this reason, every attempt was made to relax as many assumptions as possible in deriving the equations of Sec.~\ref{sec:ahop} that are used in our simulation.
As a result, this simulation includes
(1) the full solution to the coupled alkali-hybrid polarization equations, 
(2) the effect of off-resonant absorption by the \rbs D2 \& \kps D$n$ lines with realistic absorption lineshapes, 
(3) radiation trapping, which is included ``by hand" by using the factors $X_a$ \& $X'_a$,
(4) skew optical pumping \cite{chann-skew}, and
(5) effects associated with nonzero excited state populations.
The role of each of these effects, which until recently \cite{lancor-n2,hjw} have not been included in most optical pumping simulations, can be understood by generalizing Eqn.~(\ref{eq:alkpolsimple}): 
\begin{equation}
P_\al = \frac{2 \Lambda R_{01}(1-X'_\al)}{2 \Lambda R_{01}(1+X_\al) + \Gamma_\al} \label{eqn:pal-general}
\end{equation}
where $\Lambda$ is the probability of an electron spin flip (or equivalently the amount of angular momentum gained by the atom) per optical pumping cycle,
$X_\al$ \& $X'_\al$ are the total polarization independent \& dependent alkali $X$-factors respectively, 
and we have again assumed that the alkali-alkali spin exchange efficiencies are nearly unity.
In other words, Eqn.~(\ref{eqn:pal-general}) is equivalent to Eqns.~(\ref{eqn:sprb}) \& (\ref{eqn:spk})
when $\eta_\rb=\eta_\kp=1$ and the effects of off-resonant pumping, radiation trapping, skew optical pumping, and nonzero excited state populations are encoded by the parameters $\Lambda$, $X_\al$, \& $X_\al'$.

Before moving on to the main discussion, we would like to point out the importance of the assumption that $\eta_\rb, \eta_\kp \approx 1$.
In the infinite laser power limit, when the optical pumping rates overwhelm the alkali-alkali spin exchanges rates ($R_0 \gg A_\se[\kp]$ and $K_0 \gg A_\se[\rb]$),
we see from Eqns.~(\ref{eqn:etarb}) \& (\ref{eqn:etak}) that $\eta_\rb$ \& $\eta_\kp$ approach zero and the alkali polarizations in Eqns.~(\ref{eqn:sprb}) \& (\ref{eqn:spk}) become uncoupled ($P_\rb \neq P_\kp$).
Because we typically pump with laser light tuned to the \rbs D1 transition, the uncoupled alkali polarizations in this case become $P_\rb \approx 1$ and $P_\kp \approx 0$.
Therefore it is understood that the ``infinite laser power limit'' invoked in the following discussions of the limiting alkali polarization is really the ``very high laser power limit'' such that $\eta_\rb,\eta_\kp \approx 1$.

It is natural to expect that at very high laser intensities (i.e. high optical pumping rates: $R_{01} \gg \Gamma_\al$), the alkali polarization asymptotically approaches unity.
However, this is only true if there are no alkali spin-relaxation mechanisms that scale with the laser intensity.
In analogy to the \he $X$-factors mentioned earlier, we find that the alkali $X$-factors prevent the alkali polarization from saturating at unity in this limit:
\begin{equation}
P_\al = \frac{1-X'_\al}{1+X_\al + \Gamma_\al/(2\Lambda R_{01})} \rightarrow P_\infty = \frac{1-X'_\al}{1+X_\al}
\end{equation}
We are now in a position to discuss the sources, relative sizes, and importance of the $\Lambda$, $X_\al$, and $X'_\al$ terms.

First we'll consider violations of the assumptions made at the beginning of Sec.~\ref{sec:ahop}, which, if perfectly true, are equivalent to $2 \Lambda = 1$.
Because the excited state hyperfine coupling precession period ($1$ ns) is long compared to the excited state electron disorientation rate ($0.01$ ns),
very little of the angular momentum that is stored by the nuclear spin is lost to the electron while in the excited state \cite{bhaskar81}.
Recently, Lancor \& Walker \cite{lancor-n2} have argued that this lowers the average amount of angular momentum gained per optical pumping cycle and, under conditions typical in target cells, we find $2 \Lambda \approx 1 - 10^{-2}$.
The effect of nonzero multipole moments in the excited state is discussed in Ref.~\cite{singh} and it was shown that $2 \Lambda \approx 1 - 10^{-3}$.
In summary, these two effects are quite small, increase the photon demand by at most a percent, and do not ultimately limit the alkali polarization.

Nonzero values of $X_\al$ \& $X'_\al$, on the other hand, do limit the ultimate alkali polarization.
As a consequence, the transparency of the polarized alkali vapor is reduced, and, as will be discussed more in the next section, even a small alkali $X$-factor can significantly increase the photon demand.
Although off resonant absorption and skew pumping are already contained in Eqns.~(\ref{eqn:sprb}) \& (\ref{eqn:spk}), it is useful to describe them in the form of alkali $X$-factors so that their sizes can be compared with other alkali polarization-limiting factors.

Off resonant absorption of light by D2 transitions is detrimental because optical pumping of the D2 transitions pushes the equilibrium alkali polarization towards $-0.5$ as opposed to $+1.0$ for D1 transitions.
This situation is made worse by the fact that collisions with \he create short lived alkali-\he quasi-molecules that enhance the off-resonant cross section relative to a simple Lorentzian lineshape.
Assuming a van der Waals potential between the alkali atom and \hens, Walkup \etal \cite{walkup} have shown that this enhancement, far off resonance, is described by 
\begin{equation}
       g(2 \pi \Delta T_d) = \frac{\pi \sqrt{|2 \pi \Delta T_d|} }{6\times0.3380} 
\end{equation}
where $\Delta$ is the detuning of the laser frequency from the line center of the off resonant transition and $T_d$ can be thought of as the effective lifetime of the quasi-molecule.
This $T_d$ parameter quantifies the degree to which the lineshape is modified and is possibly detuning-dependent. 

In order to better understand the limits imposed solely by off resonant optical pumping, we express the limiting alkali polarization as:
\begin{equation}
P_\infty(\nu) = \frac{1-[\sigma_\rb^2(\nu) + D \sigma_\kp^2(\nu)]/[2 \sigma_\rb^1(\nu)]}{1+[\sigma_\rb^2(\nu)+D\sigma_\kp^2(\nu)]/\sigma_\rb^1(\nu)} \label{eqn:pinfty}
\end{equation} 
where we've assumed pumping with perfectly monochromatic light with frequency $\nu$.

Measurements by Romalis \etal \cite{pbRb} and more recently by Lancor \etal \cite{lancor-rb1,lancor-rb2,lan11} indicate that $|T_d|$ is of order $0.1\ \mathrm{ps}$ for alkali-\he collisions.
Using this value, we find that $(3\sigma_\al^2)/(2\sigma_\rb^1) \approx 10^{-3}$ and consequently $(3D\sigma_\al^2)/(2\sigma_\rb^1)=1$ when $D \approx 10^3$.
This $D$ scaling of the \kps D2 off resonant absorption helps explain both the calculated roll off discussed in Sec.~\ref{sec:optD} and the observed drop in $P_\infty$ with increasing $D$ reported in Ref.~\cite{PhysRevLett.91.123003}.
Using the Walkup parameterization is a simple \& alternative way to account for the effect of off resonant absorption, which was first pointed out in Ref.~\cite{lancor-rb1}.
We justify the use of this approach because it reliably reproduces the measured values of $P_\infty$ as function of $\nu$ for a pure Rb cell reported in Ref.~\cite{lancor-rb2}, when we set $T_d = 0.3\ \mathrm{ps}$ for Rb D2.
Finally, the benefits of pure \kps \& $\mathrm{Na}$ SEOP are also limited by off resonant absorption by the D2 lines, where $(3\sigma^2_\al)/(2\sigma^1_\al) \approx 0.01\ \&\ 0.1$ respectively.

The largest effect associated with nonzero excited state populations is radiation trapping.
There is a probability of a few percent that an excited alkali atom will not be quenched due to collisions with \nm  and therefore will reradiate a photon.
This photon is only partially polarized and its reabsorption by a neighboring Rb atom appears as a spin relaxation mechanism. The \nm also mixes the fine structure states very efficiently, so that both D1 and D2 light is emitted.
The simple model used for the calculation of $X_a$ \& $X'_a$ due to this reemission/reabsorption process in a spherical pumping chamber is described in Ref.~\cite{singh}.

\begin{figure}
\begin{center}
\includegraphics[width = 2.75truein]{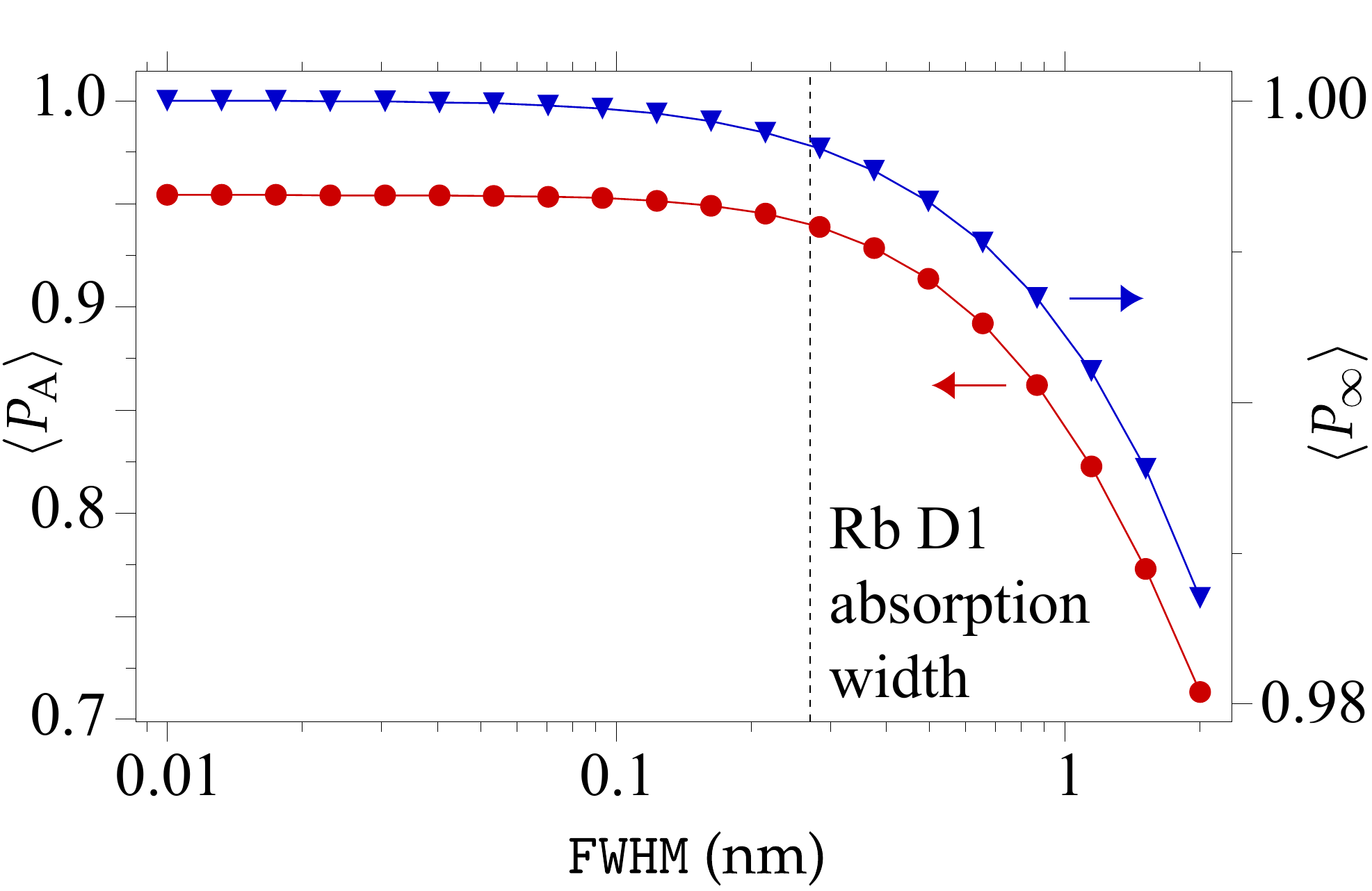}
\caption{(color online) Alkali Polarization vs. Laser Linewidth. The red circles indicated the volume averaged alkali polarization under realistic conditions using the baseline parameters for a hybrid cell.
The blue triangles indicate the best case alkali polarization averaged over a Gaussian spectral lineshape assuming $\Gamma_\al = 0$.}
\label{fig:linewidth}
\end{center}
\end{figure}

\begin{widetext}
\begin{table*}
\begin{tabular}{c|c|c|c}
\hline
\bf source	&	\bf Narrowband		&	\bf Broadband 	    &	\bf scaling	\\ 
		&	$100 \left < X_\al+X'_\al \right >_\pc $	&	$100 \left < X_\al+X'_\al \right >_\pc $ &		\\ 
\hline
\kps D2 absorption	   &	0.29	&	4.0	&	$D[^3\mathrm{He}]^{(1+n)}\mathtt{FWHM}^{(1-n)}$	\\ \addlinespace
\rbs D2 absorption   &	0.13	&	1.9	&	$[^3\mathrm{He}]^{(1+n)}\mathtt{FWHM}^{(1-n)}$	\\ \addlinespace
radiation trapping &	0.35	&	0.9	&	$1/[\mathrm{N}_2]$	\\ \addlinespace
skew pumping	   &	0.14 	&	0.1	&	$\theta^2$	\\ \addlinespace \addlinespace
\bf total		& 0.91 	& 6.9	& 	\\ \addlinespace \midrule \addlinespace
\bf $\left < P_\infty \right >_\pc$ 		& 0.99 & 0.93	& 	\\ 
\hline
\end{tabular}
\caption{Estimates for Alkali $X$-Factors.  
For a narrowband (broadband) laser, $n=1$ ($n=0$) and the linewidth is given by $\mathtt{FWHM} = \mathrm{0.2\ (2)\ nm}$. 
Under our typical laser powers, the effect of stimulated emission $(X'_A \approx 10^{-5})$ can be safely ignored.
\label{tab:alkali_x}}
\end{table*}
\end{widetext}
The typical size of each of these mechanisms, averaged over the pumping chamber volume, for the baseline parameters for a hybrid cell (with $D=6$) are listed in Tab.~\ref{tab:alkali_x}.
It can be seen that off resonant absorption is usually the dominant alkali $X$-factor and it is about an order of magnitude larger than the values at the front of the pumping chamber listed in Tab.~(\ref{tab:alkali-excited}).
This is because the on-resonant optical pumping rate decreases more quickly than the off resonant optical pumping rates as the the light travels through the cell.
As can be seen in Fig.~\ref{fig:linewidth}, a narrowband laser ($\mathtt{FWHM}=100\ \mathrm{GHz}$) is far less sensitive to the deleterious effects of off-resonant pumping than a broadband laser ($\mathtt{FWHM} = 1000\ \mathrm{GHz}$).
The quantity $\left < P_\infty \right >$ depicted as blue triangles in this figure is simply Eqn.~(\ref{eqn:pinfty}) averaged over a Gaussian spectral lineshape.
It is not too surprising that a laser linewidth that is about the size of the \rbs absorption linewidth ($0.27\ \mathrm{nm} = 130\ \mathrm{GHz}$ for $[\hens] = 6.5\ \mathrm{amg}$) is sufficiently narrow to take full advantage of narrowband pumping.

\subsection{Estimating the Photon Demand\label{sec:photonD}}
To fully polarize the entire alkali vapor, there must be a sufficiently high intensity of laser light to overcome the alkali spin relaxation rate throughout the pumping chamber.
There are two main things to consider in order to satisfy this condition.
First, the transverse intensity profile of the laser beam must be chosen carefully in order to ensure that enough light is incident across the face of the pumping chamber \cite{cummings}.
To achieve this for a spherical pumping chamber, we've found that the optimal beam radius of the laser light is about equal to the inner radius of the pumping chamber.

Second and more challenging is ensuring that enough laser power is transmitted to the back of the pumping chamber.
Because there is always some nonzero alkali spin relaxation, the laser beam is attenuated as it propagates through the alkali vapor.
As mentioned earlier, the amount of laser light absorbed by the vapor per unit time is referred to as the ``photon demand.''
To estimate this photon demand, we first rewrite Eqn.~(\ref{eqn:propeqn}) in the following way:
\begin{equation}
	\left [ \frac{1}{\Phi} \frac{d \Phi}{dz} \right ] = - \sigma_1^\rb [\rb] \left \{ (1 + X_\gamma) - P_\al (1-X'_\gamma)\right \} \label{eqn:phi-general}
\end{equation}
where $X_\gamma$ \& $X'_\gamma$ describe alkali polarization independent \& dependent light absorption mechanisms such as off-resonant pumping.
These so-called ``photon $X$-factors'' ruin the transparency of the alkali vapor even with unity alkali polarization.
The photon demand $\Delta P$ is found by integrating this equation over all laser frequencies and over the pumping chamber volume, and,
after rewriting $[\rb]$ in terms of the spin exchange rate and $\Gamma_\al$ in terms of the alkali spin-exchange efficiencies, we find:
\begin{equation}
	\frac{\Delta P}{94\ \mathrm{W}} = 
\left [ \frac{2 R_\pc}{7.62\ \mathrm{cm}} \right ]^3
\left [ \frac{\gamma_\se}{1/(3\ \mathrm{hrs})} \right ]
\left [ \frac{[\he]}{6.5\ \mathrm{amg}} \right ]
\left [ \frac{0.01}{\eta_\mathrm{op}} \right ] \ \ . \label{eqn:dPscaling}
\end{equation}
The optical pumping efficiency $\eta_\mathrm{op}$ is the ratio of the number of \he nuclei polarized to the number of photons absorbed and, 
assuming a narrowband laser and small photon \& alkali $X$-factors, it is given by:
\begin{equation}
\frac{1}{\eta_\mathrm{op}} = \frac{1}{\eta_\mathrm{ahse}}  + \frac{R_{01}}{\Gamma^\rb_\se} \left [ \frac{X_\gamma+X'_\gamma + X_\al + X'_\al}{1+D'} \right ]\ \ .  \label{eqn:etaop}
\end{equation}
The alkali-hybrid spin-exchange efficiency $\eta_\mathrm{ahse}$ is the ratio of the number of alkali-\he spin exchange collisions to the total number of collisions that result in the loss of alkali polarization and 
it is given by:
\begin{equation}
	\frac{1}{\eta_\mathrm{ahse}}  = \left < \frac{1}{\eta_\se} \right > + \frac{(k_\kp [\kp]/\Gamma^\rb_\se) + D' (k'_\rb[\rb]/\Gamma^\kp_\se)}{1+D'} \label{eqn:eta-ahse}
\end{equation}
where the first term is given by:
\begin{equation}
	\left  < \frac{1}{\eta_\se} \right > = \frac{1/\eta_\se + D'/\eta'_\se}{1+D'} \label{eqn:eta-se}
\end{equation}
and the ``pure'' efficiencies are $\eta_\se = \Gamma^\rb_\se/\Gamma_\rb$ and $\eta'_\se = \Gamma^\kp_\se/\Gamma_\kp$.
It is apparent by examining these last two equations that $\eta_\mathrm{ahse}$ is essentially the alkali density weighted average of the individual alkali-\he spin exchange efficiencies taking into account \rb-\kps collisions.
To lowest order, the photon \& alkali $X$-factor sum in the numerator of the last term in Eqn.~(\ref{eqn:etaop}) is given by:
\begin{equation}
\sum X =  \frac{3 (D K_{02} + R_{02})}{R_{01}} + X_a + X'_a + \theta^2
\end{equation}
where off resonant absorption \& skew pumping contribute to both the photon and alkali $X$-factors.

\begin{figure}
\begin{center}
\includegraphics[width = 3.3truein]{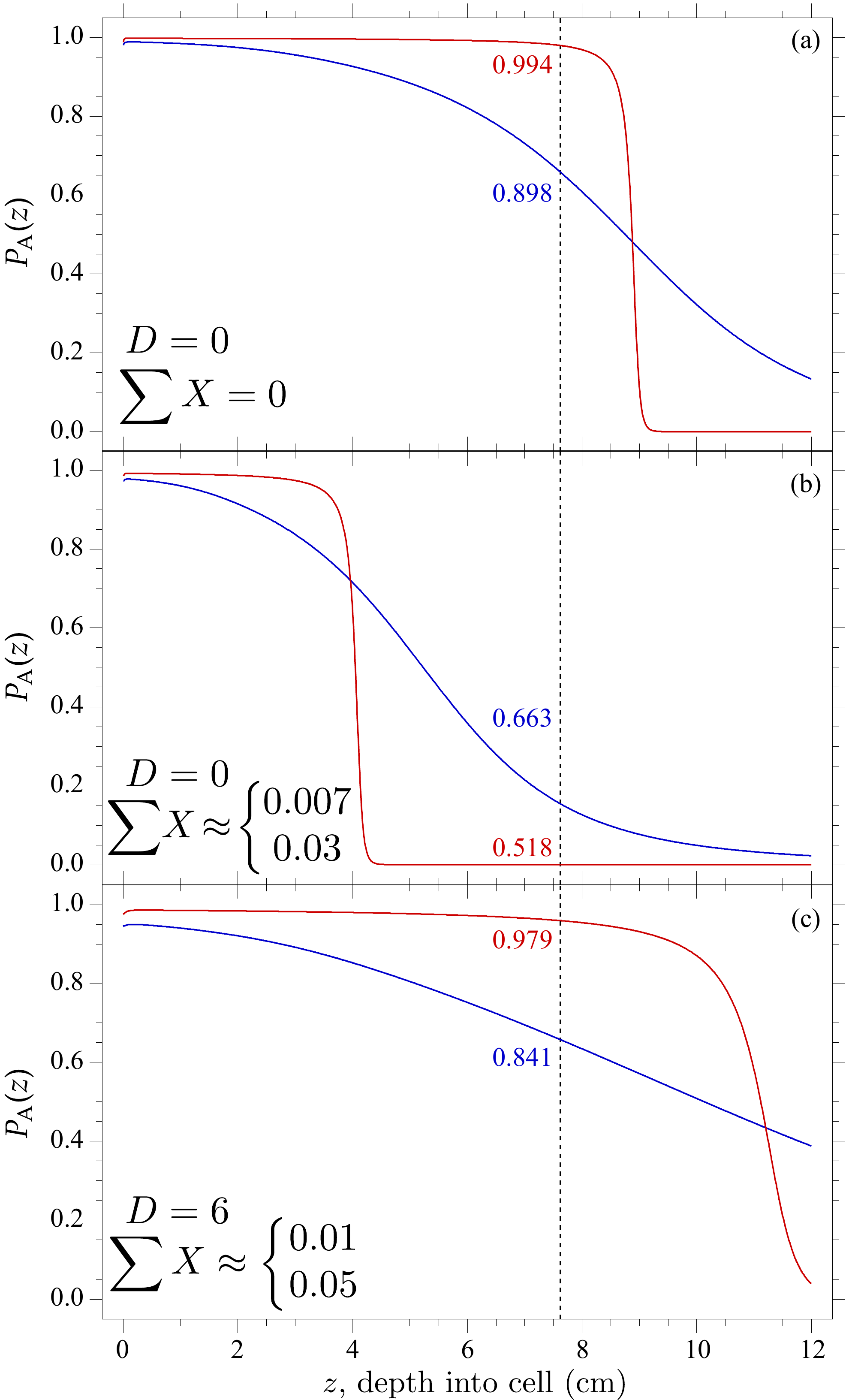}
\caption{(color online) Alkali Polarization vs. Depth into Pumping Chamber. The red (blue) curve is NB (BB) which refers to a narrowband laser with 0.2 nm (2.0 nm) linewidth, where
the corresponding numbers are the line-averaged polarization to the end of the cell (dashed line).
(a) Pure Rb SEOP with no alkali $X$-factors.
(b) Pure Rb SEOP with nonzero alkali $X$-factors.
(c) AHSEOP with nonzero alkali $X$-factors.
The upper (lower) values for $\sum X$ is the average value at the front of the cell for narrowband (broadband) pumping.
}
\label{fig:single-line}
\end{center}
\end{figure}
To put the impact of AHSEOP and narrowband lasers on the laser power requirement on a more quantitative footing, we will use the quantities listed in Tabs.~\ref{tab:baseline}, \ref{tab:alkali-rates}, \& \ref{tab:alkali-excited} throughout the following discussion.
We start by requiring that the alkali polarization at the back of the pumping chamber be $0.98$.
For $\Gamma_\al = 1.5\ \mathrm{kHz}$, this can be accomplished with 54~W of broadband light or 12~W of narrowband light.
Now we will estimate the photon demand under different scenarios and compare these estimates with the results of the full simulation depicted in Fig.~\ref{fig:single-line}.

For the first scenario, we consider pure \rbs SEOP ($D=0$) and ignore all photon \& alkali $X$-factors ($\sum X = 0$).
Using $\eta_\op = \eta_\se = 0.011$,  we find that the photon demand is 86~W.
The total estimated laser power requirement is 125~W of broadband light or 95~W of narrowband light.
We now turn our attention to the top panel of Fig.~\ref{fig:single-line} which shows the alkali polarization as a function of depth into the cell at the center of the pumping chamber for 75~W.
First, we note the difference in the alkali polarization profile between narrowband and broadband pumping.
This is because narrowband lasers have a higher density of photons at the absorption cross section of the \rbs D1 line, which results in about a factor of 5 higher optical pumping rate.
This means that as the laser penetrates the vapor, the alkali polarization will be very high, until a sharp cut off when the optical pumping rate quickly drops to zero.
In other words, where there is narrowband light, there is very high and uniform alkali polarization.
It appears that indeed 75~W of narrowband light is enough to keep the alkali vapor highly polarized throughout the pumping chamber (i.e. cell depth $< 7.6\ \mathrm{cm}$), whereas 75~W of broadband light is clearly not enough.

Unfortunately, photon \& alkali $X$-factors as small as $0.01$ significantly reduce the optical pumping efficiency because, under our conditions, $R_{01}/\Gamma^\rb_\se \approx 10^4$.
The observation that indeed many more photons were being absorbed than expected based on the measured spin exchange efficiency (i.e. $\eta_\mathrm{ahse} < \eta_\mathrm{op}$) was first made in a remarkable paper by Babcock \etal \cite{PhysRevLett.91.123003}.
We now consider pure \rbs SEOP under this more realistic scenario of nonzero photon \& alkali $X$-factors.
Although $\sum X$ is much smaller for narrowband pumping, the ratio $R_{01}/\Gamma^\rb_\se$ is larger for narrowband pumping by almost the same amount.
Consequently, we find that $\eta_\op = 0.0025$ for both narrowband \& broadband pumping and the photon demand is more than quadrupled to 380 W.
The total estimated laser power requirement is now 410~W of broadband light or 400~W of narrowband light.
Both of these are well above the 75~W used to generate the middle panel of Fig.~\ref{fig:single-line}.
Unsurprisingly, the laser light only penetrates about half way into the cell for both narrowband and broadband pumping.

Finally, we consider AHSEOP with $D=6$ and find that $\eta_\op = 0.008$ and the photon demand is significantly reduced to 88~W.
The total estimated laser power requirement is now 180~W of broadband light or 100~W of narrowband light.
The estimated narrowband power is more than the 75~W used to generate the bottom panel of Fig.~\ref{fig:single-line}.
In both cases, the laser light is able to penetrate past the end of the pumping chamber, but the alkali polarization is only very high with the narrowband light.

In order to better account for the photon demand, we've had to introduce both alkali $X$-factors in Eqn.~(\ref{eqn:pal-general}) and photon $X$-factors in Eqn.~(\ref{eqn:phi-general}).
As has been demonstrated, these terms collectively result in extra light absorption that significantly increases the photon demand.
We have made every effort to include all that is known in calculating these terms.
Although we may have not accounted for every mechanism that contributes to $\sum X$, 
we believe that Eqns.(\ref{eqn:pal-general})~\&~(\ref{eqn:phi-general}) do not need to be generalized further to fully describe optical pumping.
By comparing directly to the full simulation, we shown that Eqns.~(\ref{eqn:dPscaling}) \& (\ref{eqn:etaop}) can be used to quickly and reliably estimate the laser power requirement for future targets.
\section{Experimental Methods\label{sec:expmeth}}
\subsection{The He-3 Targets}
\subsubsection{Overview of the Targets}
\begin{figure}[htbp]
\begin{center}
\includegraphics[width = 8.6cm]{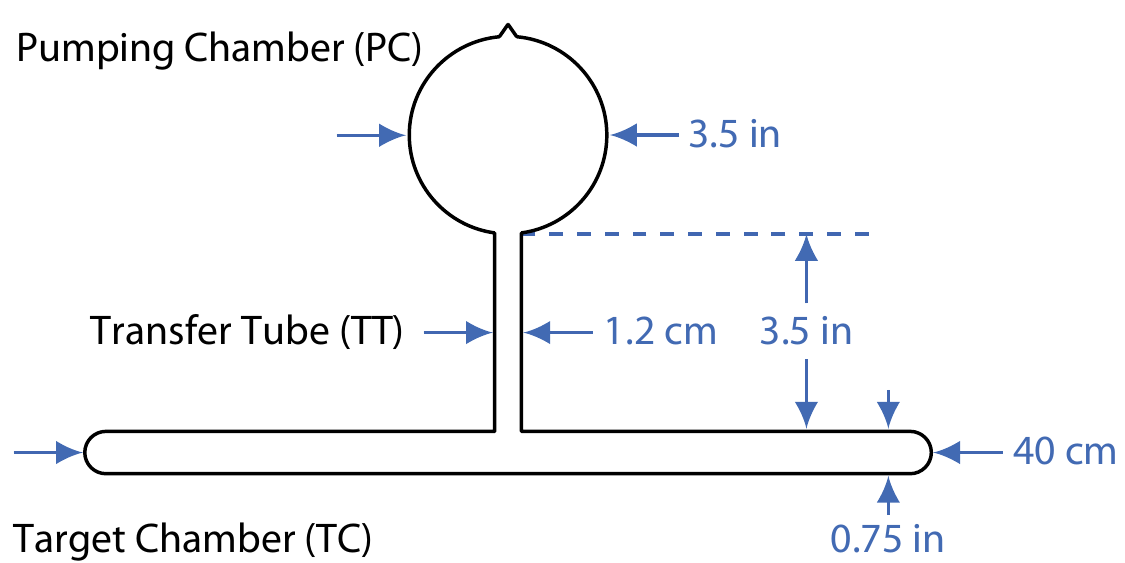}
\caption{(color online) Shown is the geometry of a two-chambered glass cell used for polarized \he targets.  The dimensions shown are typical of those used in GEN.}
\label{fig:genstyle}
\end{center}
\end{figure}

As discussed earlier, the \he target cells studied in this work included two chambers, a pumping chamber and a target chamber.
An example of the cell geometry is shown in Fig.~\ref{fig:genstyle}, which shows dimensions that correspond most closely the the targets used for GEN. All of the tubing used in the cell's construction was re-sized, a process in which the diameter of commercial glass tubing was manually altered using a glass lathe and a hand torch.
Re-sized tubing has proven to be critical to minimizing wall relaxation~\cite{new93,cates_slac_1993}, perhaps because it minimizes the prevalence of microfissures.
The cells used for the saGDH were made of Corning 1720, and all others discussed in this work were made of GE180, both of which are types of aluminosilicate glass.  

Before being sealed, the cells were attached to a glass manifold which was itself connected to a gas-handling system.
The cells were baked under vacuum at about 400$^\circ$C for roughly 48 hours.
It is believed that this process removes moisture and other contaminants that are by-products of the glass-blowing process.
We note that we have, more recently, also baked target cells at a more modest $150-200^\circ$C with apparently similar results.
After baking, the cells were filled with approximately $7$--$9$ amagats of \he gas (see Table~\ref{table:fill}) and a small amount (approximately $0.1$ amagats) of nitrogen to nonradiatively quench the optically pumped alkali atoms.
Alkali-hybrid alloys were distilled into the pumping chamber before the cell was sealed.  

\subsubsection{Creating Alkali-Hybrid Mixtures\label{sec:makemix}}
Alkali metals react violently with oxygen and water.
Consequently, to prepare a hybrid mixture, an inert atmosphere is necessary.
We employed a glove box for this purpose, which was filled with the boil-off of a liquid nitrogen dewar; although nitrogen is not totally inert, it is economical and convenient.
Because it is impossible to avoid trace amounts of oxygen and water, the air inside the glove box was passed continuously through a regeneratable purifier.
The levels of moisture and oxygen were measured with a ``light bulb test'' in which the lifetime of an exposed incandescent filament was monitored; a duration of greater than 2 hours was taken to correspond to contaminant levels less than 5 ppm~\cite{mao97}.
Details on how to calculate the masses of K and Rb required to achieve a prescribed vapor ratio as well as the reliability of the following technique is described in Ref.~\cite{spin2008}.

\begin{figure}[htbp]
 \begin{center}
\includegraphics[width = 6.5cm]{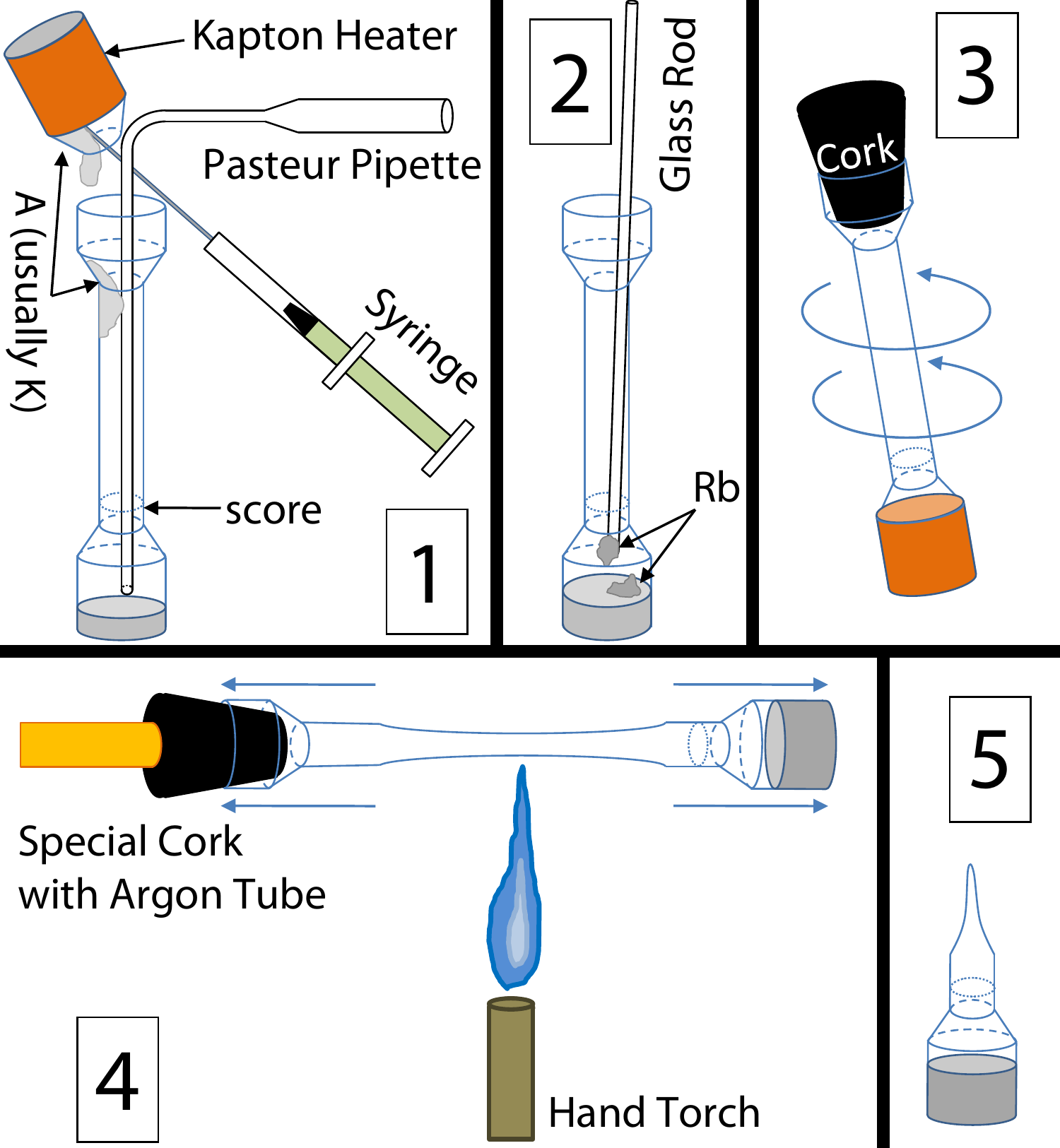}
 \end{center}
 \caption{(color online) Shown is the glove box procedure used to create alkali-hybrid mixtures:
1. Molten alkali (K,Na, or Cs) is forced out of its ampoule with a syringe (filled with \nmns).  
The Pasteur pipette is necessary as the alkali forms beads wider than the neck of the ampoule (which would become clogged).
2. Solid Rb is lowered into the ampoule.  
3. Once the desired mass ratio has been reached, the mixture is corked, heated, and swirled.  
4. The ampoule is permanently sealed under argon.  
5. The prescored ampoule can now be heated and agitated more thoroughly before being used in a target cell.}
\label{fig:GBP}
 \end{figure}
Our procedure for creating alkali-hybrid mixtures is illustrated in Fig.~\ref{fig:GBP} which is adapted from Ref.~\cite{dolph}.
These alloys were prepared by adding an appropriate amount of Rb to a larger quantity of K.
Approximately 1g of molten K was poured into a prescored ampoule with a narrow neck.
The K was then allowed to cool.
Small amounts of solid Rb (which is sticky) were lowered on the end of a glass rod into the prescored ampoule.
The Rb stuck to the K and the rod was removed.
The ampoule with the mixture was removed and weighed on a scale. 
Once the desired mass ratio had been achieved, the prescored ampoule was heated until the mixture melted.
We note that the mass ratio was chosen to provide the desired ratio of K to Rb at a particular operating temperature, and (in our later mixtures) included an adjustment to account for the fact that during distillation into the cell, the Rb moves preferentially faster than does the K.
The molten alloy was then mixed by swirling the prescored ampoule.
Finally, the mixture was corked, cooled, and removed from the glove box.
Once outside, the prescored ampoule was placed inside of an open box which had been filled with argon.
The cork was removed and replaced with a special cork.
The special cork had a hole drilled axially through it and was connected to tubing, which was filled with argon.
The tubing was then removed from the argon bottle so there was no positive pressure inside it.
Finally, the prescored ampoule was sealed along the narrow neck using a hand torch. The sealed ampoule was then heated and thoroughly agitated.

\subsubsection{Determining the $^3$He density}
One of the methods used to determine the \he density in each target involved measurements made during the cell-filling process.
A carefully calibrated volume was used, together with pressure  and temperature measurements, to determine the volume of various spaces in the gas-handling system.
With all relevant volumes determined, the total quantity of \he in the gas-handling system prior to sealing the target cell required knowing only the pressure and temperature of the system.
Once the cell had been filled and ``pulled off'' using a hand torch, the total quantity of \he remaining could be similarly determined, thus establishing the amount of \he in the cell.  
The volume of the cell was measured by determining its buoyancy force in water and applying Archimedes principle.
Knowing the quantity of $^3$He within the cell, and the cell's volume, and further assuming all parts of the cell to be at the same temperature, the $^3$He density is determined to within about 1\%.

A second technique for determining the \he density in each target involved measurements of the pressure-broadening of the D1 and D2 absorption lines of both alkali species using a scannable single-frequency laser.
Using existing accurate measurements of the pressure broadening of Rb absorption lines~\cite{pbRb}, it was possible to monitor the pressure of the sealed target cells at the level of roughly 1\%.
These data also provided a measurement of $D$ by comparing the integral of the absorption lines of the two different alkali species.
While the value of $D$ thus obtained was for the temperature at which the absorption studies were performed,
the value of $D$ at any temperature of interest could be inferred using the alkali-metal vapor pressure curves, and is shown in Table~\ref{table:CellTable} as $D_\mathrm{pb}$.
The ratio $D$ was also determined using the Faraday rotation techniques described in Section~\ref{section:farrot}, shown in Table~\ref{table:CellTable} as $D_\mathrm{fr}$, and the two methods showed good agreement within errors.  

\begin{table}\scriptsize
\begin{tabular}{|c|c|c|c|c|c|}
\hline
\multirow{2}{*}{\begin{sideways}{EXP}\end{sideways}} & Cell & Total & PC  & Fill & TC \\
&& Volume(cc) & Volume(cc) & Density(amg) & length(cm)  \\
\hline
\multirow{5}{*}{\begin{sideways}saGDH\end{sideways}} & Proteus & 235.9 & 90.8 & 6.88 & 34.3 \\
\cline{2-6}
& Peter & 208.6 & 111.3 & 8.80 & 39.4 \\
\cline{2-6}
& Penelope & 204.3 & 102.2 & 8.93 & 39.7 \\
\cline{2-6}
& Powell & 213.3 & 111.6 & 8.95 & 40.5 \\
\cline{2-6}
& Prasch & 257.7 & 114.5 & 6.94 & 35.3 \\
\hline
\hline
\multirow{9}{*}{\begin{sideways}GEN\end{sideways}} & Al & 168.4 & 90.2 & 8.91 & 38.4 \\
\cline{2-6}
& Barbara & 386.2 & 306.8 & 7.60 & 38.7 \\
\cline{2-6}
& Gloria & 378.2 & 298.8 & 7.40 & 38.4 \\
\cline{2-6}
& Anna & 386.8 & 303.7 & 8.09 & 38.7 \\
\cline{2-6}
& Dexter & 181.4 & 99.3 & 9.95 & 38.7 \\
\cline{2-6}
& Edna & 378.3 & 290.3 & 7.47 & 38.7 \\
\cline{2-6}
& Dolly & 378.3 & 293.5 & 7.42 & 38.7 \\
\cline{2-6}
& Simone & 219.5 & 118.6 & 8.17 & 37.9 \\
\cline{2-6}
& Sosa & 388.8 & 304.7 & 7.96 & 38.7 \\
\hline
\hline
\multirow{10}{*}{\begin{sideways}Transversity and $d_2^n$\end{sideways}} & Boris & 246.1 & 166.1 & 8.08 & 38.4 \\
\cline{2-6}
& Samantha & 259.0 & 176.9 & 7.97 & 38.4 \\
\cline{2-6}
& Alex & 278.3 & 193.9 & 7.73 & 39.1 \\
\cline{2-6}
& Moss & 269.8 & 184.7 & 7.92 & 38.7 \\
\cline{2-6}
& Tigger & 271.7 & 186.9 & 7.81 & 38.7 \\
\cline{2-6}
& Astral  & 251.4 & 164.9 & 8.18 & 38.4 \\
\cline{2-6}
& Stephanie & 244.3 & 164.9 & 8.10 & 38.5 \\
\cline{2-6}
& Brady & 249.9 & 169.3 & 7.88 & 38.4 \\
\cline{2-6}
& Maureen & 268.5 & 177.4 & 7.63 & 39.8 \\
\cline{2-6}
& Antoinette & 437.8 & 351.8 & 6.57 & 40.3 \\
\hline
\end{tabular}
\caption{Shown are the names, total and pumping-chamber volumes, fill densities and target-chamber lengths of the 24 target cells included in this study.  Also indicated (left-most column) are the experiments for which the targets were constructed.}
\label{table:fill}
\end{table}

The fill densities and other geometric specifications of the 24 cells we investigated are shown in Table~\ref{table:fill}.
Where possible the fill density shown is the average of the value obtained from our gas handling system and that obtained through pressure-broadening measurements.
The two methods for determining the $\mathrm{^3He}$ fill density were in agreement within uncertainties.

\subsubsection{Operation of the target cells}
The target cells were studied in an optical-pumping apparatus similar in many essential respects to the apparatus used to operate the targets during the experiments for which they were constructed.
They were heated using a forced-hot-air alumina-silicate ceramic oven, the set temperature of which is listed in Table~\ref{table:CellTable} for our various measurements.
It should be noted, however, that the set temperature is not the only relevant measure of the important temperatures affecting the operation of the cell.

Since our targets have two chambers, it was essential to know accurately the volume-averaged temperature in the pumping chamber.
This information was necessary when calculating the density of \he that was present in the target chamber during operation.
While the oven set temperature was measured using a thermocouple attached directly to the pumping chamber,
the temperature of the gas within during operation was significantly higher, by an amount $\Delta T_\mathrm{He}$, because of heating due to the lasers used for optical pumping.
We measured $\Delta T_\mathrm{He}$ by making a succession of NMR measurements with the lasers alternately turned on and off.
The difference in the measurements reflected the fact that when the lasers were turned on, the additional heating would redistribute gas from the pumping chamber to the target chamber.
We typically found values for $\Delta T_\mathrm{He}$ in the range of $20-50^\circ\rm C$, as is shown in Table~\ref{table:CellTable}.

Given the substantial values of $\Delta T_\mathrm{He}$, it is natural to ask what the relevant temperature was that determined the alkali number density.
For measurements during which Faraday rotation was used, we had a direct measurement of the alkali density, and could thus infer a temperature from expectations based on the Rb and K vapor-pressure curves listed in Ref.~\cite{CRC75}.
The difference between this {\it inferred} temperature and the oven set temperature is shown in Table~\ref{table:CellTable} as $\Delta T_\mathrm{Rb}$, and is always less than $10^\circ\rm C$.
This information provides a valuable measure of the limits inherent in estimating the alkali density based on the oven temperature alone.
We note that the inferred temperature was much closer to the oven set temperature than that of the volume averaged temperature of the gas.

The target cells were illuminated with several lasers of two distinct kinds: a spectrally broadband laser (roughly $\rm 2\,nm$ linewidth) known as a Coherent FAP (for fiber array package), manufactured by Coherent Lasers Inc., 
and a spectrally narrowed laser (roughly $\rm 0.2\,nm$ linewidth) known as a Comet system, manufactured by a subsidiary of the Newport Corporation. 
Both systems produced roughly 25~W.
Since the data we present were obtained while testing our targets, we did not always measure the exact output power.
In Table~\ref{table:CellTable}, the number of broadband lasers (B) and narrowed lasers (N) used in each test is indicated.
The average intensity is also estimated based on whatever the most current power measurements were at the time of the test.
We typically used three lasers, which is why in Sec.~\ref{sec:theory} we use 75~W as something of a standard for our simulations.

\subsection{Target-cell polarization dynamics}
The dynamics of the buildup of polarization in a double-chambered cell is somewhat involved, and has been discussed recently by Dolph \etal in Ref.~\cite{convection}.
For what follows, we have used the same notation.

\begin{figure}[htbp]
 \begin{center}
\includegraphics[width = 8.6cm]{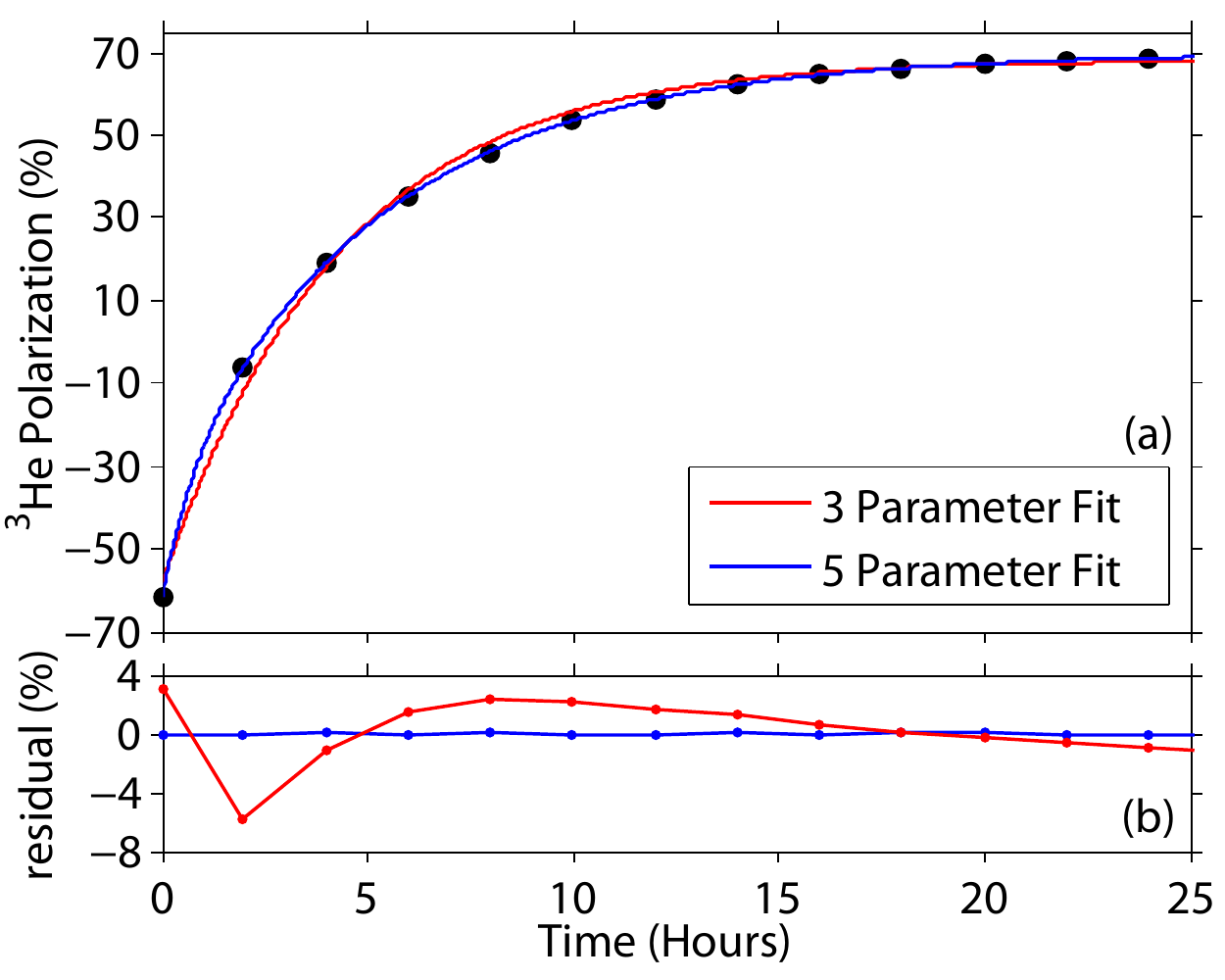}
 \end{center}
 \caption{(color online) (a) Shown is a ``spinup'' of the cell Brady in which polarization was measured as a function of time while the cell moved toward equilibrium.  The spinup data has been fit with a 3-parameter and a 5-parameter formalism (see text).
(b) The residuals of the two fits are shown.  The residual for the 3-parameter fit is large because it does not account for diffusion between the cell's two chambers.}
\label{fig:Spinup}
 \end{figure}

We monitored the accumulation of \he polarization using the NMR technique of adiabatic fast passage (AFP)~\cite{Abragam}.
These NMR measurements were calibrated using a technique in which Electron Paramagnetic Resonance (EPR) was performed on the alkali vapor in the pumping chamber, yielding an absolute polarization determination of the \he nuclei~\cite{PhysRevA.58.3004,babcock-epr}.
An example of a ``spinup,'' in which NMR measurements are successively made while a cell becomes polarized,  is shown in Fig.~\ref{fig:Spinup}.
The polarization measurements shown in Fig.~\ref{fig:Spinup}, as well all the others reported in this work, were measured in the pumping chamber of the cell.

In a single-chambered cell, a spinup is described by:
\begin{equation}
P(t) = (P^0-P^\infty)e^{-\Gamma_\mathrm{sc}t} + P^\infty
\label{eq:3parameterfit}
\end{equation}
where $P^\infty$ is the asymptotic polarization, $P^0$ is the initial polarization, and $\Gamma_\mathrm{sc} = \gamma_\mathrm{se}(1+X) + \Gamma$ is the spinup rate describing the buildup of polarization.
We use the subscript $\rm sc$ as an abbreviation for ``single chamber,'' and to distinguish it from a variable introduced below.
The quantity $\gamma_\mathrm{se}$ is the spin-exchange rate, and the quantity $X$ is used to parameterize an as-yet poorly understood spin-relaxation mechanism, first identified by Babcock \etal~\cite{PhysRevLett.96.083003}, that appears to be roughly proportional to $\gamma_\mathrm{se}$.
The quantity $\Gamma$ represents the spin relaxation rate due to all mechanisms other than spin exchange and the relaxation mechanism parameterized by $X$, and is typically measured at room temperature.
The equilibrium polarization $P^\infty$ is given by
\begin{equation}
P^\infty = {{\langle P_\al \rangle \gamma_\mathrm{se}}\over{\Gamma_\mathrm{sc}}} = {{\langle P_\al \rangle \gamma_\mathrm{se}}\over{\gamma_\mathrm{se}(1+X) + \Gamma}}\ \ 
\label{eq:psatsinglecell}
\end{equation}
where $\langle P_\al \rangle$ is the polarization of the alkali vapor averaged over the cell. 

In a double-chambered cell, the time evolution of polarization in the pumping chamber, $P_\pc(t)$, and target chamber, $P_\tc(t)$, are distinct from one another because of the time required for \he to move (mostly by diffusion) between the two chambers, and is given by:
\begin{equation}
P_\mathrm{pc}(t) = C_\mathrm{pc}e^{-\Gamma_\mathrm{f}t} + (P_\mathrm{pc}^0-P_\mathrm{pc}^\infty - C_\mathrm{pc})e^{-\Gamma_\mathrm{s}t} + P_\mathrm{pc}^\infty
\label{eqn:ppct}
\end{equation}
\begin{equation}
P_\mathrm{tc}(t) = C_\mathrm{tc}e^{-\Gamma_\mathrm{f}t} + (P_\mathrm{tc}^0-P_\mathrm{tc}^\infty - C_\mathrm{tc})e^{-\Gamma_\mathrm{s}t} + P_\mathrm{tc}^\infty
\label{eqn:ptct}
\end{equation}
where $P_\mathrm{pc}^0$ ($P_\mathrm{tc}^0$) is the initial polarization in the pumping (target) chamber, and $P_\mathrm{pc}^\infty$ ($P_\mathrm{tc}^\infty$) is the equilibrium polarization in the pumping (target) chamber.
The time constant $\Gamma_\mathrm{s}$ is the ``slow'' time constant, which essentially plays the role of $\Gamma_\mathrm{sc}$ in a single-chambered cell, and is given by
\begin{equation}
\Gamma_\mathrm{s} = \langle\gamma_\mathrm{se}\rangle(1+X) + \langle\Gamma\rangle - \delta\Gamma
\label{eqn:gammaS}
\end{equation}
where the quantity $\delta\Gamma$ contains corrections that arise because of the finite time it takes the \he to move between the target cell's two chambers.
We note that in our studies, the value of $\delta\Gamma$ was typically no more than 10\% of the size of $\Gamma_\mathrm{s}$, and never more than 15\%.  
The quantity $\langle\gamma_\mathrm{se}\rangle = f_\mathrm{pc}\,\gamma_\mathrm{se}$ is the cell-averaged spin-exchange rate, where $\gamma_\mathrm{se}$ is the spin-exchange rate in the pumping chamber, and $f_\mathrm{pc}$ is the fraction of the total number of $^3$He atoms that are in the pumping chamber.  
The quantity $\langle \Gamma \rangle$ corresponds to the quantity $\Gamma$ defined before, but is averaged over the entire target cell, thus allowing for the possibility that relaxation rates may be different in the pumping and target chambers.
The new rate that appears, $\Gamma_\mathrm{f}$ ($\mathrm{f}$ for fast) also arises because of the dynamics associated with two chambers.
Since $\Gamma_\mathrm{f}$ is, for our conditions, substantially faster than $\Gamma_\mathrm{s}$, the time dependence of $P_\mathrm{He}$ is characterized by a single rate at later times.
The quantities $\delta\Gamma$, $\Gamma_\mathrm{f}$,  $C_\mathrm{pc}$ and $C_\mathrm{tc}$ are functions of geometry, the various rates and initial conditions, but are time independent.
All of these quantities are defined in detail in Ref.~\cite{convection}.

When polarizing a target, two quantities of considerable interest include $P_\mathrm{pc}^\infty$ (or similarly $P_\mathrm{tc}^\infty$), and the rate $\Gamma_\mathrm{s}$.
In Table~\ref{table:CellTable} of Section~\ref{sec:targetproperties}, we list values of $P_\mathrm{pc}^\infty$ and $\Gamma_\mathrm{s}$ that resulted from five-parameter fits of spinup data to Eqn.~(\ref{eqn:ppct})
from each of our target cells for each temperature studied.
An example of such a five-parameter fit is shown in Fig.~\ref{fig:Spinup}, along with a three parameter fit for comparison.
The residuals to the fits, also shown in Fig.~\ref{fig:Spinup}, clearly indicate that a five-parameter fit more closely describes the data.  

Understanding {\it why} one achieves a particular value of $P_\mathrm{pc}^\infty$ requires additional measurements.
One quantity that is straightforward to measure is $\langle \Gamma \rangle_\mathrm{c}$, the cell-averaged relaxation rate with which the polarization of a particular cell will decay at room temperature.
The subscript $\mathrm{c}$ is used to distinguish this quantity from $\left < \Gamma \right >$, which is the analogous relaxation rate when the cell is at operating temperature.
Measurements of the remaining relevant parameters requires considerable additional work and represented a central effort in our target-development work.

\subsection{Faraday Rotation \label{section:farrot}}
In addition to $\langle\Gamma\rangle$, parameters that are important for understanding the limits on \he polarization include (referring to Eqn.~(\ref{eq:psatsinglecell}) for simplicity) $\langle P_\al \rangle$, 
$\gamma_\mathrm{se}$ (which is in turn is proportional to the density of alkali-metal atoms) and $X$.
A useful diagnostic for studying these parameters in a target cell is the observation of Faraday rotation~\cite{Wu:86,PhysRevA.66.032703} using a linearly polarized probe laser.  

Faraday rotation refers to the change in the orientation of the polarization axis that occurs when linearly polarized light passes through a polarized alkali vapor.
It occurs because the polarized alkali vapor exhibits circular birefringence.
For our purposes, it is sufficient to consider only the alkali-metal atom's D1 and D2 lines, in which case the Faraday rotation angle, $\phi_\mathrm{r}$ is given by:
\begin{equation}
\phi_\mathrm{r}(\nu) = \left ( \frac{r_e c}{6} \right ) P_\al \cos(\theta) [\rb] l \left \{ F_\rb(\nu) + D F_\kp(\nu) \right \} \label{eqn:farrotangle}
\end{equation}
where $l$ is the path length through the vapor, the other parameters are the same as those for the absorption lineshape given by Eqn.~(\ref{eqn:sigmaabs}) in Sec.\ref{sec:ahop},
we've assumed that the alkali D1 \& D2 oscillator strengths are $f_1 = 1/3$ \& $f_2 = 2/3$, and the function $F_\al(\nu)$ describes the frequency dependence of alkali species A, which is given by:
\begin{equation}
F_\al(\nu) = 
\frac{\nu}{\nu_1} \left [ \frac{\Delta_1}{\Delta_1^2 + \Gamma_1^2/4} \right ] - \frac{\nu}{\nu_2} \left [ \frac{\Delta_2}{\Delta_2^2 + \Gamma_2^2/4} \right ] \ \ .
\end{equation}

\begin{figure}[htbp]
\begin{center}
\includegraphics[width = 8.6cm]{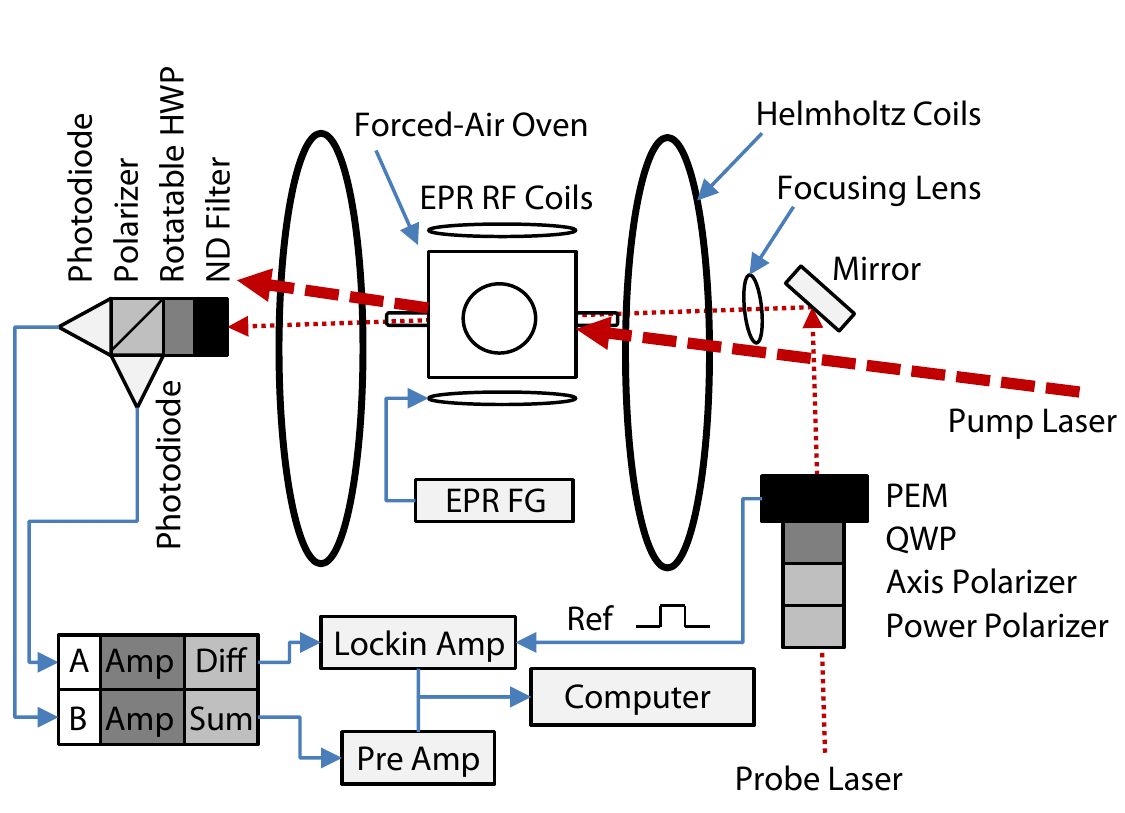}
\caption{(color online) Experimental setup for Faraday rotation studies. 
For clarity, the pump laser (used for optical pumping) has been drawn at an angle relative to the Helmholtz field.
In reality, the pump beam is nearly parallel to the Helmholtz field.  See text for details.}
\label{fig:expSetup}
\end{center}
\end{figure}

A schematic illustrating the experimental setup that was used to observe Faraday rotation is shown in Fig.~\ref{fig:expSetup}.
The probe beam first passed through two polarizing cubes. 
The first polarizing cube was used to control the probe beam power (typically 1~mW after the PEM), the second to define the linear polarization axis of the beam.
The linearly-polarized beam then passed through a quarter wave plate (QWP) and a $\rm50\,kHz$ photoelastic modulator (PEM) -- the QWP and PEM were used in conjunction with a lock-in amplifier to measure the relatively weak signals.
After the probe beam exited the cell, it passed through a neutral density (ND) filter, a rotatable half-wave plate (HWP), and a polarizing beam-splitting cube before being detected in two photodiodes.
The ND filter was used to block out room light and minimize background from the pump laser.
The polarizing cube was used to separate the linear polarization into its horizontal and vertical components.
The two photodiode signals were amplified before being added or subtracted from each other.
The HWP was used to null the difference between the photodiode signals --- before the pump lasers were turned on, the HWP was rotated until both photodiodes collected the same amount of light.
Provided the probe beam was sufficiently detuned, the ratio of the difference and the sum of the two photodiode signals, $\Delta$ (not to be confused for the detuning) and $\Sigma$ respectively, was given by
\begin{equation}
\frac{\Delta}{\Sigma} = N \sin \left (  2\phi_\mathrm{r} + 2\phi_\mathrm{misc} - 4\phi_\mathrm{HWP} \right ) 
\label{eqn:phitot}
\end{equation}
where $\phi_\mathrm{misc}$ is an alkali-polarization-independent offset, $\phi_\mathrm{HWP}$ is the HWP setting, and $N$ (which we call the ``normalization") is a function of wavelength, PEM retardation, and other electronic settings.
The normalization, $N$, and the phase $\Phi$, defined by
\begin{equation}
\Phi \equiv 2\phi_\mathrm{r} + 2\phi_\mathrm{misc}~,
\label{eqn:NormPhase}
\end{equation}
were measured (such measurements will be called ``normalizations") by monitoring $\Delta/\Sigma$ at a fixed $\phi_\mathrm{r}$ as the HWP was rotated $360^\circ$.  

Before a Faraday rotation angle was measured, the HWP was rotated such that $\Delta = 0$ (we refer to this as ``nulling" the signal).
Nulling was usually performed with the pump lasers off (which forced $P_\al = 0$). 
Under these conditions, $4\phi_\mathrm{HWP} = 2\phi_\mathrm{misc}$ and Eqn.~(\ref{eqn:phitot}) can be written as
\begin{equation}
2\phi_\mathrm{r} = \arcsin \left ( \frac{1}{N}\frac{\Delta}{\Sigma} \right ) \ \ .
\label{eqn:2phir}
\end{equation}

\begin{figure}[htbp]
\begin{center}
\includegraphics[width = 8.6cm]{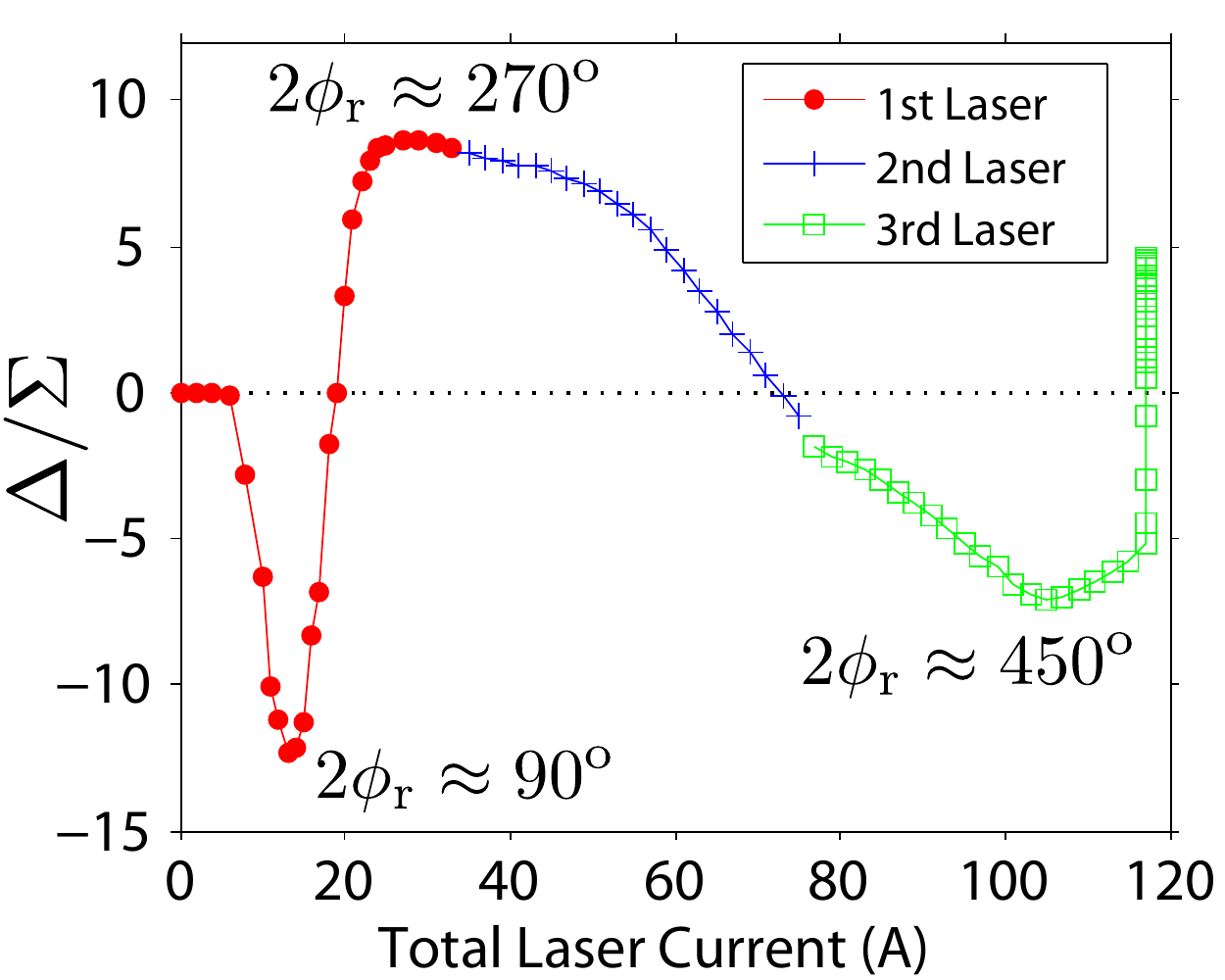}
\caption{(color online) Ramp-up scan at 781~nm and $235\,^\circ \rm C$ for the target cell Brady during a Faraday rotation study.
Shown is the quantity $\Delta/\Sigma$ as three lasers are slowly ramped up in succession thus producing 3 distinct flips and 3 zero crossings.
The irregular shape is created because the lasers heat up the cell, whose temperature is controlled by a proportional-integral-derivative (PID) feedback circuit.}
\label{fig:bradyramp}
\end{center}
\end{figure}

Because the $\arcsin$ is used to measure the rotation angle, see Eqn.~(\ref{eqn:2phir}), care was needed to be taken to distinguish which $180^\circ$ domain of $2\phi_\mathrm{r}$ was being observed.
One way to accomplish this was by slowly ramping the lasers on while monitoring the rotation angle, which increased as the increased laser power polarized the alkali vapor; we refer to this as a ``ramp-up.''
In the ramp-up shown in Fig.~\ref{fig:bradyramp}, three distinct maxima and minima (flips) and three zero crossings are visible.

A second method for checking the domain was to measure the normalization phase, $\Phi$, before and after the lasers were ramped up.
A value for $2\phi_\mathrm{r}$ can be found by taking the difference between these two numbers, see Eqn.~(\ref{eqn:NormPhase}). 
Although this method is harder to visualize, it gives a larger ($360^\circ$) domain for $2\phi_\mathrm{r}$.
This method was used to confirm that only 3 flips occurred in Fig.~\ref{fig:bradyramp}.

\begin{figure}[htbp]
\begin{center}
\includegraphics[width = 8.6cm]{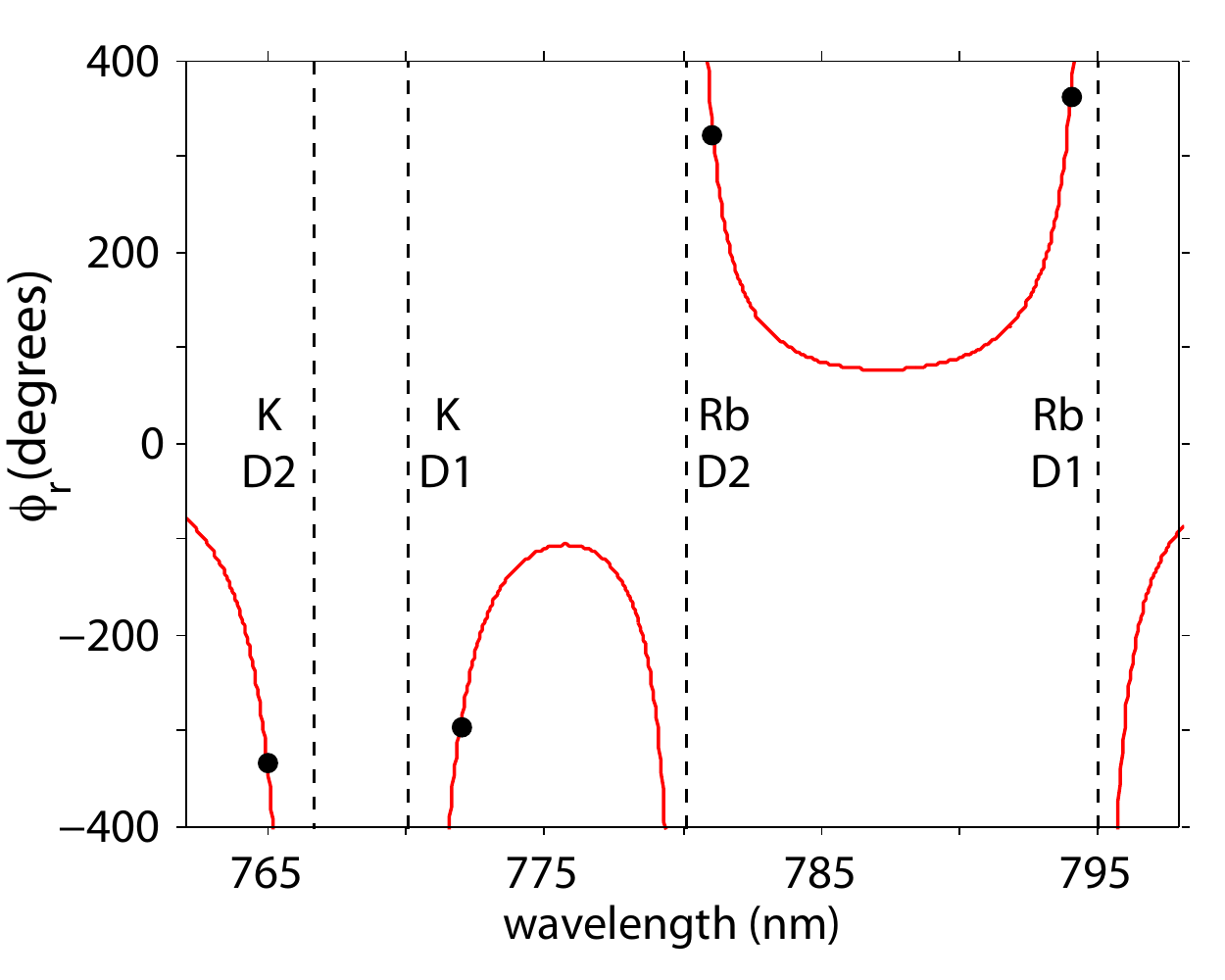}
\caption{(color online) Shown are measurements of the Faraday rotation angle for target cell Brady, at 235C, for each of four wavelengths of the probe laser.
The data are fit to Eqn.~(\ref{eqn:farrotangle}), and the vertical lines show the location of the D1 and D2 lines for both K and Rb.}
\label{fig:bradyFarrot}
\end{center}
\end{figure}

During a typical Faraday rotation measurement, $\phi_\mathrm{r}$ was measured at several probe wavelengths and fit to Eqn.~(\ref{eqn:farrotangle}).
For example, the Faraday rotation data shown in Fig.~\ref{fig:bradyFarrot} were obtained with an oven set temperature of $235^\circ$C and it was found that $P_\al[\mathrm{Rb}]l = (18.7 \pm 1.8) \times 10^{14}/\mathrm{cm}^2$, and $D = 2.6\pm0.2$.
We will discuss how we analyzed numbers such as these in subsequent sections.

\subsection{Line-Averaged Alkali Polarimetry}
\label{section:alkpol}
\begin{figure}[htbp]
\begin{tabular}{l}
\includegraphics[width = 8.6cm]{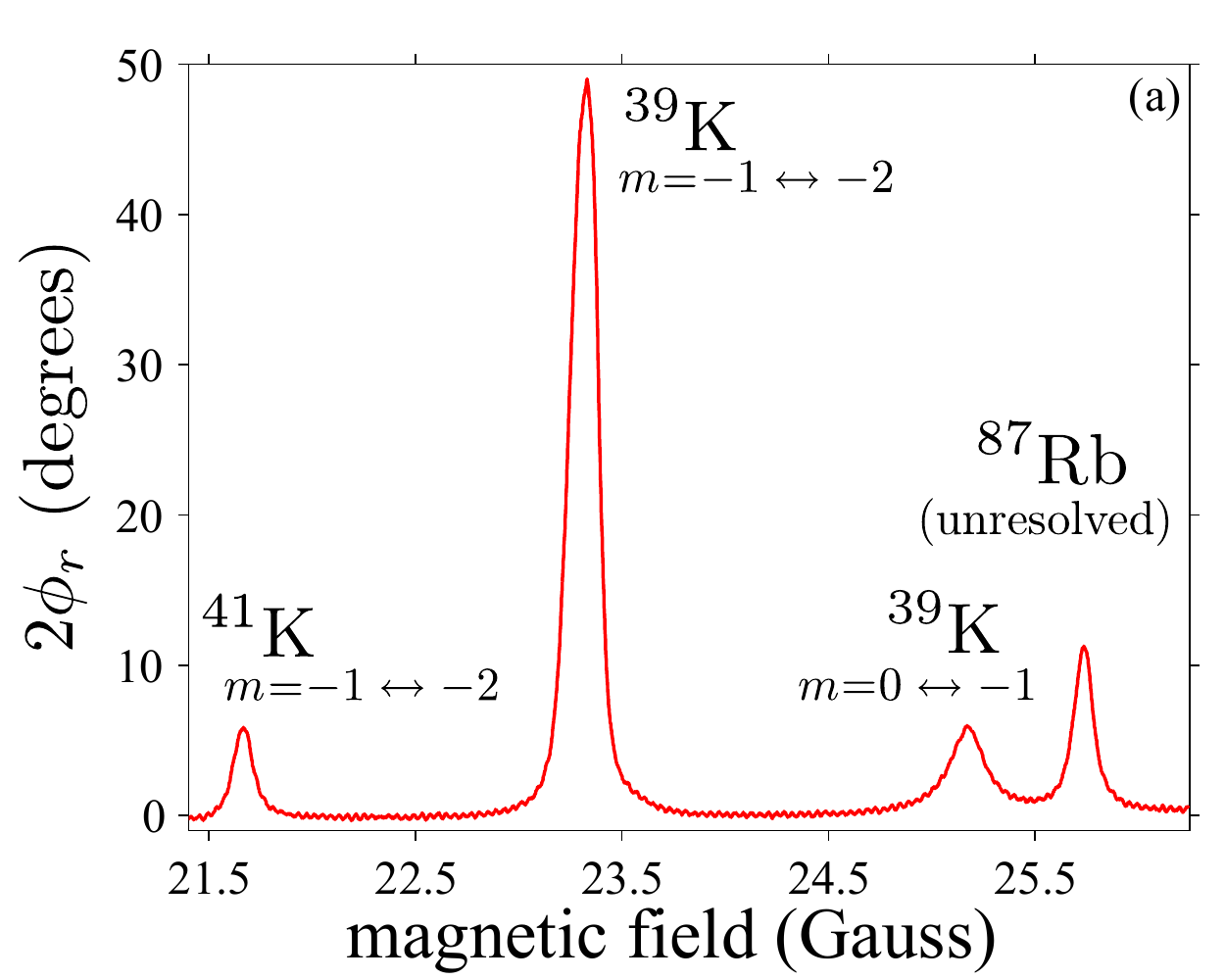} \\
\includegraphics[width = 8.6cm]{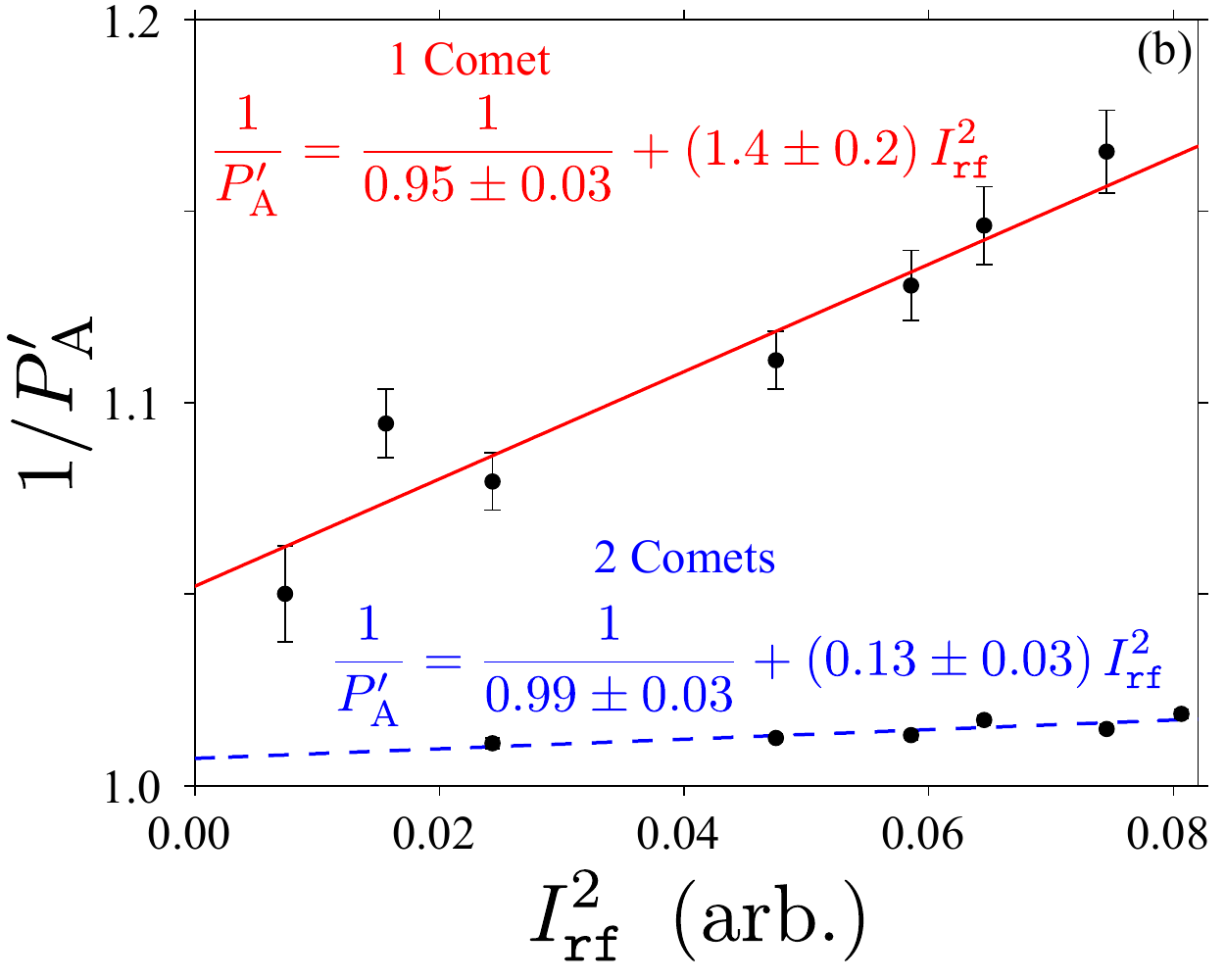}
\end{tabular}
\caption{(color online) (a) High-RF alkali polarization scans, obtained using Faraday rotation data, with 1 Comet laser for target cell Brady (probe wavelength 785nm).
With 3 lasers, the alkali polarization is so high that it is difficult to see the second $^{39}$K peak. 
(b) RF amplitude scan for 1 and 2 Comet lasers on the target cell Brady. Individual alkali polarization scans (see top figure) are plotted and then fit to Eqn.~(\ref{eqn:Prb2}).}
\label{fig:Alkspectra}
\end{figure}

In order to extract a value for $[\rb]$, it is first necessary to measure $l$ and $P_\al$.
Alkali polarization measurements can be made by probing Zeeman-transition populations~\cite{rbpolimag}.
Under our operating conditions, the populations of these sublevels are well modeled by a spin-temperature distribution~\cite{PhysRev.116.87}.
We made in situ measurements of the alkali polarization by monitoring the Faraday rotation angle while Zeeman transitions were induced~\cite{PhysRevA.66.032703}.
Such transitions depolarize the alkali-metal vapor and consequently decrease the Faraday rotation angle.
During a measurement, see Fig.~\ref{fig:Alkspectra}.A, the main holding field was swept through the Zeeman transitions, which for us, was kept at $\rm 18.2\,MHz$.
A spectrum of Lorentzian resonances was produced.
A value for the alkali polarization was extracted by comparing the areas of successive peaks.

The area under a particular peak corresponding to the transition $m \leftrightarrow m-1$ is given by
\begin{equation}
A(F,m) = A_0 \left [  F(F+1) - m(m-1) \right ] \exp \left ( \beta m \right ) 
\label{eqn:A}
\end{equation}
where $F$ is the total atomic angular momentum, $m$ is the azimuthal component, $A_0$ is a proportionality constant that is independent of $F$ \& $m$, and $\beta$ is the spin temperature given by
\begin{equation}
\beta = \log \left [ \frac{1+P_\al}{1-P_\al} \right ]
\label{eqn:beta}
\end{equation}
Under our operating conditions, transitions in the same $F$ manifold were well resolved; however, transitions involving the same $m$ were unresolved between the $F$ and $F-1$ manifolds.
Such transitions are called twin transitions.

To extract a value for $P_\al$, we compared adjacent areas in our spectra, computing the quantity
\begin{equation}
r = \frac{A(F,m) + A(F-1,m)}{A(F,m-1)+A(F-1,m-1)}
\label{eqn:R}
\end{equation}
where, since we could not resolve the twin transitions, the quantity $r$ contains contributions from both F manifolds.  
Expressions for $P_\al$ can be computed in terms of $r$ by combining Eqns.~(\ref{eqn:A})--(\ref{eqn:R}),  and are shown in Table~\ref{table:P} $P_\al$ for the different transitions. 

\begin{table}[h!]
\begin{center}
{\large
{
\renewcommand{\arraystretch}{1.6}
\begin{tabular}{|c|c|c|}
\hline
{\normalsize $r$ (transition ratio)} & {\normalsize $P_\al$ ($I = 5/2$) }&{\normalsize $P_\al$ ($I = 3/2$) }\\ 
\hline
\hline
$\frac{m=3\leftrightarrow 2}{m-1=2\leftrightarrow 1}$& $\frac{r -3/7}{r + 3/7}$ & \\
\hline
$\frac{m=2\leftrightarrow 1}{m-1=1\leftrightarrow 0}$& $\frac{r -7/9}{r + 7/9}$ & $\frac{r -1/2}{r + 1/2}$\\
\hline
$\frac{m=1\leftrightarrow 0}{m-1=0\leftrightarrow -1}$& $\frac{r -1}{r + 1}$ & $\frac{r -1}{r + 1}$\\
\hline
$\frac{m=0\leftrightarrow -1}{m-1=-1\leftrightarrow -2}$& $\frac{r -9/7}{r + 9/7}$ & $\frac{r -2}{r + 2}$\\
\hline
$\frac{m=-1\leftrightarrow -2}{m-1=-2\leftrightarrow -3}$& $\frac{r -7/3}{r + 7/3}$ & \\
\hline
\end{tabular}}
}
\end{center}
\caption{ Shown are expressions for the alkali polarization in terms of the quantity $r$ (as defined in Eqn.~(\ref{eqn:R})) for isotopes with nuclear spins $I=5/2$ (Rb-85) and $I=3/2$ (K-39 and Rb-87).}
\label{table:P}
\end{table}

We note that the alkali polarization measurement itself introduces an additional alkali-relaxation mechanism.
In the presence of RF, we can generalize the alkali polarization in Eqn.~(\ref{eq:alkpolsimple}) to include the effect of the RF as:
\begin{equation}
P_\al = \frac{R_1}{R_{01} + \Gamma_\al} \rightarrow P'_\al = \frac{R_{1}}{R_{01}+\Gamma_\al + \Gamma_\rf}
\label{eqn:Prb}
\end{equation}
where $\Gamma_\rf$ is the EPR RF depolarization rate.
It is convenient to rewrite Eqn.~(\ref{eqn:Prb}) as
\begin{equation}
\frac{1}{P'_\al} = \frac{1}{P_\al} + k I_\rf^2
\label{eqn:Prb2}
\end{equation}
where $P_{\al}$ is the alkali polarization in the absence of an RF field given by Eqn.~(\ref{eq:alkpolsimple}), 
$k$ is a proportionality constant, and it is assumed that $\Gamma_\rf \propto B_\rf^2$, the square of the magnetic field created by the RF coil, which in turn is proportional to $I_\rf^2$, the current in the RF coil.
We extrapolated to $\Gamma_\rf = 0$ by performing several sweeps with different RF field amplitudes.
Each individual sweep was fit and a value for $P'_\al(\Gamma_\rf)$ was obtained. 
The resulting polarizations were then fit to Eqn.~(\ref{eqn:Prb2}) and a value for $P_\al$ was obtained.  

Shown in Fig.~\ref{fig:Alkspectra} are data from a typical alkali polarization measurement, which was performed at an off-resonance probe wavelength of $\rm 785\,nm$. 
With one Comet laser, the extrapolated zero-RF polarization was $P_\al = 0.95\pm0.03$.
The value for two Comet lasers was $P_\al = 0.99\pm0.03$. 
With 3 Comet lasers, the alkali polarization was too high to measure, so we assume $P_\al = 0.99\pm0.03$, where the lower limit on polarization is obtained by assuming that the smaller peaks were of the order of the size of the noise.
Combining this result with the data from Fig.~\ref{fig:bradyFarrot}, we find $[\rb]l = (18.9\pm 1.9)\times 10^{14}/\mathrm{cm}^2$.

\subsection{Path Length Determination}
\begin{figure}[htbp]
\begin{tabular}{ll}
\hskip 0.1truein\includegraphics[width = 4.0cm]{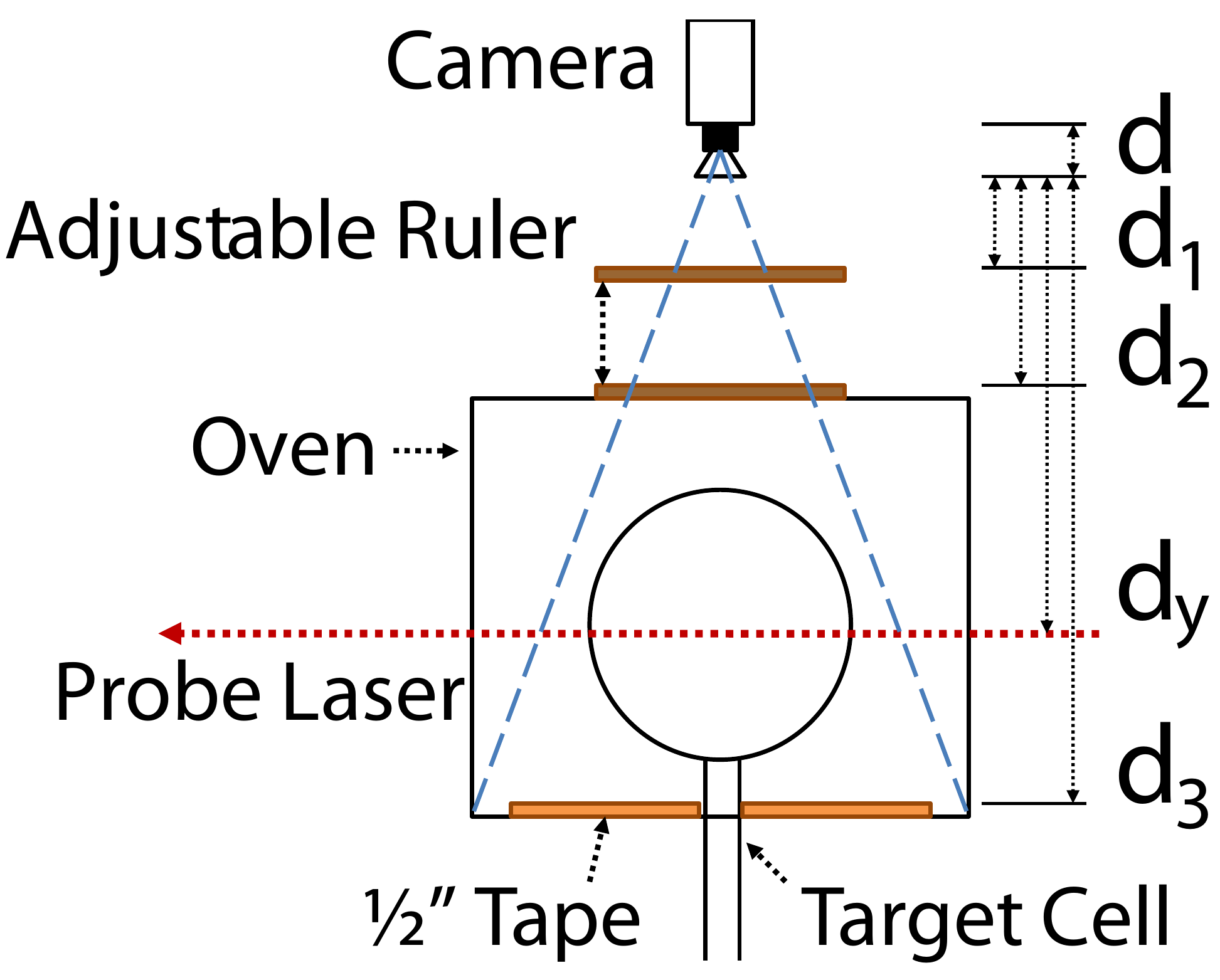}&
\hbox{\hskip -0.9truein \vbox{\vskip -0.7truein\includegraphics[width = 3.0cm]{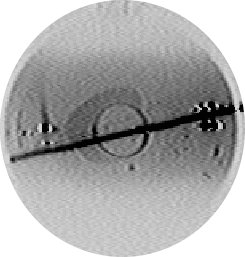}}}
\end{tabular}
\caption{(color online) Illustrated is the measurement of the path length $l$ of the probe laser during a Faraday rotation study.  At left is shown the basic setup for obtaining path-length data (see text for details), and at right is a CCD image showing the probe laser passing through the cell.}
\label{fig:pathLength}
\end{figure}

The path length of the probe laser inside the cell was difficult to measure with high accuracy.
Although measurements of the outer diameter of the pumping chamber were trivial, our cells were hand blown and had large variations in the thickness of the glass.
Moreover, it was difficult to ensure that the probe beam passed through a full diameter of the sphere and not just a chord.
Measurements of the path length of the probe laser were thus obtained using images from a CCD camera together with an appropriate calibration (see Fig.~\ref{fig:pathLength}).
The calibration was performed by calculating the ratios $c_i$ of the actual size (in cm) to the image size (in pixels) of a ruler at various distances $d_i$ from the camera.  Such a set of images was expected to obey the relation
\begin{equation}
c_i = \alpha(d_i+d)
\end{equation}
where $\alpha$ is a constant and $d$ is an unknown offset related to the distance between the front of the camera and the CCD sensor. 
Several measurements of $c_i$ were fit to a line, which yielded values for $d$ and $\alpha$. 
A value for $c_y$, the calibration constant at the location of the probe beam, was thus obtained. 
The path length image (again see Fig.~\ref{fig:pathLength}),  was obtained at modest alkali densities in the absence of the pump laser, while using a D2 filter, and with the probe laser tuned slightly off the D2 resonance. 
The path length of the probe beam for target cell Brady was found to be $(6.59\pm 0.20)\ \mathrm{cm}$.
We note that this is consistent with the typical pumping chamber wall thickness of $1.5$ mm.
Using the information from Secs.~\ref{section:farrot} and \ref{section:alkpol}, this gives $[\rb] = (2.86\pm 0.30)\times 10^{14}/\mathrm{cm}^3$. 

\subsection{Volume-Averaged Alkali Polarimetry}

One of the key parameters that determines the $^3$He polarization is the volume-averaged alkali polarization, $\langle P_A \rangle$, which is not the same as the line-averaged alkali polarization, $\pal$, whose measurement is discussed in Section~\ref{section:alkpol}.  The probe laser used for the Faraday rotation measurements was nearly parallel to direction of propagation of the optical-pumping lasers, and was well centered on the pumping chamber.  The polarization along this line was thus somewhat higher than the volume average.  To obtain the volume averaged-polarization, we calculated two corrections, the product of which is the ratio $\langle P_A \rangle/\pal$ that is listed in Table~\ref{table:CellTable}.  In all but three cases, the net correction was less than 10\%.

One of the two corrections was computed using the simulation described in Section~\ref{sec:theory} that was also used to guide us in the design of our alkali-hybrid targets. We used as inputs to the simulation the geometry of the cell, its pressure, the value of the ratio $D$ and other parameters, including information concerning the spectral content of the various optical-pumping lasers.  For the alkali density, we used the value measured using Faraday rotation.  We then compared the {\it computed} line-averaged polarization along the path of the probe laser to the {\it actual} value obtained using the methods of Section~\ref{section:alkpol}.  Generally, there was a small disagreement of a few percent. Since our line-averaged alkali polarizations had smaller errors than the errors on the alkali density, we then varied the central value of the alkali density within its error bars until the simulation successfully reproduced something very close to the measured line-averaged polarization.  Occasionally, we would also vary the central value of $D$.  Having completed the fine tuning of our inputs, the ratio of the computed values of $\langle P_\mathrm{A} \rangle$ and $\pal$ was one of our two corrections.  The error associated with this correction encompassed the full range of values obtained by varying the inputs of both the alkali density and $D$ within our experimental errors.

The second correction, not accounted for by the simulation, adjusted for portions of the cell that were not well illuminated with laser light.  In all of our tests, a small portion of the transfer tube connecting the pumping and target chambers was at elevated temperatures inside our oven, and hence had an alkali number density similar to that of the pumping chamber.  The transfer tube was not illuminated by the laser, however, meaning that approximately 1.5\% of the alkali density was not optically pumped.  
It is also the case that those portions of the cell near the edges were only poorly illuminated with imperfectly circularly polarized light.
This is because light near the edges of the pumping chamber hit the glass at a nearly glancing angle.  The net correction from poorly illuminated portions of the cell (not already accounted for by the simulation) was $0.97\pm0.02$.  The product of this factor and the factor generated using the simulation is what appears in Table~\ref{table:CellTable} as $\langle P_\mathrm{A} \rangle/\pal$.
\section{Direct measurements of target properties and performance\label{sec:targetproperties}}
Using some or all of the techniques described in the previous section, we characterized each of 24 target cells and summarize our results in Table~\ref{table:CellTable}, grouped according to the experiments for which the targets were constructed. The first group of cells, made for the saGDH experiment,  were made with Rb only.  These data from non-hybrid target cells provided something of a baseline, and thus aided in the design and construction of targets for the GEN experiment, the first of our target cells that utilized alkali-hybrid technology.  The lessons learned from the GEN cells allowed us to optimize our approach to the alkali-hybrid technology, as well as providing an opportunity to study the benefits of spectrally narrowed diode lasers.   The last group of targets was constructed for the Transversity and $d_2^n$ experiments, and show the highest performance of all.  Collectively, the data listed in Table~\ref{table:CellTable} have provided critical data for the design of the next generation of polarized $^3$He targets that will be used in experiments following the $\rm 12\,GeV$ energy upgrade of JLab that is in progress at the time of this writing.

At the simplest level, the most important two measures of the performance of a particular polarized $^3$He target are the saturation $^3$He polarization ($P_\mathrm{pc}^\infty$, because we typically monitored the pumping chamber) and the time constant that characterizes the buildup of polarization ($\Gamma_\mathrm{s}^{-1}$).  
To understand {\it why} a target achieves a certain level of performance, however, additional measurements are required.  
In the earliest group of cells studied, those built for the saGDH experiment, the only additional parameter measured was  $\langle\Gamma\rangle_\mathrm{c}^{-1}$, the time constant characterizing the cell-averaged spin-relaxation rate at room temperature.
In the later cells studied, however, we were able to study additional parameters, and achieved a significantly improved understanding of the factors influencing performance.
In this section, we focus on parameters that were measured {\it directly}.  The so-called $X$ parameter, which can be inferred from direct measurements, will be discussed in a later section.

\begin{table*}\scriptsize
\begin{center}
\begin{tabular}{|c|c|ccc|ccc|ccccc|cc|c|}
\hline
\multirow{2}{*}{\begin{sideways}{EXP}\end{sideways}}&\multirow{2}{*}{Cell} & \multirow{2}{*}{Lasers} & $I_0$ & $T_\mathrm{pc}^\mathrm{set}$ & \multirow{2}{*}{$P_\mathrm{pc}^\infty$} & $\Gamma_\mathrm{s}^{-1}$ & $\langle\Gamma\rangle_\mathrm{c}^{-1}$ & \multirow{2}{*}{$\langle P_\mathrm{A} \rangle/\pal$} & \multirow{2}{*}{$\pal$} & \multirow{2}{*}{$D_\mathrm{fr}$} & \multirow{2}{*}{$D_\mathrm{pb}$} & [Rb]$_\mathrm{fr}$ & $\Delta T_\mathrm{Rb}$ & $\Delta T_\mathrm{He}$ & \multirow{2}{*}{$X$}\\
&& & W/cm$^2$ & $^\circ$C & & hrs & hrs & & & & & $10^{14}$/cm$^3$ & $^\circ$C & $^\circ$C &\\
\hline
\hline
\multirow{5}{*}{\begin{sideways}saGDH\end{sideways}} & Proteus & 3B & 3.8 & 180 & 0.46 & 27 & 74 & - & - & 0 & 0 & - & - & - & -\\
\cline{2-16}
& Peter & 3B & 3.8 & 180 & 0.44 & 21 & 56 & - & - & 0 & 0 & - & - & - & -\\
\cline{2-16}
& Penelope & 3B & 3.8 & 180 & 0.39 & 18 & 46 & - & - & 0 & 0 & - & - & - & -\\
\cline{2-16}
& Powell & 3B & 3.8 & 180 & 0.38 & 13 & 25 & - & - & 0 & 0 & - & - & - & -\\
\cline{2-16}
& Prasch & 3B & 3.8 & 180 & 0.33 & 13 & 33 & - & - & 0 & 0 & - & - & - & -\\
\hline
\hline
\multirow{20}{*}{\begin{sideways}GEN\end{sideways}} & \multirow{2}{*}{Al} & 2.5B & 3.2 & 235 & 0.53(03) & 7.86(05) & 27.42(1.37) & - & - & - & 4.53(25) & - & - & - & - \\
& & 5B & 6.1 & 235 & 0.54(03) & 6.73(18) & 27.42(1.37) & - & - & - & 4.53(25) & - & - & - & - \\
\cline{2-16}
& \multirow{2}{*}{Barbara} & 2.5B & 1.6 & 235 & 0.37(02) & 5.50(08) & 42.95(2.15) & - & - & - & 4.80(25) & - & - & - & - \\
& & 5B & 3.1 & 235 & 0.57(03) & 4.76(63) & 42.95(2.15) & - & - & - & 4.80(25) & - & - & - & - \\
\cline{2-16}
& Gloria & 3B & 1.7 & 235 & 0.60(03) & 6.13(04) & 38.29(1.91) & - & - & - & 7.20(40) & - & - & - & - \\
\cline{2-16}
& \multirow{2}{*}{Anna} & 1B & 0.6 & 235 & 0.33(02) & 5.60(34) & 11.38(57) & - & - & - & 9.64(57) & - & - & - & - \\
& & 1.5B & 1.0 & 235 & 0.39(02) & 5.37(08) & 11.38(57) & - & - & - & 9.64(57) & - & - & - & - \\
\cline{2-16}
& \multirow{2}{*}{Dexter} & 1.5B & 1.5 & 235 & 0.47(02) & 7.58(17) & 18.45(92) & - & - & - & - & - & - & - & - \\
& & 5B & 6.1 & 235 & 0.49(02) & 6.63(12) & 18.45(92) & - & - & - & - & - & - & - & - \\
\cline{2-16}
& Edna & 3B & 2.4 & 235 & 0.56(03) & 5.71(02) & 27.42(1.37) & - & - & - & 3.63(20) & - & - & - & - \\
\cline{2-16}
& \multirow{2}{*}{Dolly} & 3B & 1.0 & 235 & 0.43(02) & 6.16(03) & 35.24(1.76) & - & - & - & 20(1.3) & - & - & - & - \\
& & 1N1B & 1.4 & 235 & 0.62(03) & 5.79(07) & 35.24(1.76) & - & - & - & 20(1.3) & - & - & 17(10) & - \\
\cline{2-16}
& \multirow{3}{*}{Simone} & 2N1B & 3.8 & 215 & 0.31(01) & 14.08(06) & 22.87(1.14) & 0.947(020)  & 0.91(05) & 10.66(54) & 8.89(45) & 0.20(02) & -7(3) & - & -0.04(12)$^\star$ \\
& & 2N1B & 3.8 & 240 & 0.48(02) & 6.89(20) & 22.87(1.14) & - & - & - & 9.76(49) & - & - & - & - \\
& & 2N1B & 3.8 & 255 & 0.58(02) & 6.45(10) & 22.87(1.14) & 0.929(023) & 0.92(05) & 12.48(83) & 10.3(52) & 0.90(09) & -4(5) & - & 0.11(06)$^\star$ \\
\cline{2-16}
& \multirow{5}{*}{Sosa} & 2N1B & 1.9 & 160 & 0.57(02) & 16.69(09) & 73.68(3.68) & 0.966(020) & 1.00(03) & 0 & 0 & 1.97(13) & 4(1) & 30(7) & 0.24(06)$^\dagger$ \\
 &  & 2N1B & 1.9 & 170 & 0.61(03) & 11.67(04) & 73.68(3.68) & 0.964(020) & 0.98(03) & 0 & 0 & 3.00(33) & 3(3) & 38(14) & 0.27(06)$^\star$ \\
 &  & 2N1B & 1.9 & 180 & 0.55(02) & 8.79(09) & 73.68(3.68) & 0.954(022) & 0.97(03) & 0 & 0 & 4.30(27) & 1(2) & 47(7) & 0.43(06)$^\dagger$ \\
 &  & 2N1B & 1.9 & 190 & 0.40(02) & 6.39(22) & 73.68(3.68) & 0.854(075) & 0.82(03) & 0 & 0 & 5.69(63) & -2(3) & 48(20) & 0.58(12)$^\star$ \\
 &  & 2N1B & 1.9 & 200 & 0.26(01) & 5.04(17) & 73.68(3.68) & - & - & 0 & 0 & - & - & 43(18) & - \\
\hline
\hline
\multirow{12}{*}{\begin{sideways}Transversity and $d_2^n\ \ \ \ \ \ \ \ \ \ $\end{sideways}} & Boris & 3B & 1.8 & 235 & 0.42(02) & 6.25(04) & 23.74(1.19) & 0.871(050) & 0.79(07) & 1.96(18) & 2.45(23) & 2.19(34) & -8(7) & - & 0.26(10)$^\star$ \\
\cline{2-16}
& \multirow{2}{*}{Samantha} & 3B & 1.8 & 235 & 0.50(02) & 6.30(13) & 36.51(1.83) & - & - & - & 4.34(23) & - & - & - & -\\
& & 3N & 2.6 & 235 & 0.68(03) & 4.62(03) & 22.13(1.11) & 0.956(020) & 0.99(03) & 4.37(10) & 4.34(23) & 1.80(10) & 7(2) & 21(10) & 0.12(05)$^\star$\\
\cline{2-16}
& Alex & 2N1B & 2.6 & 235 & 0.59(03) & 4.81(02) & 32.96(1.65) & 0.942(042) & 0.99(03) & 1.37(08) & 1.19(07) & 4.08(36) & 0(4) & 42(10) & 0.34(06)$^\dagger$ \\
\cline{2-16}
& Moss & 1N1B & 1.8 & 235 & 0.62(03) & 5.35(04) & 33.00(1.65) & - & 0.95(09) & - & 2.40(13) & - & - & 29(8) & -\\
\cline{2-16}
& Tigger & 1N1B & 1.8 & 235 & 0.51(02) & 4.89(05) & 12.62(63) & - & 0.95(09) & - & - & - & - & 23(9) & -\\
\cline{2-16}
& Astral & 2N1B & 2.6 & 235 & 0.69(03) & 6.57(12) & 48.90(2.45) & 0.954(020) & 0.99(03) & 7.09(55) & 6.21(56) & 0.97(09) & 3(5) & 25(4) & 0.17(05)$^\dagger$\\
\cline{2-16}
& Stephanie & 3N & 2.6 & 235 & 0.63(03) & 4.55(09) & 48.35(2.42) & 0.929(114) & 0.99(03) & 1.39(11) & 1.50(10) & 5.08(58) & 7(5) & 54(6) & 0.31(08)$^\star$\\
\cline{2-16}
& \multirow{3}{*}{Brady} & 1N & 0.9 & 235 & 0.62(03) & 4.82(1.08) & 33.50(1.68) & - & 0.95(03) & - & 2.36(24) & -  & - & 14(9) & -\\
& & 2N & 1.8 & 235 & 0.68(03) & 5.52(70) & 33.50(1.68) & - & 0.99(03) & - & 2.36(24) & - & - & 25(8) & -\\
& & 3N & 2.6 & 235 & 0.70(03) & 5.30(01) & 33.50(1.68) & 0.956(021) & 0.99(03) & 2.60(20) & 2.36(24) & 2.86(30) & 6(5) & 39(9) & 0.14(05)$^\dagger$\\
\cline{2-16}
& Maureen & 3N & 2.6 & 235 & 0.66(03) & 5.42(12) & 29.21(1.46) & - & 0.97(09) & - & 4.42(55) & - & - & 32(12) & -\\
\cline{2-16}
& \multirow{3}{*}{Antoinette} & 3N & 1.7 & 215 & 0.49(02) & 6.63(37) & 20.93(1.05) & 0.958(020) & 0.99(03) & 2.85(13) & - & 0.96(07)  & 0(3) & 16(8) & 0.28(08)$^\dagger$\\
& & 3N & 1.7 & 235 & 0.61(03) & 4.18(10) & 20.93(1.05) & 0.936(043) & 0.99(03) & 3.32(27) & - & 1.83(20) & 0(5) & 20(10) & 0.24(07)$^\dagger$\\
& & 3N & 1.7 & 255 & 0.41(02) & 2.66(11) & 20.93(1.05) & 0.776(099) & 0.93(10) & 3.57(23) & - & 2.88(39) & -5(6) & 33(9) & 0.55(13)$^\dagger$\\
\hline
\end{tabular}
\end{center}
\caption{Cell Performance for three sets of experiments: saGDH (top), GEN (middle), and Transversity \& $d_2^n$ (bottom). 
Within each experiment grouping, data is sorted by type of laser used (B = Broadband, N = Narrowband). 
$I_0$ is the nominal incident laser intensity at the center of the pumping chamber.
$T_\mathrm{pc}^\mathrm{set}$ is the oven set temperature.
$P_\mathrm{pc}^\infty$ is the equilibrium polarization in the pumping chamber and $\Gamma_\mathrm{s}$ is the slow time constant extracted from the five parameter fit to the polarization build up curve.
$\Gamma_\mathrm{c}$ is the cell-averaged room temperature spin relaxation rate.
$\left< P_\al \right>/P_\al^\ell$ is the volume averaged to line averaged alkali polarizaiton ratio determined from the optical pumping simulation.
$P_\al^\ell$ is the measured line averaged alkali polarization.
$D_\mathrm{fr}$ \& $D_\mathrm{pb}$ are the K to Rb density ratios determined from Faraday rotation and pressure broadening measurements.
$[\mathrm{Rb}]_\mathrm{fr}$ is the Rb number density measured from Faraday rotation.
$\Delta T_\mathrm{Rb}$ is the temperature of Rb inferred from the number density relative to the oven set temperature.
$\Delta T_\mathrm{He}$ is the temperature of $\mathrm{^3He}$ inferred from temperature tests relative to the oven set temperature.
$X$ is the best combined value for the $X$-factor.
$^\star$ indicates $X$ was measured using only spinup, alkali polarization, and Faraday rotation data.
$^\dagger$ indicates $X$ was also measured using the early-time behavior of the spinup.
}
\label{table:CellTable}
\end{table*}

\subsection{The effect of alkali-hybrid mixtures\label{sec:IVA}}
\begin{figure}[htbp]
\hskip -0.16truein \includegraphics[width = 5.4cm, angle=90]{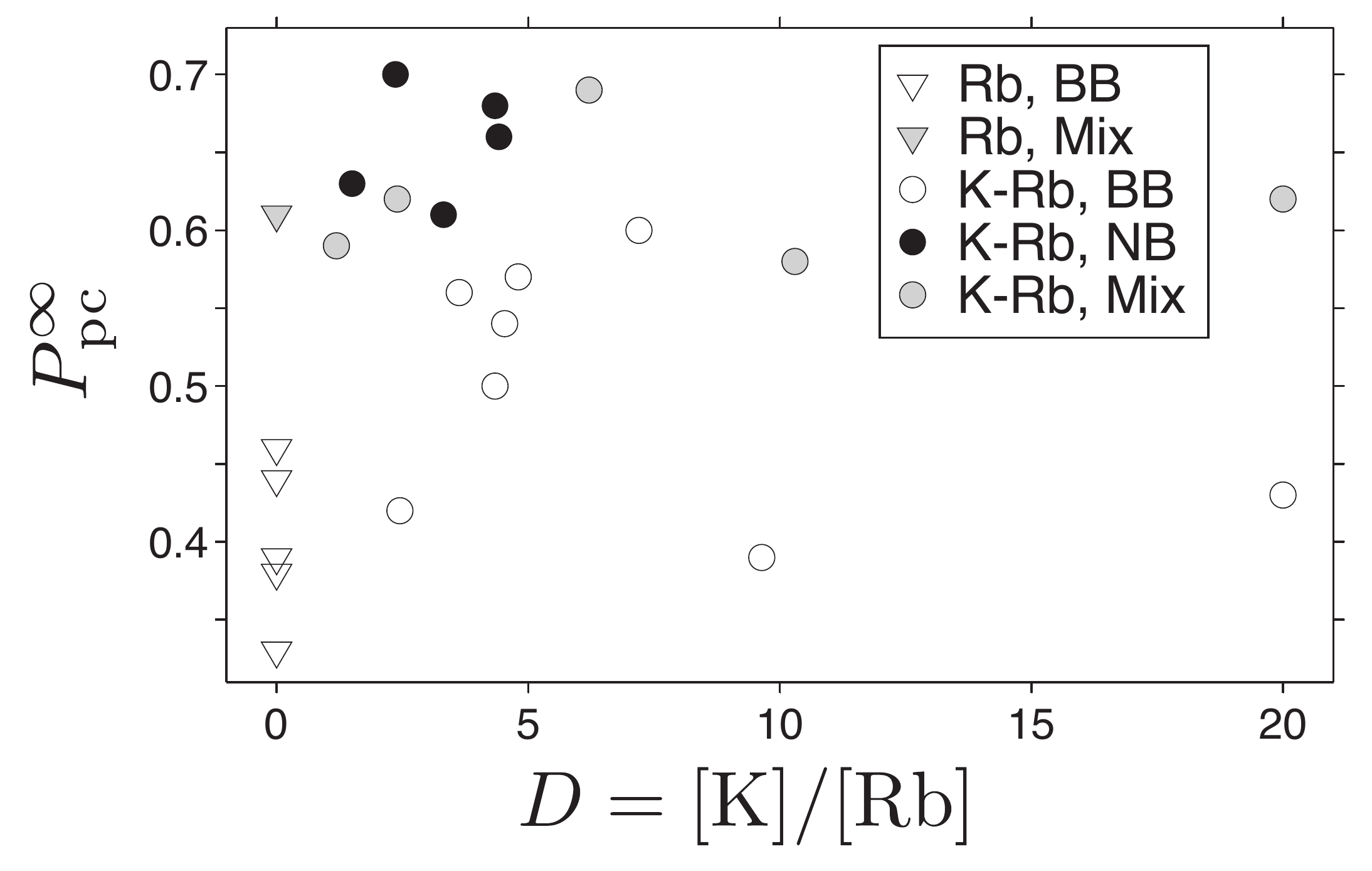}
 \vskip -0.1truein
 \caption{Maximum Achieved He-3 Polarization as a Function of $D$ for target cells included in this study.
 Open triangles (circles) correspond to pure Rb (alkali hybrid) cells pumped with broadband light.  Solid and grey data points correspond to cells pumped with all or at least some spectrally narrowed lasers.  Despite the fact that these data were acquired over an extended period of years under various different operating conditions, the value of both the alkali-hybrid mixtures as well as spectrally narrowed light is clearly evident. The values for $D$ in this table were extracted from pressure-broadening data, as more cells were measured this way.}
\label{fig:cellPerform}
 \end{figure}

The impact of using of alkali-hybrid mixtures in our target cells was dramatic, and is illustrated in Fig.~\ref{fig:cellPerform}, which plots the maximum $^3$He saturation  polarization achieved as a function of the alkali-hybrid density ratio, $D$, for each of the 24 target cells tested during our studies.  To isolate the impact of alkali-hybrid mixtures from other (laser-related) factors, it is useful to consider only those tests that used broadband laser light, which included both cells containing Rb only, shown with open triangles (that necessarily appear at $D=0$), and cells with alkali-hybrid mixtures, shown with open circles.  For Rb-only cells, the highest polarization achieved was 46\%, which corresponded to the cell Proteus as can be seen in Table~\ref{table:CellTable}.  In contrast, alkali-hybrid cells often achieved over 50\%, and in the case of Gloria, 60\%.  

It is also interesting to compare the data in Fig.~\ref{fig:cellPerform} to the simulations summarized in Fig.~\ref{fig:PAD}.  Experimentally, there were too many variables involved in each measurement to expect that we would reproduce the smooth functional dependence evident in Fig.~\ref{fig:PAD}.  Nevertheless, it is clear that the best performance observed is correlated with a relatively narrow range of $D$ values, very roughly between 3 and 7.  Interestingly, again considering only those cells pumped with broadband light, the tests of cells falling outside of that range (Boris, with $D\sim2.2$, Anna, with $D\sim9.6$ and Dolly with $D\sim20$), appear to be limited in their performance as would be expected from the simulation.

\begin{figure}[htbp]
 \begin{center}
\includegraphics[width = 5.5cm, angle=90]{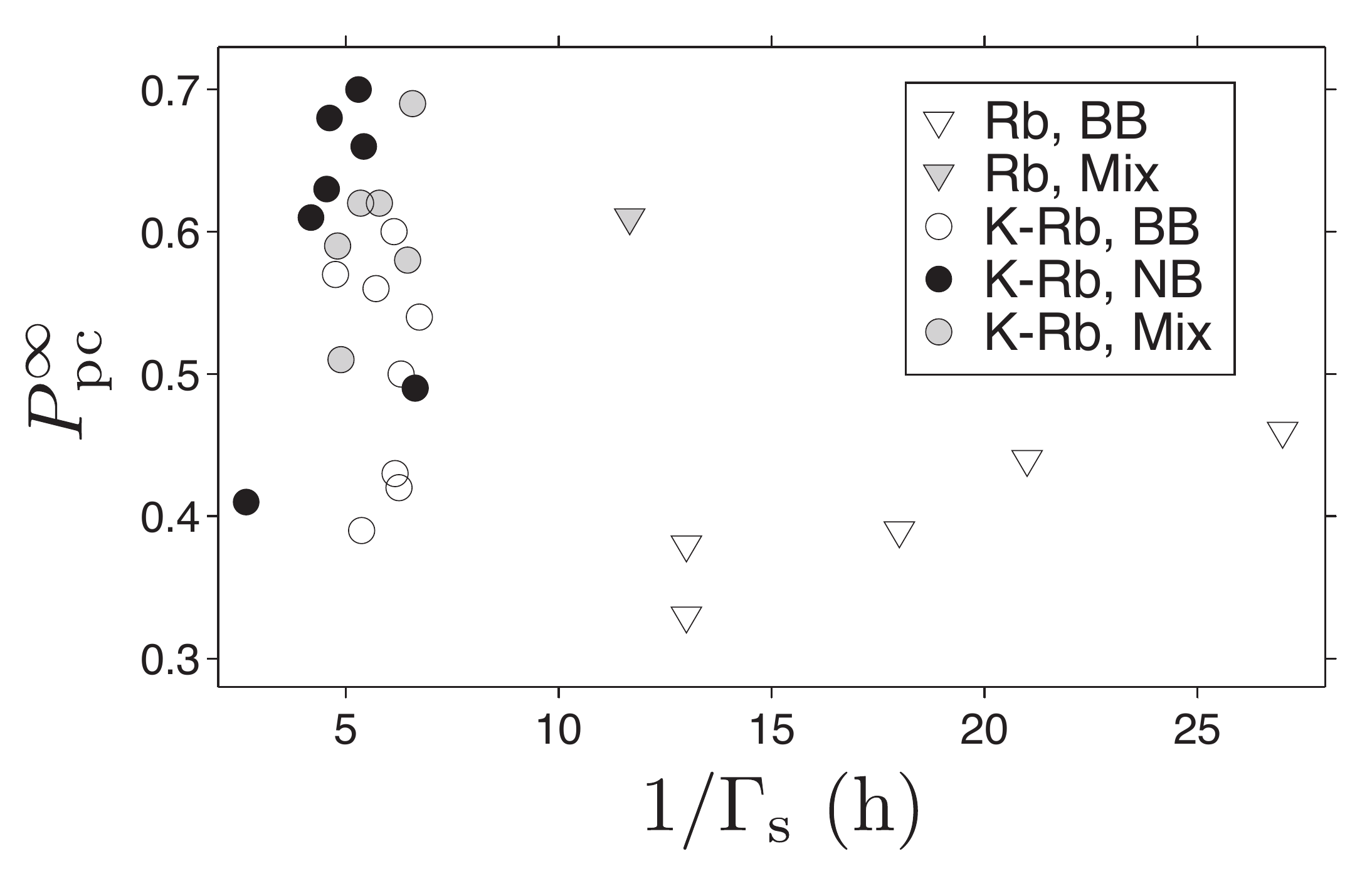}
 \end{center}
 \vskip -0.1truein
 \caption{Maximum Achieved He-3 Polarization as a function of $1/\Gamma_\mathrm{s}$ for target cells included in this study.
Triangles correspond to pure Rb cells, circles correspond to alkali-hybrid cells.  The key indicating the meaning of the different types of data points is the same as in Fig.~\ref{fig:cellPerform}.  The optimal conditions for running hybrid cells clearly correspond to smaller values of $1/\Gamma_\mathrm{s}$ than is the case for pure Rb cells.}
\label{fig:bigCellPerform}
 \end{figure}

\subsection{Impact of using narrowband Lasers}
As discussed in Section~\ref{sec:theory}, the significant benefits of alkali-hybrid SEOP arise from the more efficient use of the optical-pumping photons, which in turn makes it possible to operate at significantly higher alkali number densities than would otherwise be possible.  Higher alkali number densities result in faster spin exchange, which in turn makes it easier to overwhelm spin-relaxation mechanisms that would otherwise drive the polarization lower. This effect is clearly evident in Fig.~\ref{fig:bigCellPerform}, which shows   data on the maximum $P_\mathrm{pc}^\infty$ achieved versus $\Gamma_\mathrm{s}^{-1}$ for each cell studied under the various optical pumping conditions also considered in Fig.~\ref{fig:cellPerform}.  If, for example, we again focus our attention on the open circles and open triangles that correspond to illumination with broadband lasers, we see that the optimal values of $\Gamma_\mathrm{s}^{-1}$ for the hybrid cells were significantly smaller than was the case for the pure Rb cells, and that in all but three cases (corresponding to Dolly, Boris and Anna, the cells with $D$ outside the optimal range), they achieved higher polarizations than even the best pure Rb cell (when it was pumped with broadband lasers).  Instead of spinups characterized by time constants of around 10--20 hours, the time constants for the hybrid cells were more typically 4--6 hours.  Finally, it is worth noting that smaller values of $\Gamma_\mathrm{s}^{-1}$ also make the target less susceptible to depolarization from the electron beam.

The advantages of using spectrally-narrowed diode laser arrays are quite evident in both Figs.~\ref{fig:cellPerform} and \ref{fig:bigCellPerform}, where the highest polarizations obtained were from alkali-hybrid cells pumped using either narrow-band lasers (solid black circles) or a mix of narrow-band and broadband lasers (solid grey circles).  As discussed in Section~\ref{sec:theory}, the origins of the improved performance are twofold.  Firstly, light that is closer to the Rb D1 line center results in  higher optical pumping rates, and hence higher alkali polarizations.  Secondly, it becomes possible to use higher alkali densities (and hence higher spin-exchange rates) while still maintaining high alkali polarization. This second point is particularly evident in Fig.~\ref{fig:bigCellPerform} in which many of the smallest values of $\Gamma_\mathrm{s}^{-1}$ are associated with cells pumped with at least some narrow-band lasers.  We note that the benefits of the narrowband light are not limited to the alkali-hybrid cells.  The best performance from Rb-only cells were also obtained while using narrow-band lasers.
 
 
\subsection{Optimization and limits on polarization}
  
As the results presented in the previous two subsections make clear, significant improvements in the performance of the spin-exchange polarized $^3$He  result from using alkali-hybrid mixtures and spectrally-narrowed lasers.  For optimal performance, it is best to have the ratio of the K to Rb number densities in the general range of 3--7, and also to operate the cells with relatively small values of the spin-up time constant $\Gamma_\mathrm{s}^{-1}$.  Since the highest polarizations measured were around 70\%, however, it is clearly desirable to better understand the limits on further improvement.

As discussed earlier, Babcock \etal showed that cells used for SEOP of $^3$He suffer from a temperature-dependent spin-relaxation mechanism that can be characterized by the parameter $X$~\cite{PhysRevLett.96.083003}.  Because this relaxation increases with $\gamma_\mathrm{se}$, it cannot be overcome by running the cell ``harder" at higher temperatures.  Indeed, if we assume that $X$ is proportional to $\gamma_\mathrm{se}$, the limit of the polarization of the $^3$He at high values of   $\gamma_\mathrm{se}$ can be expressed as
\begin{equation}
\lim_{\gamma_\mathrm{se}\rightarrow \infty} P_\mathrm{He} = \lim_{\gamma_\mathrm{se}\rightarrow \infty}\frac{\langle P_\al \rangle \langle \gamma_\mathrm{se}\rangle} {\langle \gamma_\mathrm{se}\rangle(1+X) + \langle \Gamma \rangle} = \frac{\langle P_\al\rangle}{1+X}\ \ .
\end{equation}
The use of alkali-hybrid cells and the use of narrow-band lasers has made it  easier to achieve faster spin-exchange rates $\gamma_\mathrm{se}$ while maintaining high alkali polarization.  This has meant that the $X$ parameter associated with a cell has increasingly become the main limiting factor on performance.  It is thus of considerable importance to know the value of $X$ associated with a given target, and we describe our efforts to measure this parameter in Section~\ref{sec:X}.  For these studies, however, it is also important to have a good handle on the coefficients that govern spin exchange, which is the subject of the next section.
\section{The K-$^3$He spin-exchange rate constant\label{sec:kse}}
It was shown by Dolph \etal that at sufficiently small values of time, the polarization in the pumping chamber can be written
\begin{equation}
P_\mathrm{\pc} = \gamma_\mathrm{\se}\langle P_\al\rangle\!\left (t\!-\!t_0\right ) + b\!\left (t\!-\!t_0\right )^2 =  m_\pc t + b t^2 + c
\label{eq:ppc_seop_constant}
\end{equation}
where it is assumed here that the $^3$He polarization passes through zero at $t=t_0$, and we have defined the quantity $m_\pc \equiv\gamma_\mathrm{\se}\langle P_\al\rangle$  as the coefficient of the linear term in the above equation.  Indeed, measurements by Dolph \etal taken during the first 20-30 minutes of a spinup were shown to be extremely linear (see, for example, Fig.~3 of Ref.~\cite{convection}), so much so that it is hard to see the influence of the quadratic term at all.  

We have performed a set of dedicated spinups during which an NMR AFP signal was taken every 3 minutes to determine the slope $m_\pc$.  Care was taken to account for small AFP losses during measurements.  Also, since AFP measurements cause the $^3$He spins to be temporarily aligned opposite to the direction in which they are being polarized, we were careful to take into account the time during which the spins were ``anti-aligned".  We will refer to measurements of $m_\pc$ determined during such a dedicated spinup as     $m_\pc^s$, where the superscript $s$ denotes that this quantity was measured during a spinup.     

It is also possible to compute the expected value for $m_\pc$ using entirely separate measurements.  The Faraday rotation methods described in Section~\ref{section:farrot} provide us with a measure of the alkali number densities.  We further expect  
\begin{equation}
\gamma_\mathrm{\se} = k_\se^{\rb}[\rb] + k^\kp_\se[\kp]\ \ , 
\label{eq:gammase}
\end{equation}
where  $k_{\se}^{\rm Rb}$($k_{\se}^{\rm K}$) is the constant characterizing the spin-exchange rate between $^3$He and Rb (K).    Finally, we know the volume-averaged alkali polarizations from our measured line-averaged polarizations together with small corrections from our simulation.  We will refer to values of $m_{\pc}$ calculated in this manner as $m_{\pc}^F$, where the superscript $F$ denotes that this quantity was computed using the Faraday rotation data.

\begin{figure}[htbp]
\begin{center}
\includegraphics[width = 8.7cm]{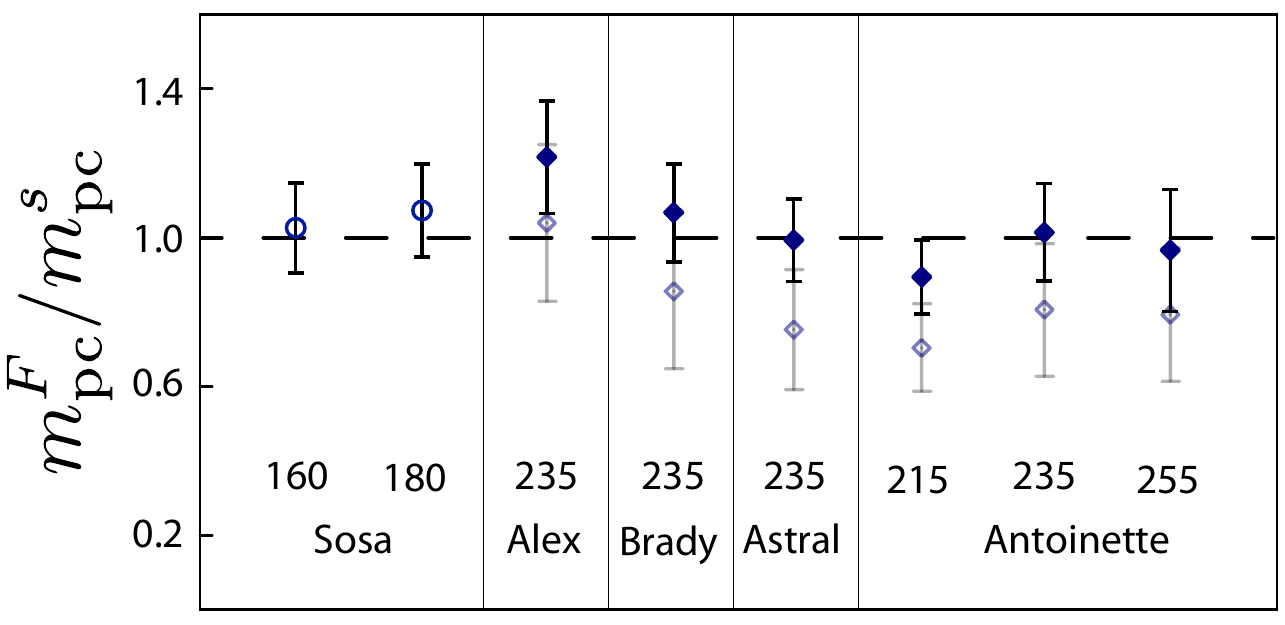}
\vskip -0.1truein
\caption{(color online) Plotted is the ratio $m_{\pc}^F/m_{\pc}^s$ for eight separate measurements, where $m_{\pc}^s$ is the slope measured at the beginning of a spinup, and $m_{\pc}^F$ is calculated using Eqn.~(\ref{eq:ppc_seop_constant}), Faraday-rotation determinations of the alkali densities, and the spin-exchange coefficients $k_{\se}^{\rm Rb}$ and $k_{\se}^{\rm K}$.  For all but the Sosa (Rb only) measurements, $k_{\se}^{\rm K}$ was treated as a free parameter, while fitting $m_{\pc}^F/m_{\pc}^s$ to unity, yielding the result of Eqn.~(\ref{eq:kseK}).}
\label{fig:consistency}
\end{center}
\end{figure}

From the above discussion, we expect the ratio $m_{\pc}^F/m_{\pc}^s$ to be equal to one, where $m_{\pc}^F$ and  $m_{\pc}^s$ are measured for a particular cell under identical conditions.  To compute this ratio, however, we must know both $k_{\se}^{\rm Rb}$, which has been measured and reported in the literature multiple times, and $k_{\se}^{\rm K}$, for which we are aware of only one measurement described in the Ph.D. thesis of Babcock~\cite{babcock}.  For $k_{\se}^{\rm Rb}$, we combine the measurement due to Baranga \etal~\cite{bar98} and the even more accurate measurements from Chann \etal~\cite{PhysRevA.66.032703} to find $k_{\se}^{\rm Rb} = (6.79\pm 0.14)\times10^{-20}\rm cm^3/s$.  We choose these particular measurements because they are insensitive to systematic effects associated with the temperature-dependent relaxation mechanism characterized by the $X$ parameter.  In Fig.~\ref{fig:consistency}, for the two measurements associated with the Rb-only cell Sosa, we show the resulting values for $m_{\pc}^F/m_{\pc}^s$ with solid circles, and the ratio is seen to be quite close to unity in both cases.

When we compute $m_{\pc}^F/m_{\pc}^s$ for cells that contain K in addition to Rb, we need a value for $k_{\se}^{\rm K}$.  The number that appears in Babcock's thesis is $(5.5\pm 0.4)\times10^{-20}\rm cm^3/s$~\cite{babcock}.  We note that this result is also referenced in a later paper by the same group, with an error that is improved by a factor of two~\cite{walker2010}. We found, however, that the resulting value for the ratio $m_{\pc}^F/m_{\pc}^s$ came out to be less than unity in all but one of the six cases we studied, as is shown with the open diamonds in Fig.~\ref{fig:consistency}.  Among other things, this causes an inconsistency between two of the methods for computing $X$ that we discuss in the next section.  We were thus led to consider fitting our alkali-hybrid data for the ratio $m_{\pc}^F/m_{\pc}^s$ to unity while treating $k_{\se}^{\rm K}$  as a free parameter.  The result, for cells Alex, Brady, Astral and Antoinette, is shown in Fig.~\ref{fig:consistency} with the solid diamonds and yields
\begin{equation}
\label{eq:kseK}
 k_{\se}^{\rm K} = (7.46\pm 0.62)\times10^{-20}\rm cm^3/s\ \ .
 \end{equation} 
 Also shown in Fig.~\ref{fig:consistency} are the oven set temperatures at which the measurements were made. The values of $D$ for each cell can be found in Table~\ref{table:CellTable}.
 
Our value for $k_\mathrm{se}^\mathrm{K}$ is significantly higher than Babcock's, and the cause is unclear.
One possibility may be temperature dependence.
Taking the oven set temperatures shown on Fig.~\ref{fig:consistency}, and adding to them the values of $\Delta T_\mathrm{He}$ as shown on Table~\ref{table:CellTable}, 
our measurements of $m_{\pc}^F/m_{\pc}^s$ were obtained over the temperature range $230-290^\circ\rm C$.
This temperature range appears to be significantly higher than that at which Babcock measured $k_\mathrm{se}^\mathrm{K}$, suggesting a possible explanation for the difference.
Both our measurement at $T \approx 260^\circ\mathrm{C}$ and Babcock's measurement at $T \approx 190^\circ\mathrm{C}$ are consistent with the temperature dependence of $k_\se^\kp$ recently calculated by Tscherbul \etal \cite{tvt2011}. With this said, we note that the coefficient that characterizes Rb-$^3$He spin exchange, $k_\mathrm{se}^\mathrm{Rb}$, does not appear to depend strongly on temperature based on the measurements of Baranaga \etal \cite{bar98} and Chann \etal \cite{PhysRevA.66.032703}. 
In the next section, we use our own determination of $k_{\se}^{\rm K}$ rather than that due to Babcock in one of the four methods we employed to measure the $X$ parameter.  We make this choice both because our determination of $k_{\se}^{\rm K}$ was made under similar operating conditions to our other measurements, and also because this choice significantly improves the internal consistency of our data, which was important for some of the other effects we studied.
\section{The X Factor\label{sec:X}}
As has already been discussed, the so-called $X$ factor is an important intrinsic property of a target cell.  Unlike most of the properties discussed in Section~\ref{sec:targetproperties}, however, which were measured fairly directly, the $X$ factor is a derived quantity, and its determination relies on our theoretical understanding of the cell's behavior, including the relevant constants that characterize spin exchange.  

\subsection{Measuring X\label{sec:measureX}}
The measurements we performed while characterizing our target cells provided sufficient data to determine $X$ in several different ways.
This provided us with a sense of the self consistency of our data, and also made it possible to combine our different determinations into an appropriately weighted average. 
Since several of the methods used to determine $X$ are performed at a single temperature, we were also able to search for a possible temperature dependence.

In the following two sections, each method used to determine $X$ relies on the cell-averaged spin relaxation rate $\left < \Gamma \right>$ at operating temperatures.
We have assumed that the only difference between $\left < \Gamma \right >$ and $\left < \Gamma \right >_\mathrm{c}$, which is measured at room temperature, is the change in the cell-averaged \hens-\hens\ dipolar spin relaxation rate when the cell is heated from 
room temperature to operating temperature.
This correction is calculated by 
\begin{equation}
\left < \Gamma \right >\! =\! \left < \Gamma \right >_\mathrm{c}\! -\! \left [ n_0\! -\! f_\pc n_\pc/\!f^d\!(t_\pc)\! -\! f_\tc n_\tc/\!f^d\!(t_\tc) \right]\!/\tau^d \label{eqn:dipcorr}
\end{equation}
where $n_0$ is the \he fill density, $n_{\pc(\tc)}$ the \he density in the pumping (target) chamber, $f_{\pc(\tc)}$ is the fraction of \he atoms in the pumping (target) chamber, 
$t_\pc = T_\pc/(296.15\ \mathrm{K})$, $t_\tc = (313.15\ \mathrm{K})/(296.15\ \mathrm{K})$, $\tau^d = 744\ \mathrm{hrs \cdot amg}$ \cite{new93}, and $f^d(t)$ is a function that parameterizes the temperature dependence of the dipolar relaxation from Appendix D.5.1 of \cite{singh}.
Under operating conditions, $\left < \Gamma \right >$ is usually only a few percent smaller than $\left < \Gamma \right >_\mathrm{c}$.
In doing this, we are implicitly assuming that the relaxation rate due to collisions with the walls is the same for the two chambers and equal to the value measured at room temperature.
This point is discussed in more detail in Sec.~\ref{sec:ptd}.
Finally, whenever the difference $\left ( \Gamma_\mathrm{s} - \left< \Gamma \right > \right )$ appears in the following sections, 
we include in $\left < \Gamma \right >$ a small additional correction to account for the measured NMR AFP losses at operating temperatures.

\subsubsection{The hot relaxation method}
\begin{figure}[htbp]
 \begin{center}
\hskip -0.01truein \includegraphics[width = 8.6cm]{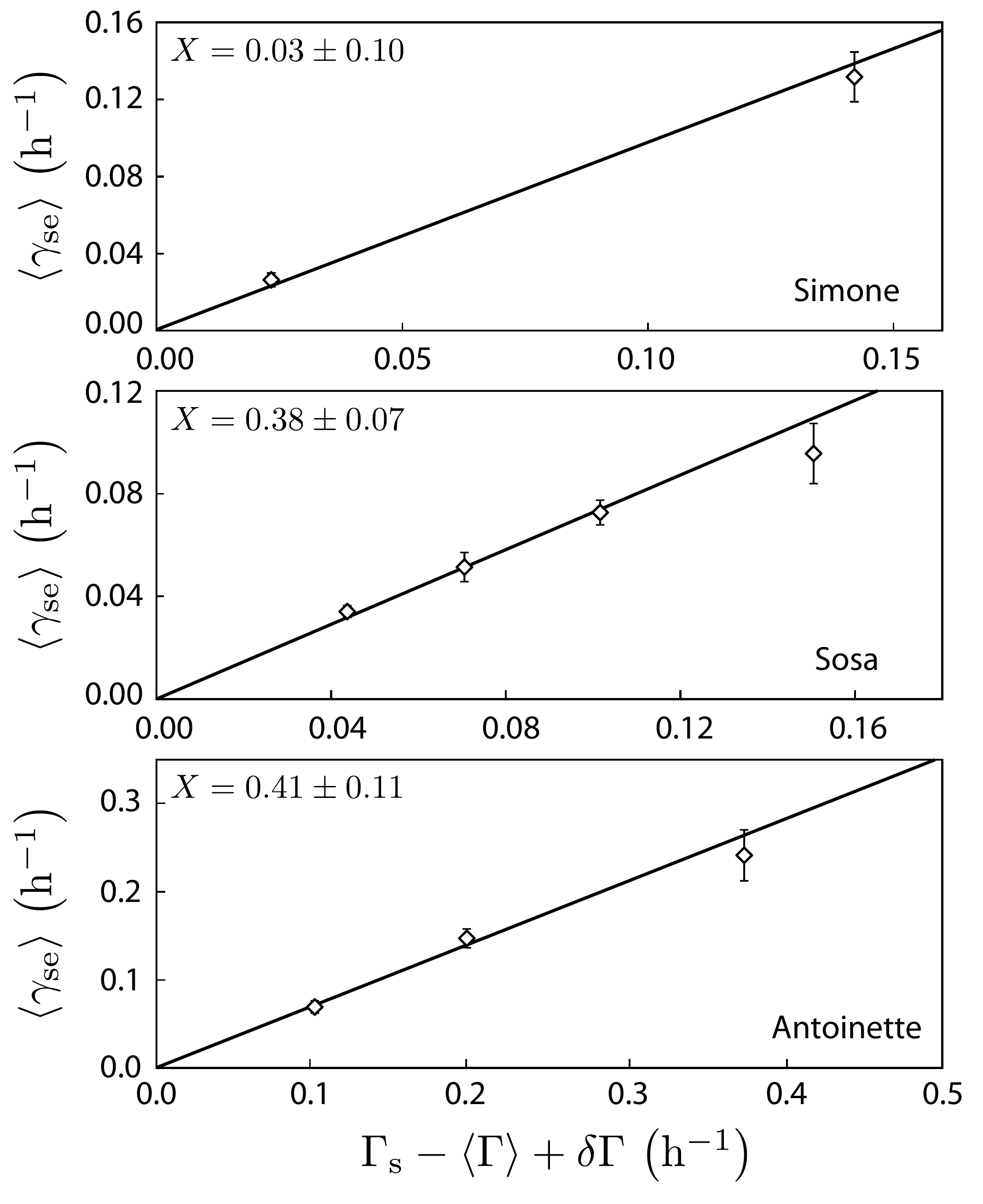}
 \end{center}
 \vskip -0.25truein
 \caption{We plot the cell-averaged spin-exchange rate, $\langle \gamma_\mathrm{\se}\rangle$, as determined using Faraday rotation and measured values of the spin-exchange constants $k_\mathrm{\se}^\mathrm{Rb}$ and $k_\mathrm{\se}^\mathrm{K}$, versus the quantity $\Gamma_\mathrm{s} - \langle \Gamma \rangle - \delta\Gamma$ for three cells, as labeled.  For $k_\mathrm{\se}^\mathrm{K}$, we use the value given in Eqn.~(\ref{eq:kseK}).  Also shown are linear fits to the data, constrained to go through zero.  The values quoted for $X$ are the inverse of slopes of the lines minus one. The errors bars on each data point are the uncorrelated errors.  The error quoted on each value of $X$ includes the uncertainty in our determination of $k_\mathrm{se}^\mathrm{K}$.}
\label{fig:ThadPlot}
 \end{figure}

The first method we describe for determining $X$ is what was described in Ref.~~\cite{PhysRevLett.96.083003} as the ``hot relaxation method."  Rearranging Eqn.~(\ref{eqn:gammaS}), we see that  $\langle\gamma_\mathrm{\se}\rangle$ can be expressed as follows:
\begin{equation}
\langle\gamma_\mathrm{\se}\rangle = {{\Gamma_\mathrm{s} - \langle\Gamma\rangle + \delta\Gamma}\over{(1+X)}}\ \ .
\label{eqn:gammaSE}
\end{equation}
To extract a value for $X$, we plot $\langle\gamma_\mathrm{\se}\rangle$ as a function of $\Gamma_\mathrm{s} - \langle\Gamma\rangle + \delta\Gamma$, and the slope of a linear fit to the data is expected to be equal to $1/(1+X)$.
In Fig.~\ref{fig:ThadPlot}, we present data collected using the hot relaxation method for three target cells.  
Also shown in In Fig.~\ref{fig:ThadPlot} are the three fits, which we note were constrained to pass through the origin.  
The resulting values of $X$ from the three fits are also shown on the figure, and, two of the three are seen to be significantly different from zero.
Because of the large uncertainty, a strong statement can not be made about Simone, the cell with the smallest $X$ value.

The quantity $\delta\Gamma$ is a relatively small correction that appears because we are working with a double-chambered cell.  It is given approximately by
\begin{equation}
\delta\Gamma \approx f_\mathrm{\pc}f_\mathrm{tc}(d_\mathrm{\pc} + d_\mathrm{tc})u^2 + \mathrm{higher\ order\ terms}
\end{equation}
where $d_\mathrm{\pc}$ and $d_\mathrm{tc}$ are transfer rates describing the probability per unit time that a particular $^3$He atom will exit the pumping and target chambers respectively, and can be computed using the geometry of the cell, its fill pressure, and information on the temperatures at which it is operated (see Dolph \etal~\cite{convection}).  The quantity $u$ is given by
\begin{equation}
u = {{\gamma_\mathrm{\se}(1+X)+\Gamma_\mathrm{\pc}-\Gamma_\mathrm{tc}}\over{d_\mathrm{\pc} + d_\mathrm{tc}}}
\end{equation}
where $\Gamma_\mathrm{\pc}$ and $\Gamma_\mathrm{tc}$ are the spin-relaxation rates in the pumping and target chambers respectively.  The awkward thing here is that $X$, the quantity we are seeking to determine using the hot relaxation method, appears in the quantity that is the abscissa of our plot.  This is not a problem, however, because as mentioned earlier, $\delta\Gamma$ is typically 10\% or less of the size of $\Gamma_\mathrm{s}$.  What we do in practice is to take $X = 0$ initially, and after finding a value for $X$,  use that value to recompute $\delta\Gamma$ and find an improved value for $X$. This process can be iterated a few times, and quickly converges to a stable value for $X$.

\subsubsection{Single Temperature Methods}
One drawback of the hot relaxation method is that it necessarily assumes that the temperature dependence of the new relaxation mechanism is identical to the temperature dependence of $\gamma_\mathrm{\se}$.  It is quite possible, however, to extract values for $X$ at a single temperature, and even search for a temperature dependence in the $X$ parameter itself.  It is also the case that making measurements such as those shown in Fig.~\ref{fig:ThadPlot} are time consuming, and we only carried out such measurements for a small subset of the target cells studied. 
All but the second method described in this section are based on the ``polarization method'' described in Ref.~\cite{PhysRevLett.96.083003}.
The only difference is that, for the third and fourth methods, we incorporated information about the product $P_\al\,\left < \gamma_\mathrm{se} \right >$ from early time measurements of the polarization buildup.
The second method described in this section is essentially a single point version of the ``hot relaxation method'' described in Ref.~\cite{PhysRevLett.96.083003}.

The first and simplest ``single-temperature" method for measuring $X$, the result from which we label as $X_1$, requires measurements of $\langle P_\al \rangle$, $P^\infty_\pc$, $\langle\Gamma\rangle$, and $\Gamma_\mathrm{s}$.  Here we start with an equation for the equilibrium $^3$He polarization in a double chambered cell:
\begin{equation}
P_{\pc}^\infty = {{\langle P_\mathrm{A} \rangle \langle \gamma_\mathrm{\se} \rangle }\over{\Gamma_\mathrm{s} + \delta\Gamma -  \delta\Gamma^{\prime}}}
\label{eq:pc_equilibrium}
\end{equation}
where $\delta\Gamma^{\prime} = f_\tc \Gamma_\mathrm{tc}^2/(\Gamma_\mathrm{tc} + d_\mathrm{tc})$.  
Eqn.~(\ref{eq:pc_equilibrium}) is essentially Eqn.~(11) from Ref.~\cite{convection} with the one difference that $\gamma_\mathrm{se}$ in the denominator has been replaced with $\gamma_\mathrm{se}(1+X)$.
We note that $\delta\Gamma^{\prime}$ is typically only a few percent of the size of $\Gamma_\mathrm{s}$.   Next we use Eqn.~(\ref{eqn:gammaSE}) for $\langle\gamma_{\se}\rangle$, substitute into Eqn.~(\ref{eq:pc_equilibrium}), and solve for $X_1$:
\begin{equation}
X_1 = \frac{\langle P_\mathrm{A}\rangle}{P_\mathrm{\pc}^\infty}\left(\frac{\Gamma_\mathrm{s} - \langle\Gamma\rangle + \delta\Gamma}{\Gamma_\mathrm{s} +  \delta\Gamma -  \delta\Gamma^{\prime}}\right) - 1\ \ .
\label{eqn:X_1}
\end{equation}
Again we have the issue that $\delta\Gamma$, which depends on $X$, appears in the expression.  We can iterate, however, beginning with $X=0$, to converge on a stable value.

The second method, the result from which we label $X_2$, requires measurements of $\langle\Gamma\rangle$, $\Gamma_\mathrm{s}$, [Rb], and $D$.
We solve Eqn.~(\ref{eqn:gammaS}) for $X$ to get 
\begin{equation}
X = \frac{\Gamma_\mathrm{s} - \langle\Gamma\rangle + \delta\Gamma}{\langle\gamma_{\se}\rangle}-1
\label{eqn:X_s2}
\end{equation}
and then plug in $\gamma_\mathrm{\se}$ from Eqn.~(\ref{eq:gammase}) to find
\begin{equation}
X_2 = \frac{\Gamma_\mathrm{s} - \langle\Gamma\rangle + \delta\Gamma}{f_{\pc}k_{\se}^{\rb}[\rb]\left(1 + D'\right)}-1\ \ .
\label{eqn:X_2}
\end{equation}
Again we have the issue with $\delta\Gamma$ which is handled the same way as with $X_1$. 
Here, however, we also have the issue of needing to know $k_{\se}^\mathrm{K}$, which as discussed earlier, is not known as accurately as is $k_{\se}^\mathrm{Rb}$. 
We have chosen in this case to use our own value for $k_{\se}^\mathrm{K}$, with its accompanying large error, since this provides the best self-consistency in our data.  

The third method, the result from which we call $X_3$, is very similar to the second method and requires measurements of $\langle P_\al \rangle$, $\langle\Gamma\rangle$, $\Gamma_\mathrm{s}$, and $m_{\pc}^s$. 
Again we start with Eqn.~(\ref{eqn:gammaS}), but now we evaluate $\left < \gamma_{\se} \right > = f_\pc m^s_\pc/\left < P_\al \right >$ from the linear term in Eqn.~(\ref{eq:ppc_seop_constant}) to get
\begin{equation}
X_3 = \left < P_\al \right > \frac{\Gamma_\mathrm{s} - \langle\Gamma\rangle + \delta\Gamma}{f_{\pc}m_{\pc}^s} - 1\ \ .
\label{eqn:X_3}
\end{equation}
For the measurements we present on the cell Sosa, which is pure Rb, $X_2$ and $X_3$ represent truly independent determinations of $X$.  In cells for which $D\ne0$, however, $X_2$ and $X_3$ are highly correlated, since we determined $k_{\se}^\mathrm{K}$ through measurements of $m_\mathrm{\pc}$.

The fourth method, the results from which we call $X_4$, requires measurements of $\langle P_\mathrm{A}\rangle$, $P_\mathrm{\pc}^\infty$, $\left < \Gamma \right >$, and $m_{\pc}^s$.
We obtain the needed expression by plugging Eqn.~(\ref{eqn:gammaS}) into Eqn.~(\ref{eq:pc_equilibrium}), solving for $X$, and evaluating $\gamma_\se$ from linear term of Eqn.~(\ref{eq:ppc_seop_constant}) to get
\begin{equation}
X_4 = \frac{\langle P_\mathrm{A} \rangle}{P_\mathrm{\pc}^\infty}-\frac{\langle P_\mathrm{A} \rangle(\langle\Gamma\rangle - \delta\Gamma^\prime)}{f_{\pc}m_{\pc}^s}-1 \ \ .
\label{eqn:X_4}
\end{equation}

In Table~\ref{table:X}, we show the values of $X$ for each cell and temperature for which we have adequate data, together with the corresponding errors.  The different values of $X$ are quite consistent with one another, even when correlations between errors are taken into account.  We note that $X_1$ is quite consistent with the other methods for determining $X$, despite being completely independent of both $m_\mathrm{\pc}$ and $k_{\se}^\mathrm{K}$.  We found that that was much less the case when we used Babcock's value for  $k_{\se}^\mathrm{K}$.  While it is a bit hard to quantify, we see this as additional evidence supporting our value for $k_{\se}^\mathrm{K}$ as being reasonable.

We also show in Table~\ref{table:X} a ``best value'' for $X$ obtained by an appropriate weighted average of either $X_1$ and $X_2$ ( referred to as $X_{12}$), or all the values of $X$ ($X_{1234}$).
The errors were assigned to the best of $X$ by taking a weighted average of the results from the different methods while taking into account the correlations among the methods.

\begin{table}\scriptsize
\begin{tabular}{|c|c|cccc|c|}
\hline
Cell & $T(\mathrm{^oC})$ & $X_1$ & $X_2$ & $X_3$ & $X_4$ & $X_{12}$/$X_{1234}$ \\
\hline
\multirow{2}{*}{Simone} & 215 & -0.02(12) & -0.10(14) & - & - & -0.04(12)\\
 & 255 & 0.13(08) & 0.08(09) & - & - & 0.11(06)\\
\hline
\multirow{4}{*}{Sosa} & 160 & 0.22(07) & 0.28(09) & 0.32(15) & 0.18(09) &0.24(06)$^\dagger$ \\
 & 170 & 0.24(07) & 0.37(15) & - & - & 0.27(06)\\
 & 180 & 0.45(08) & 0.40(09) & 0.50(17) & 0.45(09) & 0.43(06)$^\dagger$\\
 & 190 & 0.59(16) & 0.57(17) & - & - & 0.58(12)\\
\hline
\hline
Boris & 235 & 0.21(14) & 0.31(14) & - & - & 0.26(10)\\
\hline
Sam. & 235 & 0.08(06) & 0.22(09) & - & - & 0.12(05)\\
\hline
Alex & 235 & 0.34(09) & 0.35(09) & 0.63(20) & 0.29(10) &0.34(06)$^\dagger$ \\
\hline
Astral & 235 & 0.15(07) & 0.22(10) & 0.20(14) & 0.14(07) &0.17(05)$^\dagger$ \\
\hline
Steph. & 235 & 0.31(17) & 0.31(10) & - & - & 0.31(08)\\
\hline
Brady & 235 & 0.13(07) & 0.15(09) & 0.23(14) & 0.11(07) &0.14(05)$^\dagger$ \\
\hline
\multirow{3}{*}{Antoinette}
 & 215 & 0.27(09) & 0.44(17) & 0.30(19) & 0.25(11) &0.28(08)$^\dagger$ \\
 & 235 & 0.20(09) & 0.34(12) & 0.36(17) & 0.15(09) &0.24(07)$^\dagger$ \\
 & 255 & 0.55(26) & 0.54(16) & 0.50(30) & 0.56(26) &0.55(13)$^\dagger$\\
\hline
\end{tabular}
\caption{Shown are values of the $X$ factor for the indicated cells at the indicated oven set temperatures.  Either two or four separate methods (all described in the text) were used to compute $X$ in each case. The final column represents a  combined best value for $X$, which is either $X_{12}$ or $X_{1234}$, depending on whether two or four different values for $X$ were available. A $\dagger$ indicates combined values for which we were able to compute $X_{1234}$.}
\label{table:X}
\end{table}

To the best of our knowledge, there has not been a dedicated study of the $X$ factors, and the temperature-dependent relaxation mechanism they characterize, with a large number of cells using measurements of the alkali polarization since the original work by Babcock \etal~\cite{PhysRevLett.96.083003}.  The results presented in table~\ref{table:X} can thus be viewed as an independent verification of the existence of what we might call the $X$-factor mechanism.  For temperatures in the range at which we operate our targets, the $X$ factors imposed limits to the $^3$He polarization of 62--88\%. The highest polarizations measured, however, were around 70\%.  We note that we rarely pushed our targets to the highest possible temperatures because we did not want to risk damaging, or even worse, destroying a target. 

\subsection{Possible temperature dependence\label{sec:ptd}}
\begin{figure}[htbp]
\begin{center}
\hbox{\hskip -0.05truein \includegraphics[width = 8.9cm]{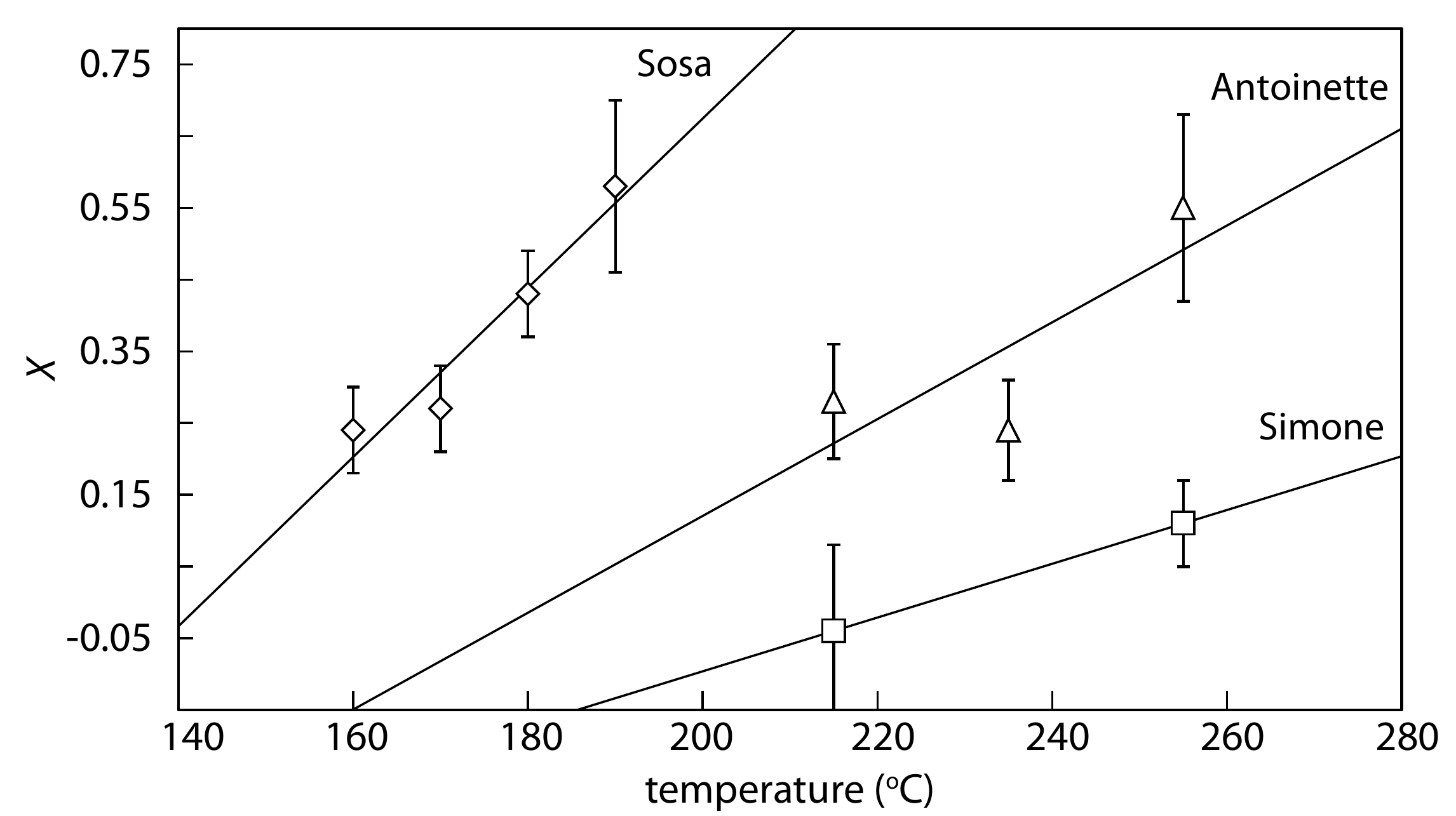}}
\end{center}
\vskip -0.1truein
\caption{Shown is the best-value for the $X$-factor (either $X_{12}$ or $X_{1234}$ from Table~\ref{table:X}) versus temperature for the three cells Sosa, Simone and Antoinette.  Also shown is a linear fit to each set of data.}
\label{fig:t_dependence}
\end{figure}

As can be seen from Table~\ref{table:X}, we have determinations of $X$ at multiple temperatures for three of our cells.  In these cases, we can thus pose the question of whether or not $X$ is constant with temperature.  Some variation, after all, would mean only that the relaxation mechanism associated with $X$ had a temperature dependence slightly different from that of $\gamma_\mathrm{\se}$, a possibility explicitly mentioned by Babcock \etal~\cite{PhysRevLett.96.083003}.  
Indeed, as can be seen in Fig.~\ref{fig:t_dependence}, for the cells Simone, Sosa (pure Rb) and Antoinette, when we plot $X$ as a function of temperature, we see what appears to be systematic variation with temperature.

If we assume a linear dependence of $X$ with temperature, we find for the case of Sosa a slope of $(0.012 \pm 0.002)/^\circ\rm C$, a slope that is six sigma away from zero.  In the case of Antoinette, the slope is $(0.007\pm0.005)/^\circ\rm C$, just over one sigma from zero.  In the case of Simone we have only two temperature points, but the trend still seems to be present.  While these data can hardly be viewed as providing conclusive evidence regarding the temperature dependence of $X$, they are certainly suggestive.

When we first noticed the temperature dependence of $X$, we were concerned that we could be viewing a systematic effect of some sort rather than physics.  For example, since our cells had two chambers, the distribution of the gas between those two chambers (as reflected in $f_\mathrm{pc}$ and $f_\mathrm{tc}$) changes at different temperatures.  We found, however, that the uncertainties in $f_\mathrm{pc}$ and $f_\mathrm{tc}$ had only negligible effects on the observed temperature dependence of $X$.  In fact, as we analyzed our data, the identification and elimination of various systematic effects only caused the temperature dependence to become more pronounced while simultaneously decreasing the scatter in our data.   We mention below two of the potential systematic effects that we considered.

One possible systematic is associated with the fact that the calculation of $X$ requires knowing $\langle \Gamma \rangle$ at operating temperatures, which is the sum of cell-averaged dipolar relaxation rate $\left < \Gamma^d \right >$ and 
the cell-averaged wall relaxation rate $\left < \Gamma^w \right >$.
At the beginning of Sec.~\ref{sec:measureX}, we discussed how we accounted for the temperature and density dependences of the dipolar relaxation rate by using Eqn.~(\ref{eqn:dipcorr}).
The wall relaxation rate is more subtle.
Following Dolph \etal \cite{convection}, we note that the cell-averaged wall relaxation rate can be expressed as $\left < \Gamma^w \right > = \Gamma_\pc^w \left ( R f_\tc + f_\pc \right )$, where $\Gamma_\pc^w$ is the wall
relaxation rate in the pumping chamber and $R$ is the ratio of the target to pumping chamber wall relaxation rates.
As a starting point, we set $\Gamma_\pc^w$ equal to the room temperature wall relaxation rate given by $\left < \Gamma \right >_\mathrm{c} - n_0/\tau^d$ where $n_0$ \& $\tau^d$ were defined in Sec.~\ref{sec:measureX}.
We then considered a range of the ratio $R$ between one and a value close to three.  
We found that the resulting variation in the values of $X$ was fairly minimal, resulting in the average value of $X_{1234}$, for example, decreasing by an average amount of 0.04 (absolute).
Again, there was essentially no effect on the temperature dependence of $X$. 
We note that in Table~\ref{table:X}, we have chosen to take $R=1$, which is equivalent to $\left < \Gamma^w \right > = \Gamma_\pc^w = \left < \Gamma \right >_c - n_0/\tau^d$.
This particular choice was motivated by some recent studies of convection-driven target cells of the sort described in Ref.~\cite{convection}, 
in which it was possible to obtain a measure of the relative size of $\Gamma_\pc$ and $\Gamma_\tc$ by observing the ratio of the polarizations in the target and pumping chamber, $P_\tc/P_\pc$, 
as a function of both time and the speed with which convection is driven.

We also considered the possibility that our choice of a value for $k_\mathrm{se}^\mathrm{K}$ might introduce an apparent temperature dependence for $X$.  In particular, we considered both our own value for $k_\mathrm{se}^\mathrm{K}$ from Eqn.~(\ref{eq:kseK}), as well as that due to Babcock~\cite{babcock}.  We found that the temperature dependence of $X$ was present with either choice,  although our values for $X$ were significantly more self consistent when using our own value for $k_\mathrm{se}^\mathrm{K}$.  Perhaps a more interesting question is whether a temperature dependence  in $k_\mathrm{se}^\mathrm{K}$ could cause the apparent temperature dependence in $X$.  Naively, this seems to be ruled out by the fact that a temperature-dependent value for $k_\mathrm{se}^\mathrm{K}$ would produce different behaviors in our calculated values of $X_1$, $X_2$, $X_3$ and $X_4$.  As can be seen in Table~\ref{table:X}, the temperature dependence of the different calculations of $X$ appear to be similar to one another within errors.  

One possible temperature-dependent contribution to the $X$-factor is anisotropic spin exchange \cite{walker2010}.
Recently, Tscherbul \etal \cite{tvt2011} have calculated that anisotropic spin exchange contributes about only about $0.03$ to the $X$-factor due to \kp-\he spin-exchange collisions for our operating temperatures.
Although their calculations indicate that the anisotropic spin exchange contribution to the $X$-factor has a temperature dependence, it is very small in the temperature range relevant to our measurements ($\approx 10^{-4}/^\circ\mathrm{C}$) above $T=463\ \mathrm{K}$.

In summary, we were unable to identify a plausible explanation for our observations other than an actual temperature dependence in $X$.  While we hesitate to suggest that our observations are conclusive, they certainly provide motivation for further study.  If $X$ factors increase with increasing temperature (as suggested by Fig.~\ref{fig:t_dependence}), the limits that they impose on $^3$He polarization may be more severe than had previously been assumed.  
\section{Conclusion\label{sec:con}}
\begin{figure}[htbp]
\includegraphics[width = 8.6cm]{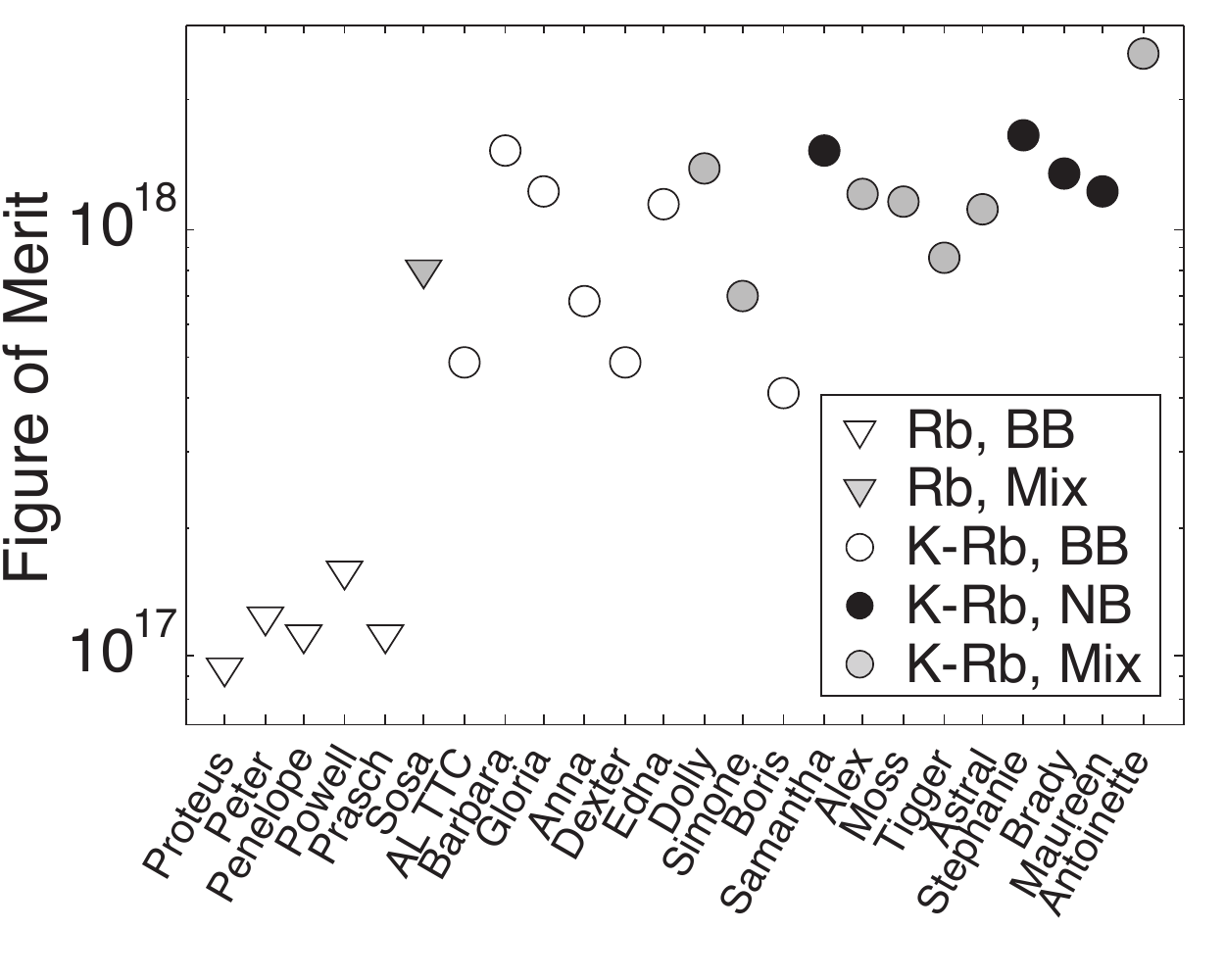}
\vskip -0.1truein
\caption{Shown for all 24 cells included in our study is the best value actually measured for the Figure of Merit equal to $V\,\rho\,\Gamma_\mathrm{s}(P_\mathrm{pc}^\infty)^2$ where $\Gamma_\mathrm{s}$ and $P_\mathrm{pc}^\infty$ are given in Table~\ref{table:CellTable}, and $V$ and $\rho$ are the  volume and fill density of the cell respectively, both given by Table~\ref{table:fill}.  Notice that the vertical axis is logarithmic.}
\label{FigureOfMeritA}
\end{figure}

We have presented data obtained while developing polarized $^3$He targets, based on spin-exchange optical pumping, for four separate experiments at Jefferson Laboratory in Newport News.  Data are included from 24 glass target cells, and they clearly demonstrate the substantial gains that were made possible through the use of hybrid mixtures of Rb and K, and the use of spectrally-narrowed high-power diode laser arrays.  One measure of these gains is the figure of merit discussed in the introduction, $\mathcal{L}^\mathcal{N}$, that essentially represents the number of spins polarized per second, weighted by the square of polarization.  In Fig.~\ref{FigureOfMeritA} we plot $\mathcal{L}^\mathcal{N}$ for all 24 of the target cells studied.  The points, labeled according to the cell in which $\mathcal{L}^\mathcal{N}$ was measured, are arranged in roughly chronological order according to when $\mathcal{L}^\mathcal{N}$ was measured. The reader should note the logarithmic scale.  

We have also presented results from a numerical simulation of optical pumping that incorporated several important effects that have recently been established as being quite important.
These simulations provided an improved understanding of how it is that hybrid alkali mixtures and spectrally narrowed lasers contribute to better target performance, and guided us in optimizing various design and operational parameters, including the optimal ratio of the K number density to the Rb number density.
The simulations were also useful in connecting the ``line-averaged" alkali polarization, which we measured experimentally, to the ``volume-averaged" alkali polarization, which is the critical quantity in determining the ultimate polarization of the $^3$He.  
Having benchmarked our simulation against real data, we also have a valuable tool for the ongoing design of future targets.  
For the few cases where the volume-averaged alkali polarization was much less than the line-averaged alkali polarization ($\left < P_\al \right>/P_\al^\ell < 0.9$), the line-averaged alkali polarization from experiment was, with large uncertainties, systematically lower than those calculated from the model. We further note that a new study \cite{chen-new} seems to indicate that the EPR technique for measuring line-averaged alkali polarizations overestimates the true volume-averaged alkali polarization \footnote{We thank the Editors of J. App. Phys. for bringing this new study to our attention.}. More rigorous tests of our simulation under a more diverse set of operating conditions are necessary. These further studies are planned and include comparisons with cells with lower $^3\mathrm{He}$ densities \& higher $D$ and dedicated measurements of $P_\mathrm{He}$ as a function of temperature.

In nine of the target cells studied, we have performed a careful determination of the $X$ factors that characterize the as-yet poorly understood relaxation mechanism that limits the maximum polarization of $^3$He targets polarized using SEOP.  We believe that this is the first careful study of $X$ factors since the work of Babcock \etal~\cite{PhysRevLett.96.083003}, and we report  unambiguous evidence confirming the $X$-factor mechanism as a dominant limiting factor in the $^3$He polarization we have achieved.  We furthermore see hints of a non-zero temperature dependence of the $X$ factor itself, although a definitive confirmation of this would require additional work.

Finally, in the course of our studies, we have made a measurement of $k_\mathrm{se}^\mathrm{K}$, the coefficient that characterizes \kp--\he spin-exchange.  We find a value somewhat larger than that found by Babcock~\cite{babcock}, which could be explained
by a temperature dependence of $k_\se^\mathrm{K}$.

When compared to the first liter-scale polarized $^3$He targets used in electron scattering at SLAC~\cite{PhysRevLett.71.959},  we report herein an increase in the FOMs $\mathcal{L}^\mathrm{eff}$ and $\mathcal{L}^\mathcal{N}$ (defined in the introduction) of 16 and 35 respectively.  While our primary motivation in these studies  was the development and construction of $^3$He targets for four experiments, we have nevertheless obtained data that are of considerable value to those using spin-exchange optical pumping for various applications.  The studies presented here also provide a critical foundation for the next generation of spin-exchange polarized $^3$He targets that are under development for future experiments at JLab.

\begin{acknowledgments}
This work was supported by the U.S. Department of Energy (DOE), Office of Science, Office of Nuclear Physics under Contract No. DE-FG02-01ER41168 at the University of Virginia and
under Contract No. DE-FG02-96ER41003 at the College of William \& Mary.
One of the authors (J.T.S.) would like to thank the Directorate of Argonne National Laboratory for generously supporting his postdoctoral fellowship while at ANL.
\end{acknowledgments}


\end{document}